\documentclass[12pt]{article}

\newcommand{\rem}[1]{}
\newcommand{\remv}[1]{}

\newcommand{\gbl}{[\![\;}
\newcommand{\gbr}{\;]\!]}

\newcommand{\commentstarts}{\begin{centering}
\hspace{-1pt}\vrule\vrule
\begin{minipage}[t]{0.03\linewidth}
\hspace{0.025\linewidth}
\end{minipage}
\begin{minipage}[t]{0.95\linewidth}}
\newcommand{\commentends}{\end{minipage}
\end{centering}
\vspace{7pt}
}
\newcommand{\afbox}[1]{
  \setbox1=\hbox{\hspace{-3pt}\scriptsize $ #1 $\hspace{-3pt}}
  \setbox2=\vbox{\hsize=\wd1 \vspace{-2pt} \box1 \vspace{-2pt}}
  \fbox{\box2}
}

\usepackage{amsmath}
\usepackage{latexsym}
\newcommand{\be}{\begin{equation}}
\newcommand{\ee}{\end{equation}}
\newcommand{\bea}{\begin{eqnarray}}
\newcommand{\eea}{\end{eqnarray}}
\newcommand{\nn}{\nonumber \\}
\newcommand{\half}{\frac{1}{2}}
\newcommand{\p}{\varphi_1}
\newcommand{\pp}{\varphi_2}

\usepackage{xcolor}
\definecolor[named]{Andrei}{rgb}{0.80,0.21,0}
\usepackage[letterpaper]{geometry}
\begin{document}

\begin{titlepage}
\begin{flushright}
IF-USP 1652\\
IPMU10-0070\\
ITEP-TH-20/10
\end{flushright}

\begin{center}
{\Large\bf $ $ \\ $ $ \\
Notes on $\beta$-deformations of the pure spinor superstring 
in $AdS_5\times S^5$
}\\
\bigskip\bigskip\bigskip
{\large Oscar A. Bedoya${}^{\dag}$, L. Ibiapina Bevil\'{a}qua${}^{\dag}$, Andrei Mikhailov${}^{\ddag}$ \\
and Victor O. Rivelles${}^{\dag}$}
\\
\bigskip
{\it ${}^\dag$ Instituto de F\'{i}sica, Universidade de S\~{a}o Paulo, \\ 
C.P. 66.318 CEP 05315-970, S\~{a}o Paulo, SP, Brasil \\ \& \\
${}^\ddag$ Institute for the Physics and Mathematics of the Universe \\
University of Tokyo, Kashiwa, Chiba 277-8582, Japan \\
\& \\
${}^\ddag$ Institute for Theoretical and 
Experimental Physics, \\
117259, Bol. Cheremushkinskaya, 25, 
Moscow, Russia}\\

\vskip 1cm
\end{center}

\begin{abstract}
We study the properties of the vertex operator for the $\beta$-deformation of the
superstring in $AdS_5\times S^5$ in the pure spinor formalism. We discuss the
action of supersymmetry on the infinitesimal $\beta$-deformation, the application
of the homological perturbation theory, and the relation between the worldsheet
description and the spacetime supergravity description.
\end{abstract}

\end{titlepage}

\newpage

\tableofcontents

\section{Introduction}

Historically, the development of the pure spinor formalism was mostly
concentrated on the special case of flat space. But in fact the flat
space case is a degenerate case.  In many ways the general
background is qualitatively different, the flat space being a special
degenerate limit. The general, ``typical'' background has a non-degenerate
Ramond-Ramond bispinor field. Among such non-degenerate examples the most
symmetric one is $AdS_5\times S^5$. Therefore the study of this background
is important for the string theory in general.  

During the last several years,  continuous progress has been made in this
direction. One of the observations made recently in \cite{Berkovits:2008ga} is
that the pure spinor Lagrangian is invariant under the action of the global
symmetry group $PSU(2,2|4)$. This is in contrast with the case of flat space,
where the Lagrangian is invariant only up to total derivatives. This observation
was generalized in \cite{Mikhailov:2009rx} where it was argued that 
the vertex operators for massless supergravity states can be
 chosen in a $PSU(2,2|4)$-covariant way. 

At this time there are two explicit examples of vertices: the
vertex for the zero mode of the dilaton (the descent of the Lagrangian) introduced
in \cite{Berkovits:2008qc} and the vertex for the $\beta$-deformation 
introduced in \cite{Mikhailov:2009rx}. In this paper we will study the vertex for the
$\beta$-deformation. We will be mostly concerned with the following subjects: 
\begin{itemize}
\item how the supersymmetry acts on $\beta$-deformations
\item extension of an infinitesimal $\beta$-deformation
   to a finite $\beta$-deformation; the homological
   perturbation theory
\item the space-time picture
\end{itemize}
First steps towards the pure spinor description of the $\beta$-deformed
$AdS_5\times S^5$ were made in \cite{Grassi:2006tj}, although our approach
is somewhat different\footnote{The authors of \cite{Grassi:2006tj} followed
the method of twisted boundary conditions previously used in
\cite{Frolov:2005ty,Frolov:2005dj,Frolov:2005iq} in the context
of Green-Schwarz approach. We are  using a more straightforward
approach, using the vertex operator and
 the homological perturbation theory.
}. 

We will now briefly outline our paper.

\subsection{Deformations of the pure spinor action}
\label{sec:IntroDeformations}
The Type IIB string worldsheet theory, in the pure spinor formulation,
has the following structure:
\begin{enumerate}
\item An action $S$ which is assumed to be local and conformally invariant;
\item A pair of BRST operators $Q_L$ and $Q_R$ with the properties:
\[
Q_L^2 = Q_R^2 = \{Q_L, Q_R\} = 0.
\]
The ``total'' BRST operator $Q$ is the sum of $Q_L$ and $Q_R$:
\[
Q = Q_L + Q_R \;;
\]
\item Two ghost number operators $\mbox{gh}_L$ and $\mbox{gh}_R$, such
   that $\mbox{gh}_L(Q_L) = 1$, $\mbox{gh}_L(Q_R)=0$, $\mbox{gh}_R(Q_L)=0$,
   and $\mbox{gh}_R(Q_R) = 1$;
\item The composite $b$-ghost $b_{++}$, $b_{--}$, which satisfy:
\[
\{Q, b_{++}\}=T_{++} \quad , \quad \{Q, b_{--}\} = T_{--}.
\]
\end{enumerate}
Given a worldsheet theory with these axioms satisfied, we ask ourselves:
how can such a theory be deformed?
It turns out that the infinitesimal deformations are parametrized by
{\em integrated vertex operators}\footnote{The subindex $1$ in $V^{(2)}_1$ 
show that this is the 1-st infinitesimal deformation, and the
superindex $(2)$ indicates a 2-form.} $V^{(2)}_1$:
\begin{eqnarray}
S & = & S_0 + \varepsilon \int V^{(2)}_1
\nonumber
\\[3pt]
Q & = & Q_0 + \varepsilon Q_1,
\end{eqnarray}
where $S_0$ is the undeformed original action, invariant under $Q_0$.

The integrated vertex operator should be a total derivative under the original BRST transformation:
\begin{equation}\label{ConditionAtFirstOrder}
Q_0V^{(2)}_1 \simeq d(\mbox{smth}).
\end{equation}
where $\simeq$ means that ``equals on-shell''. 
The condition (\ref{ConditionAtFirstOrder}) guarantees that the
deformed action is BRST-invariant at the first order; notice that the BRST 
transformation itself gets deformed, unless (\ref{ConditionAtFirstOrder}) 
is satisfied off-shell (which is usually not the case).

Generally speaking, given the first infinitesimal deformation $V_1^{(2)}$,
it should be possible to construct the series:
\begin{eqnarray}\label{PerturbedActionSeries}
S_{exact} & = & 
S_0 + \varepsilon \int V_1^{(2)} + \varepsilon^2 \int V_2^{(2)} + \ldots
\nonumber
\\[3pt]
Q_{exact} & = &
Q_0 + \varepsilon Q_1 + \varepsilon^2 Q_2 + \ldots
\end{eqnarray}
and obtain the full deformed theory.

\subsection{Special case of $\beta$-deformation}
In this paper we will consider an example: the so-called 
{\em $\beta$-deformations}. These deformations were introduced 
in field theories by Leigh and Strassler in \cite{Leigh:1995ep}. 

\subsubsection{First order in $\varepsilon$}
Let us first consider the $\beta$-deformation at the linearized level.
In the pure spinor formalism the corresponding vertex
operator has a very simple form \cite{Mikhailov:2009rx}:
\begin{equation}\label{FirstOrderV}
V^{(2)}_1 = {1\over 2} B^{ab}  j_{[a}\wedge j_{b]},
\end{equation}
where $j_a$ are the conserved currents corresponding to the global symmetries,
and  $B^{ab}$ is a  constant antisymmetric tensor, the parameter of the 
deformation\footnote{Here we consider the full supermultiplet of
the linearized $\beta$-deformations. To the best of our knowledge, the orbits
of the $\beta$-deformations under the supersymmetry have not been
previously studied. But there is a construction of the deformations
of the AdS part of $AdS_5\times S^5$ in 
\cite{McLoughlin:2006cg,Swanson:2007dh}, which must be related to the
deformations of the sphere by the supersymmetry.}:
\begin{equation}\label{SpaceOfLinearizedBetaDeformations}
B\in ({\bf g} \wedge {\bf g})_0/{\bf g},
\end{equation}
where: 
\begin{itemize}
\item ${\bf g}=psu(2,2|4)$ is the global symmetry algebra; indices $a,b$ enumerate
   the generators of ${\bf g}$
\item the subindex $0$ means that the ``inner commutator'' is zero, 
   see Eq. (\ref{ggZero})
\item the subspace ${\bf g}\subset ({\bf g}\wedge {\bf g})_0$ 
   is generated by ${f_{a}}^{bc} t_b\wedge t_c$ (for $B\in {\bf g} \subset {\bf g}\wedge {\bf g}$ we find that
   (\ref{FirstOrderV}) is a total derivative); in other words we consider $B_1$
   and $B_2$ equivalent if:
   \begin{equation}\label{EquivalenceRelationIntro}
      B_1^{ab} - B_2^{ab} = {f^{ab}}_c G^c
   \end{equation}
\end{itemize}
We want to construct the series of the form 
(\ref{PerturbedActionSeries}) so that $Q_{exact}$ is a symmetry of $S_{exact}$
and $Q_{exact}^2=0$. It follows from the general principles of string theory,
that this should be always possible starting from the first order
$V_1^{(2)}$ given by (\ref{FirstOrderV}). 

\subsubsection{Second order in $\varepsilon$}
The second order correction $V_2^{(2)}$ depends on $B$ quadratically.
It turns out that the dependence of $V_2^{(2)}$ on $B$ is rather subtle.
Notice that the space of linearized $\beta$-deformations 
(\ref{SpaceOfLinearizedBetaDeformations}) is fibered by the orbits
of $PSU(2,2|4)$. The structure of $V_2^{(2)}$ depends on which orbit
$B$ belongs to.  The formula for $V_2^{(2)}$ is relatively simple
when $B$ satisfies a certain quadratic equation. This equation
says that the Schouten-Nijenhuis-Gerstenhaber bracket $\gbl B,B \gbr$
is equivalent to zero. The standard definition of this bracket is:
\begin{align}\label{DefinitionOfSchouten}
\gbl B,B \gbr \in & \; {\bf g}\wedge {\bf g} \wedge {\bf g}
\\
\gbl B,B \gbr^{abc} = & \; B^{e[a}{f^b}_{ef}B^{c]f}
\end{align}
However this definition does not respect the equivalence relation
(\ref{EquivalenceRelationIntro}). The construction which does respect
this equivalence relation is this one:
\begin{align}
\gbl B,B \gbr \mbox{ mod } & L_{\Delta} 
\\
\mbox{where } L_{\Delta} \mbox{ is } & \mbox{generated by } 
{f^{[ab}}_m A^{|m|c]}
\end{align}
In our terminology, the $\beta$-deformation is called {\em real} if:
\begin{equation}\label{SchoutenBracketInLDelta}
\gbl B,B \gbr \in L_{\Delta}
\end{equation}
We distinguish  $\beta$-deformations
 of the following three types: {\em real}, {\em complex}, 
and {\em obstructed}.

\subsection{Types of $\beta$-deformations}
\subsubsection{Real $\beta$-deformations}
\label{sec:IntroRealBeta}
In the special case when $B\in \Lambda^2{\bf su}(4)$, 
the condition (\ref{SchoutenBracketInLDelta}) for real $\beta$-deformations is
equivalent to:
\begin{equation}\label{SchoutenBracketZero}
\gbl B,B \gbr =0.
\end{equation}
We explicitly constructed $V_2^{(2)}$ for real $\beta$-deformations in 
Section \ref{sec:ConstructionOfV2Integrated}  Eqs. (\ref{FormulaForV2}), (\ref{V2WithAntifields}). 
We find that $V_2^{(2)}$ 
 is a polynomial function of the
currents $j$ and the group element $g$.
For such $B$, we {\em conjecture} that the polynomial dependence
of $V_n^{(2)}$ on the currents and the group element will persist
at higher orders. This agrees with the formula for the obstruction
suggested in \cite{Aharony:2002hx}. 
In fact, we suspect\footnote{because it was proven in
\cite{Aharony:2002hx} that there the obstruction to the
existence of a polynomial solution only appears at the third
order in $\epsilon$, thus we expect to have problems only 
 with $V^{(2)}_n$ for $n>2$} 
that $V_2^{(2)}$ is always a polynomial function,
but we point out that the formula is much simpler when 
(\ref{SchoutenBracketInLDelta}) is satisfied.
In fact we do not even know the explicit
formula in the case when $B$ does not satisfy (\ref{SchoutenBracketInLDelta}). 

Notice that for the real Maldacena-Lunin \cite{Lunin:2005jy} solutions   
$B$ satisfies a stronger condition:
\begin{equation}\label{StrongerCondition}
B^{ea}{f^b}_{ef}B^{cf}=0
\end{equation}
(no antisymmetrization of $abc$). 

Does (\ref{SchoutenBracketZero}) imply (\ref{StrongerCondition})?
The condition (\ref{SchoutenBracketZero}) can  be interpreted as a classical
Yang-Baxter equation for the $r$-matrix $r=B$ \cite{MR583806}.  
If this condition is satisfied,
then the antisymmetric tensor $B$ defines a left-invariant Poisson structure
on the supergroup $PSU(2,2|4)$.  In this context the solutions of
(\ref{SchoutenBracketZero}) have been previously studied in the mathematical
literature. For a compact Lie group (such as $SU(4)$) it was proven in
\cite{CahenGuttRawnsley:1994} that (\ref{SchoutenBracketZero}) implies that
$B$ lies in the exterior product of an abelian subalgebra ${\bf a}\subset {\bf g}$:
\begin{equation}\label{AbelianB}
B\subset {\bf a}\wedge {\bf a}.
\end{equation}
This means that in this case (\ref{SchoutenBracketZero}) implies 
(\ref{StrongerCondition}).
However, for non-compact groups (such as $SU(2,2)$) there are more general
solutions. Solutions of the rank 8 for $SU(2,2)$ were constructed in
\cite{Agaoka:2000}. 

Therefore the results of \cite{Agaoka:2000} suggest that there are
solutions more general than those considered in \cite{Lunin:2005jy}, corresponding
to the deformation of the AdS part of $AdS_5\times S^5$.
But at this time we have not proven that such solutions would not
be obstructed at the cubic and higher orders.

\subsubsection{Obstructed $\beta$-deformations}
What happens for a general $B$? 

Generally speaking, {\em any} solution
of the linearized supergravity can be ``repaired'' to the full
exact nonlinear solution, if we dress it appropriately with
the corrections to self-interaction. In other words, it is always
possible to construct the series of the form (\ref{PerturbedActionSeries})
order by order in $\varepsilon$. 
But for a generic $B$ the nonlinear solution will not be a polynomial
in the current and the group element. 
In particular, the solution for a general $B$ will not be periodic
in the global time of $AdS_5$. In other words, the nonlinear solution
will not be a universal cover of anything (while $AdS_5$ was a universal
cover of the hyperboloid). 

\subsubsection{Complex $\beta$-deformations}
It is natural to ask the following question: what is the condition
on $B$ necessary and sufficient for the nonlinear solution 
to be, order by order in $\varepsilon$, of the polynomial type?

At this time, we do not have a full answer to this question in our approach.

The condition (\ref{SchoutenBracketInLDelta}) is probably sufficient, although we have
only proven this at the order $\varepsilon^2$. But it is not
necessary. It appears too strong. For some $B$ violating 
(\ref{SchoutenBracketInLDelta}) there are still polynomial solutions. This can be seen using the
solution-generating technique of \cite{Lunin:2005jy}. As we will review in Sections
\ref{sec:ComplexStrucure} and \ref{sec:ComplexStructureIn45} the space
of linearized $\beta$-deformations has a complex structure, {\it i.e.}
there is an operator $\cal I$ commuting with the $PSU(2,2|4)$ such that
${\cal I}^2=-1$. The results of \cite{Lunin:2005jy} imply that if $B$
corresponds to a polynomial solution then $e^{{\cal I}\phi}B$ also
corresponds to a polynomial solution. But the action of ${\cal I}$
violates the condition (\ref{SchoutenBracketInLDelta}). In our terminology, the complex $\beta$-deformations
are those which can be connected to the real $\beta$-deformations
by $e^{{\cal I}\phi}$.

The analysis of \cite{Aharony:2002hx} implies that the obstruction
first appears at the third order of perturbation theory 
({\it i.e.} $\varepsilon^3$). The authors of \cite{Aharony:2002hx} suggest the
formula for the obstruction, which we review and supersymmetrize
in Section \ref{sec:ComparisonWithAKY} --- see Eq. (\ref{FromAKY}) and its supersymmetric
generalization (\ref{SupersymmetricAKY}). 
It is interesting that this obstruction can
{\em almost} be expressed in terms of the Schouten bracket,
but not quite --- see Section \ref{sec:SUSYExtensionOfAKY}.
It appears to us that there are cases when $B$ satisfies 
(\ref{SupersymmetricAKY}) but is not in the orbit $e^{{\cal I}\phi}B$
of the real $B$. This means, provided  that  $B$ satisfying
(\ref{SupersymmetricAKY}) are indeed unobstructed, that not all of these solutions
can be obtained by the solution-generating trick of \cite{Lunin:2005jy}
from the real solutions. 

\paragraph     {Complex $\beta$-deformations receive $\alpha'$-corrections}
The known results from the field theory side
\cite{Elmetti:2006gr} combined with the AdS/CFT correspondence imply that
the complex solutions receive accumulating\footnote{these $\alpha'$ 
corrections are ``accumulating'' in the sense that the corrected
background is not periodic in the global time of AdS; the deviation from
periodicity corresponds to the anomalous dimension on the field theory
side} 
$\alpha'$-corrections 
\cite{Frolov:2005ty}. This suggests that there should be a proof
of finiteness to all orders in $\alpha'$ which works for $B$ satisfying
(\ref{SchoutenBracketInLDelta})
and does not work for the complex $\beta$-deformation.

The picture presented in the current literature 
\cite{Aharony:2002hx,Lunin:2005jy,Frolov:2005ty,Elmetti:2006gr}
(as we understand it) is the following:
\begin{itemize}
\item real $\beta$-deformations are periodic in global time, including
   the $\alpha'$ corrections
\item complex $\beta$-deformations are classically periodic, but receive
   accumulating $\alpha'$-corrections quantum mechanically
\item the  $\beta$-deformations  which we call ``obstructed'' 
   are not periodic even classically
\end{itemize}

\paragraph     {Brief review of the literature}
As pointed out in \cite{Lunin:2005jy}, one can generate the
Lunin-Maldacena background by performing a TsT chain of transformation on
the $AdS_5\times S^5$ background:  a T-duality transformation
along one angular variable $\p$, a shift in another angular variable
$\pp$ and again a T-duality along $\p$. The parameter of the deformation
is introduced by the shift and it is therefore real. On the other hand,
the $\beta$-deformed field theory is allowed to have complex $\beta$
\cite{Leigh:1995ep}. In order to generate a complex parameter in the dual
geometry, we have to apply S-duality before and after the TsT chain, so
we would have a STsTS chain (for further discussion, see
\cite{Lunin:2007zza}).

Note that, since we perform twice the S-duality transformation, the
original and the deformed solutions are in the same coupling
regime. However, as it was pointed out by Frolov in 
\cite{Frolov:2005dj}, the
S-duality step departs from the world-sheet treatment, as opposed to
T-duality. Indeed, as it is discussed in Frolov, Roiban and Tseytlin's
paper \cite{Frolov:2005ty}, the T-duality can be implemented directly at the level
of the world-sheet, so the starting point may be the classical
Green-Schwarz action on $AdS_5\times S^5$. They also say that there are
no good reason to believe that the S-dual background will not be deformed
by $\alpha'/R^2$ corrections, while TsT does not introduce any
correction. Indeed, while T-duality and a
coordinate shift preserve the 2d conformal invariance of the string
theory,  with S-duality things are very different and we may need to
modify the classical superstring action by extra $\alpha'/R^2$ correction
terms in order to ensure its quantum 2d conformal invariance.



   

\subsection{Plan of the paper}


In Section \ref{sec:NotationsAndIdentities} we list basic
formulas for the pure spinor superstring in $AdS_5\times S^5$ and briefly discuss
the descent procedure.  Then in Section \ref{sec:VertexForBetaDeformation} we introduce
the vertex operator which corresponds to the $\beta$-deformation at the linearized
level. In Section \ref{sec:AdditionalConstraintOnB} we discuss an additional
constraint on the parameter $B$, which we don't understand as well as we
would want to.  We then discuss the symmetries of the vertex in 
Section \ref{sec:SymmetriesOfTheVertex}, with a surprising conclusion that
our vertex is not strictly speaking covariant. 
In Section \ref{sec:GeneralDeformationTheory} we discuss the general
deformation theory of the  classical worldsheet action, and in Section
\ref{sec:ApplyingToBeta} apply it to the $\beta$-deformation. In particular, in Sections 
\ref{sec:ConstructionOfV2Integrated} and \ref{sec:ClassicalValuesOfAntifields}
we obtain the explicit formula for the second order correction $V^{(2)}_2$.
 It turns out that the Schouten
bracket $\gbl B,B \gbr$ plays an important role, and we further study its
properties in Section \ref{sec:PropertiesOfSchouten}.
In Section \ref{sec:AboutComplex} we discuss the complex $\beta$-deformations
and the equation for the obstruction proposed by Aharony, Kol and
Yankielowicz in \cite{Aharony:2002hx}; we discuss the supersymmetric
generalization of their formula.
 In Section
\ref{sec:ReadingSupergravityFields} we show (at the linearized level) that the
target space supergravity fields of our worldsheet theory agree with the known
supergravity description of the $\beta$-deformation. In Section
\ref{sec:RelationToAAF} we explain (at the linearized level) the relation between our
approach and the approach of 
\cite{Frolov:2005ty,Frolov:2005dj,Frolov:2005iq,Grassi:2006tj}
which uses the twisted boundary conditions.  In Section
\ref{sec:RelationBetweenNSNSAndRR} we discuss an interesting general relation
between the RR fields and the NSNS fields for the $\beta$-deformed background.

\subsection{Open questions}
We will list here several open questions:
\begin{enumerate}
\item The constraint on the internal commutator described in 
   Section \ref{sec:AdditionalConstraintOnB} has to be explained.
\item Even under the condition (\ref{SchoutenBracketInLDelta}) we have only
   constructed the action up to the second order in $\epsilon$. It should
   be possible to find an explicit expression for $V_n^{(2)}$ for $n>2$.
\item Is it true that the condition for the existence of the classical 
   periodic solution is given by Eq. (\ref{FromAKY}) of Section 
   \ref{sec:ComparisonWithAKY}? We must understand the relation to the covariant
   subcomplex of \cite{Mikhailov:2009rx} suggested in 
   Section \ref{sec:GhostNumberThreeAlwaysQExact}.
\item It would be nice to prove the nonrenormalization theorem
   for real $\beta$-deformations without invoking the $TsT$-transformations
   argument of \cite{Lunin:2007zza}. 
   In the language of twisted boundary conditions, which
   we review in Section \ref{sec:RelationToAAF}, how do we derive
   (\ref{SchoutenBracketInLDelta})? It has to be related to the BRST
   symmetry of the twisted boundary conditions. 
\item Solutions of \cite{Agaoka:2000} mentioned 
   in Section \ref{sec:IntroRealBeta} have to be studied explicitly.
   Are they obstructed at the higher orders of $\varepsilon$?
\end{enumerate}

\section{Notations and various identities}
\label{sec:NotationsAndIdentities}
In this Section we will use  
\cite{Berkovits:2004jw,Berkovits:2004xu,Mikhailov:2007mr}.
\subsection{Notations}
\label{sec:Notations}

The global symmetry algebra is ${\bf g} = psu(2,2|4)$.
It has the ${\bf Z}_4$ grading ${\bf g}={\bf g}_0 + {\bf g}_1 +
{\bf g}_2 + {\bf g}_3$. Various worldsheet fields
take values in ${\bf g}$; the lower index will denote the
${\bf Z}_4$ grade of the field. For example, consider this field:
\[
w_{1+}
\]
The index $1$ means that it takes values in ${\bf g}_1$, and the
index $+$ means that it has the conformal dimension $(1,0)$;
similarly the field $w_{3-}$ has the conformal dimension
$(0,1)$ and takes values in ${\bf g}_3$.

The pure spinor ghosts are $\lambda_3$ and $\tilde{\lambda}_1$.
The tilde over $\tilde{\lambda}_1$ is redundant; it is 
to stress that this field would be right-moving in the free
field limit. We will then sometimes write $\lambda$ and $\tilde\lambda$ for short.

The corresponding conjugate momenta are 
$w_{1+}$ and $\tilde{w}_{3-}$;
the kinetic term in the action is given by
Eq. (\ref{OurKineticTerm}) below. Once again, our
notations are highly excessive because there is no
such things as for example $w_{1-}$. Therefore, we could
have just written $w_+$ and $w_-$ instead of 
$w_{1+}$ and $\tilde{w}_{3-}$.

The pure spinor action is constructed out of the right-invariant current
\begin{equation}
\label{ric}
J =-d gg^{-1},
\end{equation}
which is invariant under $g\rightarrow gH$, with $H \in PSU(2,2|4)$ being a global parameter.

We use notations from Section 2 of \cite{Mikhailov:2007mr}.
As in that paper, the spectral parameter of the Lax operator
will be denoted $z$. The Lax equation is
\begin{equation}
\left[ \partial_+ + J_+[z]\;,\; \partial_- + J_-[z] \right] = 0,
\end{equation}
where
\begin{align}
J_+[z]  = & J_{0+} - N_+ + z^{-1} J_{3+} + z^{-2} J_{2+} 
+ z^{-3} J_{1+} + z^{-4} N_+
\label{JPlusSpectral}
\\
J_-[z]  = & J_{0-} - N_- + z J_{1-} + z^2 J_{2-} 
+ z^3 J_{3-} + z^4 N_-.
\label{JMinusSpectral}
\end{align}
The Lax connection $J_{\pm}[z]$ defined above should not be confused with the
current $J_{\pm}=-\partial_\pm gg^{-1}$ on the right-hand side.   The current does not depend on the Lax parameter,
while the Lax connection does.

We will also introduce $l$ by
\begin{equation}
l=\log z.
\end{equation}
Then the density of the global conserved charges can be written as
\begin{equation}
j= \left. g^{-1} \frac{dJ}{ dl}\right|_{l=0} g.
\end{equation}
Following Berkovits and Howe, we also denote $z,\overline{z}$ the worldsheet coordinates. 
We think that the meaning of $z$ will be always clear from the context. 

For any $x\in {\bf g}$ we denote:
\begin{equation}
x_a = \mbox{Str}(xt_a)
\end{equation}
In particular:
\begin{equation}
j_a = \mbox{Str}\left( 
\left. g^{-1} \frac{dJ}{ dl}\right|_{l=0} g
\; t_a \right)
\end{equation}

\subsection{Various identities}
\label{sec:VariousIdentities}
In calculations involving supersymmetry, it is very
convenient (and in fact suggested by the definitions) to introduce
a set of sufficiently many formal anticommuting constant parameters:
\begin{equation}\label{EpsilonParameters}
\epsilon,\; \epsilon', \; \epsilon'', \; \ldots
\end{equation}
For example, the superalgebra ${\bf gl}(m|n; {\bf C})$ consists of the
block matrices 
$\left( \begin{array}{cc} 
a_{00} & a_{01} \cr
a_{10} & a_{11} 
\end{array} \right)$  where $a_{00}$, $a_{10}$, $a_{01}$, $a_{11}$ 
are $m\times m$, $m\times n$, $n\times m$ and $n\times n$ matrices,
respectively, and moreover $a_{00}$ and $a_{11}$ are constructed from
polynomials over ${\bf C}$ involving even number of  $\epsilon$-parameters
(\ref{EpsilonParameters}) 
and $a_{10}$ and $a_{01}$ involve odd number of $\epsilon$. The commutator
is then the usual commutator $[x,y] = xy-yx$. We will use such parameters
in our notations; they should not be confused with the small parameter
in (\ref{PerturbedActionSeries}) which is denoted $\varepsilon$.

The BRST charge\footnote{In this section we will consider
BRST transformations in the undeformed theory on $AdS_5\times S^5$;
we will omit the index $0$ and write $Q$ instead of $Q_0$, to
simplify the notations. All the formulas in this section are about
the undeformed theory.} acts on the group element $g$ as
\begin{equation}\label{PostulatedQ}
\epsilon Q \; g 
= (\epsilon \lambda_3 + \epsilon \tilde{\lambda}_1) \; g
\end{equation}
and, because of (\ref{ric}) and (\ref{JPlusSpectral})-(\ref{JMinusSpectral}),
\begin{eqnarray}\label{QJ}
\epsilon Q\; J = -D(\epsilon\lambda + \epsilon \tilde{\lambda}) \\
\epsilon Q \; z\partial_z J =
D(\epsilon\lambda -\epsilon\tilde{\lambda}) 
-[z \partial_z J \;, \;\epsilon\lambda + \epsilon\tilde{\lambda}],
\end{eqnarray}
where
\begin{equation}
D=\partial + [J, \ \ \ ].
\end{equation}
It will be useful to define the ``composite'' ghosts 
\begin{equation}
\Lambda(\epsilon) = g^{-1}(\epsilon \lambda - \epsilon \tilde{\lambda}) g, \qquad \overline\Lambda(\epsilon) = g^{-1}(\epsilon \lambda + \epsilon \tilde{\lambda}) g.
\end{equation}
Remind that $Q=Q_L + Q_R$, and denote $\overline{Q} = Q_L-Q_R$. We then have
\begin{eqnarray}
\epsilon Qg = g\overline\Lambda(\epsilon), & \qquad & \epsilon \overline Qg = g\Lambda(\epsilon), \\
\epsilon Qg^{-1} = -\overline\Lambda(\epsilon)g^{-1}, & \qquad & \epsilon \overline Qg^{-1} = -\Lambda(\epsilon)g^{-1}.
\end{eqnarray}

Because of the pure spinor constraint, $\epsilon\lambda\epsilon'\lambda = \epsilon\tilde\lambda\epsilon'\tilde\lambda = 0$ , and thus
\begin{equation}
\Lambda(\epsilon)\Lambda(\epsilon') = - \overline\Lambda(\epsilon)\overline\Lambda(\epsilon'), \qquad
\Lambda(\epsilon)\overline\Lambda(\epsilon') = - \overline\Lambda(\epsilon)\Lambda(\epsilon')
\end{equation}
(note also that the above expressions are antisymmetric under $\epsilon \leftrightarrow \epsilon'$). 

Using the identities above, we show that
\begin{eqnarray}
\epsilon Q \overline\Lambda(\epsilon') & = & (\epsilon Q g^{-1}) (\epsilon'\lambda + \epsilon'\tilde\lambda)g + g^{-1} (\epsilon'\lambda + \epsilon'\tilde\lambda)(\epsilon Q g) = \nonumber \\
& = & -\overline\Lambda(\epsilon) \overline\Lambda(\epsilon') + \overline\Lambda(\epsilon') \overline\Lambda(\epsilon) = \nonumber \\
& = &  - [\overline\Lambda(\epsilon), \overline\Lambda(\epsilon')] =
\nonumber \\
& = & [\Lambda(\epsilon), \Lambda(\epsilon')] = - \epsilon \overline Q \Lambda(\epsilon'), \\
\epsilon Q \Lambda(\epsilon') & = & - [\overline\Lambda(\epsilon), \Lambda(\epsilon')] = 0, \\
\epsilon Q \Lambda(\epsilon') & = & - [\Lambda(\epsilon), \overline\Lambda(\epsilon')] = 0.
\end{eqnarray}

We may also write the above equations in $psu(2,2|4)$ components: 
\begin{equation}
\Lambda_a = \mbox{Str}(\Lambda t_a)
\end{equation}
\begin{equation}
\epsilon\overline{Q}\Lambda_a(\epsilon') = - {f_a}^{bc} \Lambda_b (\epsilon)\Lambda_c (\epsilon')\;,\;\;
Q \Lambda_a = 0\;,\;\;
\epsilon Q\overline{\Lambda}_a (\epsilon') = - {f_a}^{bc} \overline{\Lambda}_b(\epsilon) \overline{\Lambda}_c(\epsilon') \;,\;\;
\overline{Q}\overline{\Lambda}_a = 0.
\end{equation}
There are constraints on $\Lambda$ and $\overline{\Lambda}$:
\begin{equation}
{f_a}^{bc} \Lambda_b \overline{\Lambda}_c = 0\;,\;\;
{f_a}^{bc} ( \Lambda_b \Lambda_c + \overline{\Lambda}_b \overline{\Lambda}_c ) = 0.
\end{equation}
The ``composite'' ghosts satisfy also
\begin{equation}\label{QAnnihilatesOneForm}
Q\left( d\overline\Lambda(\epsilon') -
2 [ g^{-1} z\partial_z J g \;,\;
\Lambda(\epsilon') ]\right)=0,
\end{equation}
since
\begin{eqnarray}
d\; [\Lambda(\epsilon), \Lambda (\epsilon')] & = & d\; [g^{-1}(\epsilon\lambda - \epsilon\tilde{\lambda})g\;,\;
        g^{-1}(\epsilon'\lambda - \epsilon'\tilde{\lambda})g] =
\nonumber \\
& = & 2\; [g^{-1} D(\epsilon\lambda - \epsilon\tilde{\lambda}) g\;,\;
          g^{-1}  (\epsilon'\lambda - \epsilon'\tilde{\lambda}) g] =
\nonumber \\
& = & 2\; \left[g^{-1}\epsilon Q(z\partial_z J) g + 
          g^{-1} [z\partial_z J, 
                   \epsilon\lambda + \epsilon\tilde{\lambda}g]
\;,\;\;
         g^{-1}(\epsilon'\lambda - \epsilon'\tilde{\lambda}) g\right] =
\nonumber \\
& = & 2\epsilon Q [ g^{-1} z\partial_z J g \;,\;
g^{-1}(\epsilon'\lambda - \epsilon'\tilde{\lambda}) g ]
\nonumber \\
& = & 2\epsilon Q [ g^{-1} z\partial_z J g \;,\;
\Lambda (\epsilon')].
\end{eqnarray}
Finally, since the conserved charge $j$ satisfies
\be
j = g^{-1} z\partial_z J g, \ee
then,
\bea
\epsilon Q j & = & d(g^{-1} \epsilon \overline{Q} g),
\label{QOfSmallJ}
\\
\epsilon \overline{Q} j & = & d(g^{-1} \epsilon Q g);
\eea
and note that Eq. (\ref{QAnnihilatesOneForm}) can be written as
\begin{equation}
\Omega = 
\epsilon \overline{Q} j -  2 [j, g^{-1} \epsilon \overline{Q} g]
\in \mbox{Ker}(Q).
\end{equation}

\subsection{BRST transformation of $S_0$}
\begin{eqnarray}
S_0 =&&   {R^2\over \pi} \int d^2 z\, \hbox{Str} \Big( {1\over 2} J_{2+}J_{2-} +
{3\over 4} J_{1+}J_{3-} +{1\over 4} J_{3+}J_{1-}  \nonumber \\[5pt]
&&   \qquad      +
w_{1+}\partial_-\lambda_3 + w_{3-}\partial_+\lambda_1 +N_{0+}J_{0-}
+N_{0-}J_{0+}-N_{0+}N_{0-} - w^{\star}_{1+} w^{\star}_{3-}  \Big)  \,,
\label{TheAction}
\end{eqnarray}
the ghost currents are:
\begin{equation}
N_{0+}=-\{w_{1+},\lambda_3\} \,,\qquad N_{0-}=-\{w_{3-},\lambda_1\} \,,
\end{equation}
The BRST transformations of the currents are:
\begin{eqnarray}
\epsilon Q_L\; J_{3} & = & - D_{0}\epsilon\lambda_3    \nonumber
\\[1pt]
\epsilon Q_L\; J_{2} & = & - [J_{3},\epsilon\lambda_3]  \nonumber
\\[1pt]
\epsilon Q_L\; J_{1} & = & - [J_{2},\epsilon\lambda_3]  \nonumber
\\[1pt]
\epsilon Q_L\; J_{0} & = & - [J_{1},\epsilon\lambda_3]
\end{eqnarray}
Therefore the BRST transformation of the first row is:
\begin{eqnarray}
\mbox{Str} &&
   -{1\over 2} [J_{3+},\epsilon\lambda_3] J_{2-} 
   -{1\over 2} J_{2+} [J_{3-},\epsilon\lambda_3] 
\nonumber\\ 
&&
   -{3\over 4} [J_{2+},\epsilon\lambda_3] J_{3-} 
   -{3\over 4} J_{1+} (D_{0-} \epsilon\lambda_3) 
\nonumber\\
&&
   -{1\over 4} (D_{0+}\epsilon\lambda_3) J_{1-}    
   -{1\over 4} J_{3+} [J_{2-},\epsilon\lambda_3]
\end{eqnarray}
We transform this as follows:
\begin{eqnarray}
\mbox{Str} &&
-{1\over 4}D_{0+}(\epsilon\lambda_3 J_{1-})
+{1\over 4}D_{0-}(\epsilon\lambda_3 J_{1+})
+{1\over 4}\epsilon\lambda_3 (D_{0+}J_{1-} - D_{0-}J_{1+})
\nonumber \\
&& 
- J_{1+} (D_{0-} \epsilon\lambda_3) 
\nonumber \\
&&
-{1\over 2} [J_{3+},\epsilon\lambda_3] J_{2-} 
-{1\over 2} J_{2+} [J_{3-},\epsilon\lambda_3] 
\nonumber \\
&&
-{3\over 4} [J_{2+},\epsilon\lambda_3] J_{3-}
-{1\over 4} J_{3+} [J_{2-},\epsilon\lambda_3]
\end{eqnarray}

Let us use the Maurer-Cartan equation:
\begin{equation}
D_{0+}J_{1-} - D_{0-}J_{1+} + [J_{3+},J_{2-}] + [J_{2+},J_{3-}] = 0
\end{equation}
We get:
\begin{eqnarray}
\mbox{Str} && 
-{1\over 4}\partial_{+}(\epsilon\lambda_3 J_{1-})
+{1\over 4}\partial_{-}(\epsilon\lambda_3 J_{1+})
\nonumber \\
&& 
- J_{1+} (D_{0-} \epsilon\lambda_3) 
\end{eqnarray}
The term $-J_{1+}D_{0-}\epsilon\lambda_3$ cancels with the variation
of $w_{1+}D_{0-}\lambda_3$ in the second row. We conclude:
\begin{equation}
\epsilon Q_L {\cal L} = 
-{1\over 4} d\; \mbox{Str}(\epsilon\lambda_3\; J_1)
\end{equation}
\begin{equation}\label{QLIsTotalDerivative}
\epsilon Q {\cal L} = -{1\over 4} d\;
\mbox{Str}(\epsilon\lambda_3 J_1 - \epsilon\lambda_1 J_3 )
\end{equation}
This equation is the first step of the descent procedure for 
the Lagrangian itself. The second step is:
\begin{eqnarray}
&&
\epsilon'Q \;\mbox{Str}(\epsilon\lambda_3 J_1 - \epsilon\lambda_1 J_3) =
\nonumber
\\ 
& = &
\mbox{Str}(\epsilon\lambda_3 D_0 \epsilon'\lambda_1 -
\epsilon\lambda_1 D_0 \epsilon'\lambda_3) =
\nonumber
\\
& = &
\epsilon\epsilon'\; d\; \mbox{Str}(\lambda_3\lambda_1)
\end{eqnarray}
Notice that Eq. (\ref{QLIsTotalDerivative}) can be rewritten
in the following way:
\begin{equation}
\epsilon Q {\cal L} = -{1\over 4}d\;
\mbox{Str}\left(\epsilon\Lambda g^{-1}dg\right)
\end{equation}
Using the notations of Section \ref{sec:DefinesI}:
\begin{equation}
I_{\epsilon Q_0}^{(1)} = - {1\over 4} 
\mbox{Str}\left(\epsilon\Lambda g^{-1}dg\right)
\end{equation}

\subsection{Adding antifields}
The $Q$ defined so far is only nilpotent on-shell. Indeed, we get:
\begin{equation}
Q w_{1+} = - J_{1+} \;\; , \;\;
Q J_{1+} = - D_{0+}\tilde{\lambda}_1 - [J_{2+},\lambda_3]
\end{equation}
\remv{check}\rem{(link: "verify_Qw\.jpg" "beta" "mirage")}
To make $Q$ nilpotent off-shell we have to introduce,
following \cite{Berkovits:2007rj}, the fermionic antifields
$w^{\star}_{1+}$ and $w^{\star}_{3-}$ satisfying the constraints: 
\begin{equation}\label{ConstraintsOnAntifields}
\{ \tilde{\lambda}_1, w_{1+}^{\star} \} = \{ \lambda_3, \tilde{w}_{3-}^{\star} \} = 0
\end{equation}
and modify the BRST transformations: 
\begin{eqnarray}
Q w_{1+} = - J_{1+} - w^{\star}_{1+} 
&\hspace{10pt}&
Q \tilde{w}_{3-} = - J_{3-} - \tilde{w}^{\star}_{3-} 
\nonumber
\\
Q w^{\star}_{1+} = D_{0+} \tilde{\lambda}_1 - [N_+,\tilde{\lambda}_1] 
&\hspace{10pt}&
Q \tilde{w}^{\star}_{3-} = D_{0-}\lambda_3 - [N_-,\lambda_3]
\label{BRSTonw}
\end{eqnarray}
\remv{check w star}\rem{(link: "verify_Qw_star\.jpg" "beta" "mirage")}
\remv{chech w}\rem{(start-process "mirage" "*scratch*" "mirage" 
"photos/verify_Qw.jpg" )}
\remv{BerkovitsVafa}\rem{(start-process "evince" "*scratch*" "evince" 
"../../Papers/Berkovits/AdS/BerkovitsVafa.pdf" "--page-label=9" )}
\remv{AKSZ}\rem{(start-process "evince" "*scratch*" "evince" 
"../../Papers/A.S.Schwarz/aksz.pdf" "--page-label=12" )}
\remv{Losev}\rem{
(start-process "gthumb" "*scratch*" "gthumb" "photos/losev-2009-09-16.jpg")}
With this modification, we obtain:
\begin{equation}
Q^2 w_{1+} = [( J_{2+} + \{w_{1+},\tilde{\lambda}_1\}),\lambda_3] 
+ [\{\lambda_3,\tilde{\lambda}_1\}, w_{1+}]
\end{equation}
which is a combination of the Lorentz gauge transformation
and the pure-spinor-constraint gauge transformation of $w$.

Now we would like to modify the currents to include antifields.
We propose:
\begin{equation}
\widehat{j}_+ = g^{-1} \left( \left.{dJ_+\over dl}\right|_{l=0} 
- 4 w^{\star}_{1+} \right)g
\end{equation}
\begin{equation}
\widehat{j}_- = g^{-1} \left( \left.{dJ_-\over dl}\right|_{l=0} 
+ 4 w^{\star}_{3-} \right)g
\end{equation}
We need the off-shell version of Eq. (\ref{QJ}), and with  $Q$ modified
according to (\ref{BRSTonw}):
\begin{align}
\epsilon Q\; J_+(z)  = & 
-D_+^{[z]}\left( {1\over z}\epsilon\lambda_3 +  z\epsilon\lambda_1\right)
+\left(z - {1\over z^3}\right) 
(D_{0+}\epsilon\lambda_1 - [N_+, \epsilon\lambda_1]) +
\\
&
+\left( 1 - {1\over z^4} \right) [w_{1+}^{\star} , \epsilon\lambda_3]
\nonumber
\\[5pt]
\epsilon Q\; J_-(z)  = & 
-D_-^{[z]}\left( {1\over z}\epsilon\lambda_3 +  z\epsilon\lambda_1\right)
+\left({1\over z} - z^3 \right) 
(D_{0-}\epsilon\lambda_3 - [N_-, \epsilon\lambda_3]) +
\\
&
+\left( 1 - z^4 \right) [w_{3-}^{\star} , \epsilon\lambda_1]
\end{align}
\remv{VerifyQj}\rem{(start-process "mirage" "*scratch*" "mirage" 
"photos/verify_Qj.jpg" )}
\remv{check}\rem{(link: "verify_part-of-Qj\.jpg" "beta" "mirage")}
This means that:
\begin{equation}
\epsilon Q\; \widehat{j} =
d\left( g^{-1}(\epsilon\lambda_3 - \epsilon\lambda_1)g \right) 
\end{equation}

\section{Vertex corresponding to $\beta$-deformation}
\label{sec:VertexForBetaDeformation}
\subsection{The vertex and its descent}
\label{sec:VertexAndItsDescent}
As discussed in Section \ref{sec:VariousIdentities}, we introduce
a set of formal anticommuting constants $\epsilon,\epsilon',\epsilon'',\ldots$.
\paragraph     {Definition of the vertex}
It was proposed in \cite{Mikhailov:2009rx}
 that the unintegrated vertex corresponding to the 
$\beta$-deformation is given by this expression:
\begin{equation}\label{VBeta}
V^{\rm beta}_{ab}(\epsilon,\epsilon') = 
(g^{-1}\epsilon(\lambda_3-\lambda_1)g)_a \;
(g^{-1}\epsilon'(\lambda_3-\lambda_1)g)_b
\end{equation}
Here the indices $a$ and $b$ enumerate the adjoint representation
of $psu(2,2|4)$. Notice that (\ref{VBeta}) is antisymmetric under
the exchange of $a$ and $b$. Therefore this vertex is
in the antisymmetric product of two adjoint representations
of $psu(2,2|4)$. We will parametrize the $\beta$-deformations by
a constant antisymmetric tensor $B^{ab}$:
\begin{equation}\label{VParametrizedByB}
V[B](\epsilon,\epsilon') = B^{ab}\;
(g^{-1}\epsilon(\lambda_3-\lambda_1)g)_a \;
(g^{-1}\epsilon'(\lambda_3-\lambda_1)g)_b
\end{equation}

\paragraph     {Equivalence relation}
The antisymmetric product of two adjoint representations is not
an irreducible representation. In particular, it has a subspace
consisting of $B^{ab}$ of the form: $B^{ab} = f^{ab}_c A^c$.
It turns out that such $B$ corresponds to BRST exact vertices:
\begin{eqnarray}
{f^{ab}}_c V^{\rm beta}_{ab} =
[(g^{-1}\epsilon(\lambda_3-\lambda_1)g),
(g^{-1}\epsilon'(\lambda_3-\lambda_1)g)]_c =
\nonumber
\\
= \epsilon Q_{BRST} (g^{-1} \epsilon'(\lambda_3 + \lambda_1) g)_c
\end{eqnarray}
Therefore the tensors $B^{ab}$ and $B^{ab} + f^{ab}_c A^c$ give the
same $\beta$-deformation:
\begin{equation}\label{EquivalenceRelation}
B^{ab} \;\simeq \; B^{ab} + f^{ab}_c A^c
\end{equation}
We will explain in Section \ref{sec:BProportionalToStructureConstants}
that the gauge transformation (\ref{EquivalenceRelation}) should be
accompanied by the change of variables (field redefinition). 
This is because the corresponding integrated vertex is only invariant
on-shell. 

\paragraph     {Descent procedure and integrated vertex}
The deformation of the action corresponding to (\ref{VBeta}) follows
from the standard descent procedure.  Let us denote:
\begin{equation}
\Lambda_a(\epsilon) = (g^{-1} \epsilon( \lambda_3 - \lambda_1 ) g)_a
\end{equation}
The operator $\Lambda_a(\epsilon)$ corresponds to the local conserved currents in the following
sense:
\begin{equation}\label{DescentFromLambdaStep1}
d\Lambda_a(\epsilon) = \epsilon Q (j_a)
\end{equation}
where $j_{a\pm}(\tau^+,\tau^-)$ is the density of the local
conserved charge corresponding to the global symmetries. Therefore:
\begin{equation}\label{DescentFromLambdaStep2}
d(\Lambda_{[a}(\epsilon) \Lambda_{b]}(\epsilon')) =
2\epsilon Q j_{[a} \Lambda_{b]}(\epsilon')
\end{equation}
and:
\begin{equation}\label{DescentFromLambdaStep3}
d(j_{[a}\Lambda_{b]}(\epsilon)) =
-\frac{1}{ 2} \epsilon Q (j_{[a}\wedge j_{b]})
\end{equation}
We conclude that for any constant antisymmetric matrix $B^{ab}$
we can infinitesimally deform the worldsheet action as follows:
\begin{equation}\label{LinearizedBeta}
S\to S + {1\over 2} B^{ab} \int j_{[a}\wedge j_{b]}
\end{equation}
\paragraph     {Summary of the descent procedure:}
\begin{equation}\label{DescentRelation}
(d+Q)\left( 
V^{(0)}_1[B](\epsilon,\epsilon') +
\epsilon V^{(1)}_1[B](\epsilon') +
\epsilon\epsilon' V^{(2)}_1[B]
\right) = 0
\end{equation}
where
\begin{eqnarray}
&& V^{(0)}_1[B](\epsilon,\epsilon') = 
{1\over 2} B^{ab}\;(g^{-1}\epsilon(\lambda_3-\lambda_1)g)_a \;
(g^{-1}\epsilon'(\lambda_3-\lambda_1)g)_b 
\nonumber
\\[4pt]
&& V^{(1)}_1[B](\epsilon')  = 
B^{ab}\; j_a \; (g^{-1}\epsilon'(\lambda_3-\lambda_1)g)_b
\nonumber
\\[4pt]
&& V^{(2)}_1[B]  =  {1\over 2} B^{ab}\; j_a \wedge j_b
\end{eqnarray}
In Eq. (\ref{DescentRelation}) we assume that $d$ commutes with $\epsilon$.

\paragraph     {``Bosonic'' example}
Consider for example $B^{ab}$ in the directions of $S^5$.  We get:
\begin{equation}\label{BInSDirections}
S\to S + B^{[kl][mn]} \left(
\int X_{[k} dX_{l]} \wedge X_{[m} dX_{n]}
+\ldots \right)
\end{equation}
where $X_j$ describes the embedding of $S^5$ into ${\bf R}^6$:
\begin{equation}\label{SphereEmbedding}
X_1^2 + X_2^2 + \ldots + X_6^2 = 1
\end{equation}
and dots denote $\theta$-dependent terms. These $\theta$-dependent
terms appear because  $j_a$ includes $\theta$. 
The subspace
${\bf g}\subset {\bf g}\wedge {\bf g}$ corresponds to $B$ of the
following form:
\begin{equation}
B^{[kl][mn]} = \delta^{km} A^{ln} - \delta^{lm} A^{kn}
+ \delta^{ln} A^{km} - \delta^{kn} A^{lm}
\end{equation}
where $A^{mn}$ is antisymmetric matrix; then the corresponding
deformation of the Lagrangian is a total derivative
$d(A^{mn}X_m dX_n)$. The complementary space has real dimension
90, it corresponds to the representation $\bf 45_C$ of $so(6)$.

\subsection{What happens to the integrated vertex when 
$B^{ab}$ is proportional to ${f^{ab}}_c A^c$?}
\label{sec:BProportionalToStructureConstants}
\paragraph     {In this case the integrated vertex becomes a total derivative}
Indeed, consider the
descent procedure. When $B^{ab}$ is proportional
to the structure constant, this means that the
vertex operator is of the form 
$  \left[g^{-1}\epsilon \frac{d\lambda}{ dl}g\; , \;
         g^{-1}\epsilon'\frac{d\lambda}{ dl}g\right]  $. Here we use:
\begin{equation}
l=\log z
\end{equation}
--- see  Section \ref{sec:Notations}. 

We want to apply the descent procedure and obtain the corresponding
integrated vertex operator. The first step is to take the 
derivative of our unintegrated vertex and see that it is BRST exact:

\begin{equation}
d\;\left[g^{-1}\epsilon\frac{d\lambda}{  dl}g,\;
    g^{-1}\epsilon'\frac{d\lambda}{ dl}g\right]
 = 
- 2\epsilon Q \left[ g^{-1} \frac{d J}{ dl} g \;,\;
g^{-1}\epsilon' \frac{d\lambda}{ dl}g \right]
\end{equation}
But now a special thing happens; on the right hand side $Q$ is
taken of the expression which is $d$ of something plus $Q$ of 
something:
\begin{equation}\label{QOfSmthPlusDOfSmth}
- 2 \left[ g^{-1} \frac{d J}{ dl} g \;,\;
g^{-1}\epsilon\frac{d\lambda}{ dl} g \right] 
 = 
\epsilon Q\left( g^{-1}\frac{d^2 J}{ dl^2} g \right) +
d(g^{-1}\epsilon\lambda g)
\nonumber
\end{equation}
This formula can be derived as follows:
\begin{align}
\epsilon Q \left( g^{-1} \frac{d^2J}{ dl^2} g\right) 
 = &
  g^{-1} \left[ \frac{d^2 J}{ dl^2},\; \epsilon \lambda \right] g
- g^{-1} \frac{d^2}{ dl^2} (D\epsilon\lambda) g =
\nonumber 
\\
 = & 
-2g^{-1} \left[ \frac{dJ}{ dl},\; \frac{d\epsilon\lambda}{ dl} \right] g
-d(g^{-1} \epsilon\lambda g)
\label{DescentOfCommutatorMiddleStep}
\end{align}
The second (and the last) step of the descent procedure is to
take the $d$ of $- 2 \left[ g^{-1} \frac{d J}{ dl} g \;,\;
g^{-1}\epsilon\frac{d\lambda}{ dl} g \right]$ and see that it is
$Q$ of some expression, which is then the corresponding integrated
vertex operator. But we can see directly from (\ref{QOfSmthPlusDOfSmth})
that in fact $d$ of  $- 2 \left[ g^{-1} \frac{d J}{ dl} g \;,\;
g^{-1}\epsilon\frac{d\lambda}{ dl} g \right]$  is equal to $Q$ of
$d\left( g^{-1}\frac{d^2 J}{ dl^2} g \right)$. This means that, indeed,
the corresponding integrated operator is a total derivative.

This can be easily seen explicitly. The integrated vertex operator
is $[j\wedge, j] = 
- g^{-1} \left[ \frac{dJ}{ dl} \wedge,\; \frac{dJ}{ dl} \right] g$.
We observe:
\begin{eqnarray}
d\left(g^{-1} \frac{d^2J}{ dl^2} g\right) & = &
g^{-1} D \frac{d^2J}{ dl^2} g = 
\nonumber
\\
& = &
- g^{-1} \left[ \frac{dJ}{ dl} \wedge,\; \frac{dJ}{ dl} \right] g 
+ {1\over 2} g^{-1} \left( {d^2\over dl^2} DJ \right) g
\end{eqnarray}
Notice that the second term on the right hand side 
$ {1\over 2} g^{-1} \left( {d^2\over dl^2} DJ \right) g $ is proportional
to the equations of motion. Therefore, this term should be
canceled by an infinitesimal field redefinition. 

\paragraph     {Field redefinition}
More precisely, 
$\mbox{Str}\left(t_a g^{-1} \left( {d^2\over dl^2} DJ \right) g\right)$ is the result of the variation of the action
with respect to the infinitesimal left shift of $g$ by
$ 8(gt_ag^{-1})_1 - 8(gt_ag^{-1})_3 $, plus some variation of $\lambda$ and $w$. 
Let us denote this vector field ${\cal X}_a$:
\begin{equation}
{\cal X}_a S = 
\mbox{Str}\left(t_a g^{-1} \left( {d^2\over dl^2} DJ \right) g\right)
\end{equation}
(where $S$ is the action).
We will not need the explicit form of ${\cal X}_a$ in this paper. 

\noindent
We conclude that:
\begin{itemize}
\item the infinitesimal gauge transformation $B^{ab}\to B^{ab} + {f^{ab}}_c A^c$ changes
   the vertex by a total derivative plus terms which can be absorbed
   into an infinitesimal field transformation corresponding to the
   vector field ${\cal X}_aA^a$
\end{itemize}

\section{Constraint on the internal commutator}
\label{sec:AdditionalConstraintOnB}
\subsection{Additional constraint on $B$}
The vertex (\ref{VParametrizedByB}) cannot as such be the right description
of the $\beta$-deformation because it gives extra states which are not
present in the supergravity description. For example, consider
$B$ of the form:
\begin{equation}\label{BadB}
B^{ab} = 
\left\{
\begin{array}{cl}
f^{ab}_c A^c & \mbox{ if both $a$ and $b$ are even (bosonic) indices }
\cr
0 & \mbox{ otherwise}
\end{array}
\right.
\end{equation}
where $A \in so(6) \subset {\bf psu}(2,2|4)$. The corresponding linearized excitation
of $AdS_5\times S^5$ is constant in the AdS directions, and transforms
in the adjoint representation of $so(6)$ (rotations of $S^5$).
But there is no such state in the supergravity spectrum \cite{Kim:1985ez}.

\paragraph     {Conjecture}
It is necessary for the consistency of the deformed worldsheet theory that
the vertex $V^{(0)}$ is given by a primary operator. 
This condition was investigated in \cite{Berkovits:2000yr}; it was found
that the double pole of the vertex operator with the energy-momentum
tensor is proportional to the action of the Laplacian on $psu(2,2|4)$.
In our case, when $V^{(0)} = B^{ab}\Lambda_a \Lambda_b$, 
this is proportional to:
\begin{equation}
B^{ab}{f_{ab}}^c {f_c}^{de} \Lambda_d \Lambda_e =
Q (B^{ab} {f_{ab}}^c \overline{\Lambda}_c)
\end{equation}
Therefore if $B^{ab} {f_{ab}}^c \neq 0$ then the unintegrated vertex
operator is not a conformal primary of the weight zero.
Therefore we must impose this condition on $B$:
\begin{equation}
B^{ab}{f_{ab}}^c = 0
\end{equation}

\paragraph     {The descent of the anomalous dimension}
The anomalous dimension of $V^{(0)}$ is $Q_0$-exact:
\begin{equation}
\Delta V^{(0)} = Q_0 U^{(0)}
\end{equation}
Let us act on this by $d$, and then use that $dV^{(0)} = Q_0V^{(1)}$:
\begin{align}
(\Delta - 1) dV^{(0)} = &\; Q_0 d U^{(0)}
\\
(\Delta - 1) Q_0 V^{(1)} = &\; Q_0 d U^{(0)}
\end{align}
Then use that there is no $Q_0$-cohomology in the conformal
dimension 1 and ghost number 1. Therefore exists $U^{(1)}$ such that:
\begin{align}
(\Delta - 1) V^{(1)}  = &\; dU^{(0)} + Q U^{(1)}
\label{DescentOfAnomDim}
\\
(\Delta - 2) dV^{(1)} = &\; Q dU^{(1)}
\\
(\Delta - 2) V^{(2)}  = &\; dU^{(1)}
\end{align}
Therefore:
\begin{equation}
\left[\begin{array}{c}
\mbox{anomalous dimension} \cr
\mbox{of the unintegrated vertex} \cr
\mbox{is BRST exact}\end{array}\right]
\Rightarrow
\left[\begin{array}{c}
\mbox{anomalous dimension} \cr
\mbox{of the integrated vertex} \cr
\mbox{is a total derivative}\end{array}\right]
\end{equation}
In our case Eq. (\ref{DescentOfAnomDim}) is $\log \epsilon$ times
Eq. (\ref{DescentOfCommutatorMiddleStep}); {\it i.e.} $U^{(1)}$ is proportional to
$g^{-1} \frac{d^2J}{ dl^2} g$.
\remv{Explanation}\rem{Because $dU_0$ should be equal to
the renormgroup of $V^{(1)}$ plus $Q$ of something.
Notice that the renormgroup of $V^{(1)}$ should not contain
the derivative of $\Lambda$, at least I don't see from which
diagramm a term with the derivative of $\Lambda$ would come.
The only possibility appears as stated.}

\paragraph     {Renormalization of the integrated vertex}
We can demonstrate that the integrated vertex has a nonzero
anomalous dimension in the case when $B^{ab}{f_{ab}}^c \neq 0$. Let us pick
a point in $AdS_5\times S^5$ and consider the
near flat space expansion around this fixed point as in 
\cite{Mikhailov:2007mr}. This means that we write  $g=e^{\vartheta/R}e^{x/R}$
and expand around the selected point $x = \vartheta = 0$. Suppose that
the only nonzero components of $B$ are in ${\bf g}_0\wedge {\bf g}_0$ where
${\bf g}_0$ corresponds to the rotations around the selected point. In other
words $B$ is $B^{[\mu\nu][\rho\sigma]}$. Then the integrated vertex
$V^{(2)}$ contains the terms:
\begin{equation}
B^{[\mu\nu][\rho\sigma]} \; x_{\mu} dx_{\nu} \;\wedge\; x_{\rho} dx_{\sigma}
\end{equation}
The log divergence comes from the contraction of $x_{\mu}$ and
$x_{\rho}$:
\begin{equation}
\log\varepsilon \; 
g_{\mu\rho} B^{[\mu\nu][\rho\sigma]} dx_{\nu} \;\wedge \; dx_{\sigma}
\end{equation}
This is indeed proportional to $B^{ab} {f_{ab}}^c$.

\remv{Berkovits and Chandia}\rem{
(start-process "gv" "*scratch*" "gv" 
"/home/andrei/a/Papers/Berkovits/AdS/BerkovitsChandia.pdf" "-page=11")}
\remv{Berkovits and Howe on anomalies}\rem{
(start-process "gv" "*scratch*" "gv" 
"/home/andrei/a/Papers/Berkovits/BerkovitsHowe.pdf" "-page=25" )
}
\remv{Log divergencies and conformal invariance}\rem{
(start-process "evince" "*scratch*" "evince" 
"/home/andrei/a/Work/rtt/Revise/operevised2.ps" "--page-label=17" )
}

\subsection{Example of an unintegrated vertex violating the constraint}
Pick a constant $A\in {\bf g}_2$ and  consider the following vertex operators:
\begin{align}
V_A(\epsilon,\epsilon') = & \; 
\mbox{Str}\left(\; A
   \left[ \left(g^{-1}(\epsilon\lambda_3 - \epsilon\lambda_1) g\right)_2 \;,\; 
   \left(g^{-1}(\epsilon'\lambda_3 - \epsilon'\lambda_1) g\right)_0 \right]
          \;\right)
\label{DefV20}
\end{align}
This operator corresponds to the following $B^{ab}$:
\begin{equation}
B^{ab} = \left\{
\begin{array}{lr}
{f^{ab}}_c A^c &
\begin{array}{r}
\mbox{ for } a \mbox{ in } {\bf g}_2 \mbox{ and } b \mbox{ in } {\bf g}_0 \cr
\mbox{  or } a \mbox{ in } {\bf g}_2 \mbox{ and } b \mbox{ in } {\bf g}_0 
\end{array} \cr
& \cr
0 & \mbox{ otherwise }
\end{array}
\right.
\end{equation}
The internal commutator is:
\begin{equation}
B^{ab} {f_{ab}}^c = C_{so(6)} A^c \neq 0
\end{equation}
where $C_{so(6)}$ is the adjoint Casimir of $so(6)$.
Therefore the internal commutator constraint is not satisfied
for this vertex. There is no such state in Type IIB SUGRA on 
$AdS_5\times S^5$. 

\subsection{What happens in the flat space limit}
In order to better understand this vertex we will consider its flat
space limit. We will use the flat space expansion similar to the
one used in \cite{Mikhailov:2007mr}. We will write:
\begin{equation}
g = e^{x_2 + \theta_3 + \theta_1}
\end{equation}
and consider $x$ and $\theta$ small. To reproduce the flat space BRST
operator we consider the ``flat space scaling'':
\begin{equation}\label{FlatSpaceScaling}
 x\simeq R^{-2}\;,\;\; \theta\simeq R^{-1}\;,\;\; \lambda\simeq R^{-1}
\end{equation}
$R$ is the radius of the AdS space entering the action as in 
(\ref{TheAction}). 

{\small
\commentstarts
Notice that there are two differences with \cite{Mikhailov:2007mr};
\cite{Mikhailov:2007mr} used a different gauge 
$g= e^{\theta} e^x$; also that paper used the ``uniform'' scaling
$x\simeq \theta\simeq \lambda\simeq R^{-1}$  
which is different from the ``flat space'' scaling 
(\ref{FlatSpaceScaling}) which we use here. In the flat space limit
the ``flat space scaling'' gives the BRST operator 
$\lambda^{\alpha}\left( {\partial\over\partial\theta^{\alpha}} +
\Gamma^m_{\alpha\beta} \theta^{\beta} {\partial\over\partial x^m} \right)$,
which is the correct BRST operator in flat space. While the ``uniform''
scaling $x\simeq \theta\simeq \lambda\simeq R^{-1}$ used in \cite{Mikhailov:2007mr} gives 
$\lambda^{\alpha}{\partial \over \partial\theta^{\alpha}}$.
\commentends
}

\noindent
With these notations the vertex operator becomes a function of
$x,\theta,\lambda$. The BRST operator in terms of $x,\theta,\lambda$ is calculated
in Appendix \ref{sec:NearFlatBRST}. The expansion of the vertex
(\ref{DefV20}) starts with the following terms:
\begin{align}
V_A(\epsilon,\epsilon') = \; \mbox{Str}\Big(A\;\Big( &
[[\theta_3,\epsilon\lambda_3],[\theta_1,\epsilon'\lambda_3]] +
[[\theta_1,\epsilon\lambda_1],[\theta_3,\epsilon'\lambda_1]] -
\nonumber
\\ 
-\; & [[\theta_3,\epsilon\lambda_1],[\theta_3,\epsilon'\lambda_3]] 
- [[\theta_1,\epsilon\lambda_1],[\theta_1,\epsilon'\lambda_3]] 
+ \ldots \Big) \Big)
\end{align}
where $\ldots$ stands for the terms of higher order in $1/R$ expansion.

We used a Mathematica program to recast $V_A(\epsilon,\epsilon')$ in various
BRST-equivalent forms. It turns out that $V_A(\epsilon,\epsilon')$ is
BRST equivalent to the following expression:
\begin{align}
\mbox{Str}\Big( A \Big( 
& - {8\over 9} [\;[\theta_3,[\theta_3,\epsilon\lambda_3]]\;,\;
                  [\theta_3,[\theta_3,\epsilon'\lambda_3]]\;] -
\nonumber
\\  
& - {8\over 9} [\;[\theta_1,[\theta_1,\epsilon\lambda_1]]\;,\;
[\theta_1,[\theta_1,\epsilon'\lambda_1]]\;] +\ldots \Big) \Big)
\label{Picture2002}
\end{align}
On the other hand, $V_A(\epsilon,\epsilon')$ is also equivalent to this:
\begin{align}
\mbox{Str}\Big( A \Big(\;
   4\; &[\;\; x\;\; , \;\; [\;[\theta_1 ,\epsilon  \lambda_1 ]\;,\;
               [\theta_3 ,\epsilon' \lambda_3 ]\;] \;\; ] -
\nonumber \\  
   - &  [\; [\theta_1 ,\epsilon \lambda_1] \;,\;
        [\theta_3 , [\theta_3 , [\theta_3 ,\epsilon'\lambda_3]]]\; ] -
\nonumber \\  
   - &  [\; [\theta_3 ,\epsilon \lambda_3] \;,\; 
        [\theta_1 , [\theta_1 , [\theta_1 ,\epsilon'\lambda_1]]]\; ] 
     +\ldots \Big) \Big)
\label{Picture11}
\end{align}
In both (\ref{Picture2002}) and (\ref{Picture11})
 $\ldots$ stands for  terms of the order $R^{-8}$ and higher in $1/R$ expansion.

We will call (\ref{Picture2002}) ``the (2,0)+(0,2)-gauge'' and
(\ref{Picture11}) ``the (1,1)-gauge''.

\paragraph{Flat space notations} 
Eqs. (\ref{Picture2002}) and (\ref{Picture11}) are written in terms
of the algebraic structures of $psu(2,2|4)$, the commutator and the
supertrace. It is possible to rewrite them using the gamma-matrices.

\vspace{7pt}

\noindent
The (2,0)+(0,2) gauge expression (\ref{Picture2002}) reads:
\begin{align}
- & {8\over 9} (\theta_3\Gamma_{klm}\theta_3)
(\theta_3 \Gamma^k \epsilon\lambda_3) 
 \overline{A}^l 
(\theta_3 \Gamma^m \epsilon'\lambda_3) -
\nonumber \\ 
- & {8\over 9} (\theta_1\Gamma_{klm}\theta_1)
(\theta_1 \Gamma^k \epsilon\lambda_1) 
 \overline{A}^l 
(\theta_1 \Gamma^m \epsilon'\lambda_1) + \ldots
\end{align}
where we denoted:
\begin{equation}
\overline{A}^l = \left\{
\begin{array}{rl}
 A^l  & \mbox{ if } l \in \{ 0, \ldots, 4 \} \cr
-A^l  & \mbox{ if } l \in \{ 5, \ldots, 9 \} 
\end{array} 
\right.
\end{equation}
On the other hand, the (1,1) expression (\ref{Picture11}) reads:
\begin{align}
& 2\; (\overline{A}_{[m} x_{n]} + A_{[m}\overline{x}_{n]})\;
(\theta_1 \Gamma^m \epsilon\lambda_1) (\theta_3 \Gamma^n \epsilon'\lambda_3) 
-  \nonumber 
\\
& 
- A_n(\theta_1 \Gamma_m \epsilon\lambda_1) 
              (\theta_3 \overline{\Gamma}^{[m} \Gamma^{n]} \Gamma^l \theta_3) 
              (\theta_3 \Gamma_l \epsilon'\lambda_3)
-  \nonumber
\\ 
&
- A_n(\theta_3 \Gamma_m \epsilon\lambda_3) 
              (\theta_1 \overline{\Gamma}^{[m} \Gamma^{n]} \Gamma^l \theta_1) 
              (\theta_1 \Gamma_l \epsilon'\lambda_1)
\end{align}
We leave the target-space interpretation of these states (even in
flat space) as an open question. 

\section{Symmetries of the vertex}
\label{sec:SymmetriesOfTheVertex}
In this section we will collect the necessary fact from the representation
theory and discuss the symmetries of our vertex.
\subsection{Representation theory}
\label{sec:RepresentationTheory}
We will denote: 
\begin{align}
{\bf g} = &\; psu(2,2|4) 
\\
\widehat{\bf g}  = &\; su(2,2|4)
\\
\widehat{\bf g}' = &\;  u(2,2|4)
\end{align}
\subsubsection{Matrix notations}
Elements of $\widehat{g}'$ are block matrices:
\begin{equation}
u^a_b =
\left( \begin{array}{cc}
u^i_j & u^i_{\alpha} \cr
u^{\alpha}_i & u^{\alpha}_{\beta} 
\end{array} \right)
\end{equation}
satisfying some hermiticity property. The precise form of the hermiticity
property will not be important for us. The indices $i,j,k,\ldots$ correspond to
the fundamental representation of the $su(4)$ and the indices $\alpha,\beta,\gamma,\ldots$
to the fundamental of the $su(2,2)$. The letters $a,b,c,\ldots$
stand for either $i,j,k,\ldots$ or $\alpha,\beta,\gamma,\ldots$.

\subsubsection{Exterior product of two adjoint representations}
Let us consider the exterior product of two adjoint representations 
of $\widehat{\bf g}'$:
\begin{equation}
\widehat{\bf g}' \wedge \widehat{\bf g}'
\end{equation}
In matrix notations, this is the space of matrices $b^{ac}_{bd}$ 
satisfying the antisymmetry property:
\begin{equation}
b^{ac}_{bd} = (-)^{(\bar{a} + \bar{b})(\bar{c} + \bar{d}) + 1} b^{ca}_{db}
\end{equation}
where $\bar{I}$ is $0$ if $I$ is the index of $su(2,2)$ and $1$ if $I$
is the index of $su(4)$. 

The difference between ${\bf g}$ and $\widehat{\bf g}'$ is in the central charge $c$ 
and the differentiation $s$. The central charge is the unit $8\times 8$ matrix, and
the differentiation is $\mbox{diag}(1,1,1,1,-1,-1,-1,-1)$.

Let ${\bf R} s$ denote the 1-dimensional linear space spanned by 
the differentiation and ${\bf R} c$ the 1-dimensional linear space spanned by the central
charge. Let us consider $\widehat{\bf g}' \wedge \widehat{\bf g}'$ as a representation
of ${\bf g}$. We observe that $\widehat{\bf g}' = {\bf g} + {\bf R}_s + {\bf R}_c$. 
Therefore:
\begin{equation}\label{DecompositionOfWedgeProduct}
\widehat{\bf g}' \wedge \widehat{\bf g}' = {\bf g} \wedge {\bf g} 
+ {\bf R}_s\otimes {\bf g} + {\bf R}_c\otimes {\bf g} + {\bf R}_s\otimes {\bf R}_c
\end{equation}
We observe the following facts about ${\bf g}\wedge {\bf g}$:
\begin{enumerate}
\item The representation ${\bf g}\wedge {\bf g}$ is not irreducible, because it
   contains two invariant subspaces:
\begin{align}
({\bf g}\wedge {\bf g})_0 & 
\;\; \mbox{consisting of} \;\; \sum_I x_I\wedge y_I 
\;\; \mbox{such that} \;\; \sum_I [x_I,y_I] = 0
\label{ggZero}
\\
{\bf g}\subset ({\bf g}\wedge {\bf g})_0 &\;\;\mbox{spanned by}\;\; f_a^{bc}t_b\wedge t_c
\label{gDiagonal}
\end{align}

\item We therefore have two exact sequences:
\begin{align}
0 \rightarrow {\bf g} \rightarrow ({\bf g}\wedge {\bf g})_0 
\rightarrow ({\bf g}\wedge {\bf g})_0 / {\bf g} \rightarrow 0
\label{FactorSpaceExactSequence}
\\
0 
\rightarrow ({\bf g}\wedge {\bf g})_0 
\rightarrow ({\bf g}\wedge {\bf g})
\rightarrow {\bf g} \rightarrow 0
\end{align}
Both of them {\em do not split}. This means that there is no complementary
subspace to (\ref{gDiagonal}) in $({\bf g}\wedge {\bf g})_0$ and no complementary
subspace to (\ref{ggZero}) in ${\bf g}\wedge {\bf g}$.
\end{enumerate}
We introduce an ``inner commutator'' map $F$:
\begin{equation}
F\;:\; {\bf g}\wedge {\bf g} \to {\bf g}\;,
\qquad 
F(\sum_I x_I\wedge y_I) = \sum_I [x_I,y_I] 
\end{equation}
With this notation $({\bf g}\wedge {\bf g})_0 = \mbox{Ker} \; F$.
Notice that $(x,y) = \mbox{Str}(xy)$ is a nondegenerate symmetric
scalar product, but this scalar product is not positive definite. 
Let $F^*$ denote the conjugate to $F$ with respect to this scalar product.
Because the Casimir operators vanish in the adjoint representation
we have:
\begin{equation}
FF^* = 0
\end{equation}
On the other hand,
\begin{equation}
F^*F \; : \;
{\bf g}\wedge {\bf g}\; \to \; ({\bf g} \wedge {\bf g})_0
\end{equation}
is non-zero. In fact $F^*F$ can be identified with the action of the
quadratic Casimir of ${\bf psu}(2,2|4)$ on ${\bf g}\wedge {\bf g}$:
\begin{equation}
\Delta_2 = C^{ab} t_a t_b = F^*F
\end{equation}
Note that $\Delta_2$ on ${\bf g}\wedge {\bf g}$ is nilpotent: $(\Delta_2)^2 = 0$. We conclude that:
\begin{itemize}
\item The space of linearized $\beta$-deformations 
$ ({\bf g}\wedge {\bf g})_0/{\bf g} $ can  be identified
with $\mbox{Ker}\;\Delta_2 \over \mbox{Im}\; \Delta_2$.
\end{itemize}

\subsubsection{Complex structure}
\label{sec:ComplexStrucure}
Notice that $\widehat{\bf g}' \wedge \widehat{\bf g}'$
has a complex structure, which acts as a multiplication by $i$ and the
exchange of the upper indices:
\begin{equation}
{\cal I}\; b^{IK}_{JL} = i \; b^{KI}_{JL}
\end{equation}
Notice that ${\cal I}^2 = -1$. Let us discuss the action of ${\cal I}$ on the
decomposition (\ref{DecompositionOfWedgeProduct}). We get:
\begin{align}
{\cal I}({\bf R}_c\otimes {\bf g}) = &\; 
({\bf g}\subset {\bf g}\wedge {\bf g})
\\
{\cal I}({\bf g}\subset {\bf g}\wedge {\bf g}) = &\;
({\bf R}_c\otimes {\bf g})
\end{align}
Generally speaking ${\cal I}(x\wedge y)$ has a component in ${\bf R}_s\otimes {\bf g}$, but when 
restricted on $({\bf g}\wedge {\bf g})_0$ it lands into  $({\bf g}\wedge {\bf g})_0 + {\bf R}_c\otimes {\bf g}$. 
We conclude that:
\begin{itemize}
\item the operation ${\cal I}$ induces a complex structure
on the space of linearized $\beta$-deformations 
$ ({\bf g}\wedge {\bf g})_0/{\bf g} $
\end{itemize}

\subsection{Our vertex is not covariant}
\label{sec:NotCovariant}
The linearized $\beta$-deformations transform in the following representation
of ${\bf g}={\bf psu}(2,2|4)$:
\begin{equation}
({\bf g}\wedge {\bf g})_0/{\bf g}
\end{equation}
But the $B$ tensor satisfying $ B^{ab}{f_{ab}}^c = 0 $ is in $({\bf g}\wedge {\bf g})_0$. The object
transforming in $({\bf g}\wedge {\bf g})_0/{\bf g}$ is the equivalence
class of $B^{ab} \simeq B^{ab} + {f^{ab}}_c A^c$. Let us denote this equivalence class $[B]$.
The short exact sequence
(\ref{FactorSpaceExactSequence}) does not split. Therefore
it is not possible to pick a representative for $B$ in the
equivalence class $[B]$ in a way consistent with the supersymmetry.

It was argued in \cite{Mikhailov:2009rx} that there is always a way
to choose the vertex covariantly, {\it i.e.} in a way consistent
with the supersymmetry. However the proof   used the assumption
that the representation of ${\bf g}$ in which the state transforms
has a sufficiently large spin when restricted to $so(6)\subset {\bf g}$.
The $\beta$-deformation is a low-spin case, so there is an obstacle
to choosing the vertex in a covariant way.

\section{General deformation theory}
\label{sec:GeneralDeformationTheory}
\subsection{Some general notations}
\label{sec:DefinesI}
We will consider the deformation of the Lagrangian of the following form:
\begin{equation}
{\cal L}^{(2)}_{deformed} = {\cal L}^{(2)} + \varepsilon V_1^{(2)} + 
\varepsilon^2 V_2^{(2)} + \ldots
\end{equation}
Here the upper index ${}^{(2)}$ indicates that the object is a 2-form ({\it e.g.}
the action density ${\cal L}^{(2)}$).
When we say that some equation is valid ``on-shell'' we will generally speaking
mean on-shell with respect to the undeformed Lagrangian ${\cal L}^{(2)}$.  We will
write:
\begin{equation}
F\simeq 0
\end{equation}
when $F$ is zero up to the equations of motion of ${\cal L}^{(2)}$. Given some
infinitesimal field transformation $\xi$ we will define $I^{(1)}_{\xi}$ by the following
formula:
\begin{equation}\label{DefinesI}
\xi . {\cal L}^{(2)} \simeq d I^{(1)}_{\xi}
\end{equation}
This equation holds on-shell, but we will consider it in situations where it
actually defines $I^{(1)}_{\xi}$ also off-shell. The only ambiguity would be to
add to $I^{(1)}_{\xi}$ some local conserved current, but for those $\xi$ which we
need there will be no local conserved currents with appropriate symmetries.  We
conclude that:
\begin{itemize}
\item for every vector field $\xi$ there is a 1-form $I_{\xi}^{(1)}$ defined by
   (\ref{DefinesI}); it is defined up to $d$ of something
\end{itemize}

\subsection{Deforming with integrated vertex operator}
\label{sec:DeformingWithIntegrated}
Let us return to the descent procedure discussed in Section
\ref{sec:VertexAndItsDescent} Eq. (\ref{DescentRelation}). 
The general relation is:
\begin{equation}
(d+Q) \left( V^{(0)} + V^{(1)} + V^{(2)} \right) = 0
\end{equation}
Given the  vertex operator $V$, we can
perturb the action by 
adding to it  $\int V^{(2)}$:
\begin{equation}\label{FirstOrderDeformation}
\int [dg\; d\lambda\; dw]\; e^{\int{\cal L}^{(2)}} \longrightarrow
\int [dg\; d\lambda\; dw]\; 
e^{\int\left({\cal L}^{(2)} + \varepsilon V_1^{(2)} + \ldots \right)}
\end{equation}
where $\varepsilon$ is an infinitesimally small parameter.  (The lower index
$1$ in $V_1^{(2)}$ is to indicate that $V_1^{(2)}$ is the coefficient
of the first power of $\varepsilon$.)

\subsubsection{BRST invariance at the first order in $\varepsilon$}
 The vertex
operator $V_1^{(2)}$ in (\ref{FirstOrderDeformation}) 
should be such that {\em on-shell} $Q_0 V^{(2)}_1$ is a total
derivative. Generally speaking this means that there exists an odd vector field,
which we will call $Q_1$, and a $1$-form $X_1^{(1)}$ such that:
\begin{equation}\label{DefQ1WithRHS}
Q_0 V_1^{(2)} + Q_1 {\cal L}^{(2)} = d X_1^{(1)}
\end{equation}
{\small
\paragraph     {Comment 1: }
Notice that Eq. (\ref{DefQ1WithRHS}) determines
$Q_1$ up to an infinitesimal transformation of ghost
number one which leaves the action invariant. We assume that there
are no such infinitesimal transformations except for $Q_0$
(in other words the BRST symmetry is the only symmetry with the 
ghost number one). Under this assumption the  infinitesimal 
transformation $Q_1$ is defined by
(\ref{DefQ1WithRHS})  unambiguously. 

\paragraph     {Comment 2: }
The combined $Q_0 + \varepsilon Q_1$ is a symmetry of the
action off-shell. But is this a nilpotent symmetry?
In other words, is it true that $\{Q_0,Q_1\}=0$? In fact this is true,
for the following reason. Observe that if $\{Q_0,Q_1\}$ is not zero, then
it would be a symmetry of the unperturbed action:
\begin{equation}
\{Q_0,Q_1\} S_0 = -Q_0^2V^{(2)} = 0
\end{equation}
Under the assumption that there are no symmetries of the ghost number 2,
we conclude that $\{Q_0,Q_1\}$ should be zero\footnote{See footnote on
p. 7 of \cite{Berkovits:2004xu}. A.M. would like to thank V.~Puletti for
a discussion about this.}. Therefore $Q_0 + \varepsilon Q_1$
is automatically nilpotent.
}

\vspace{9pt}

\subsubsection{BRST invariance at the second order in $\varepsilon$}
Eq. (\ref{DefQ1WithRHS}) guarantees that the deformation (\ref{FirstOrderDeformation})
exists at the first order in $\varepsilon$. Similarly, the consistency condition at the second
order in $\varepsilon$ is:
\begin{equation}\label{ConsistencyAtTheSecondOrder}
Q_1V_1^{(2)} + Q_0 V_2^{(2)} + Q_2{\cal L}^{(2)} = d X_2^{(1)}
\end{equation}
This equation is a definition of $Q_2$ and $V_2^{(2)}$; the existence of
$Q_2$ and $V_2^{(2)}$ satisfying (\ref{ConsistencyAtTheSecondOrder})
is the consistency condition. But it is more convenient to describe
the consistency condition in terms of the dimension zero operators
(unintegrated vertices). We will now translate the consistency
condition (\ref{ConsistencyAtTheSecondOrder}) from the dimension
two language to the dimension zero language. 

\subsubsection{The descent of $Q_1 V_1^{(2)}$}
Let us act on (\ref{DefQ1WithRHS}) with $Q_1$. 
We get\footnote{The only reason why this equation would not hold off-shell is
 the use of $Q_1^2 {\cal L} = dI^{(1)}_{Q_1^2}$.}:
\begin{equation}\label{Q0Q1OnV1}
-Q_0 Q_1 V_1^{(2)} \simeq d( Q_1X_1^{(1)} - I_{Q_1^2}^{(1)} )
\end{equation}
Therefore $ d (Q_0Q_1X_1^{(1)} - Q_0I_{Q_1^2}^{(1)}) \simeq 0 $.
In fact this is also true off-shell because there are no local 
conserved charges of the ghost number three:
\begin{equation}\label{DefinesW20}
Q_0 ( Q_1X_1^{(1)} - I^{(1)}_{Q_1^2} ) = d W_2^{(0)}
\end{equation}
This equation is the definition of $W_2^{(0)}$. An alternative
notation for $W_2^{(0)}$ could be $-(Q_1V_1^{(2)})^{(0)}$.  

\paragraph     {Condition on $W_2^{(0)}$}
Now we want to derive a constraint on $W_2^0$ following from
(\ref{ConsistencyAtTheSecondOrder}). On the left hand side of (\ref{Q0Q1OnV1}),
let us replace:
\[
Q_1V_1^{(2)} \; \rightarrow \; dX_2^{(1)} - Q_0 V_2^{(2)} - Q_2 {\cal L}
\]
as follows from (\ref{ConsistencyAtTheSecondOrder}), and
use $Q_2{\cal L} \simeq dI^{(1)}_{Q_2}$. We get:
\begin{equation}
d( - Q_0 X_2^{(1)} + Q_0I^{(1)}_{Q_2} ) \simeq d( Q_1X_1^{(1)} - I_{Q_1^2}^{(1)} )
\end{equation}
This implies:
\begin{equation}
 - Q_0 X_2^{(1)} + Q_0I^{(1)}_{Q_2} = Q_1X_1^{(1)} - I_{Q_1^2}^{(1)} + 
d(\mbox{smth})
\end{equation}
Therefore in this case $W_2^{(0)}$ defined by (\ref{DefinesW20})
is $Q_0$-exact:
\begin{equation}\label{W20IsExact}
W_2^{(0)} = Q_0(\mbox{smth})
\end{equation}

\subsubsection{Going back} 
Now suppose that (\ref{W20IsExact}) is satisfied:
\[
W_2^{(0)} = Q_0 T^{(0)}_2
\]
Then we get:
\begin{equation}\label{Q0OfQ1X1MinusSmth}
Q_0 ( Q_1X_1^{(1)} - I^{(1)}_{Q_1^2} - d T_2^{(0)} ) = 0
\end{equation}
Let us \underline{assume} that the following is true:
\begin{itemize}
\item the  cohomology of $Q_0$ on 1-forms of the  ghost number 2
   is trivial
\end{itemize}
Then (\ref{Q0OfQ1X1MinusSmth})
 implies the existence of $X_2^{(1)}$ such that:
\begin{equation}
Q_1X_1^{(1)} - I^{(1)}_{Q_1^2} = d T_2^{(0)} - Q_0 X^{(1)}_2
\end{equation}
Let us compare this to (\ref{Q0Q1OnV1}): 
$-Q_0 Q_1 V_1^{(2)} \simeq d( Q_1X_1^{(1)} - I_{Q_1^2}^{(1)} )$. We get:
\begin{equation}\label{Q0Q1V12Zero}
Q_0 ( Q_1 V_1^{(2)} - dX_2^{(1)} )  \simeq  0
\end{equation}
Let us \underline{assume} that:
\begin{itemize} 
\item the covariant cohomology of $Q_0$ on 2-forms of the
   ghost number 1 is trivial 
\end{itemize}
Then (\ref{Q0Q1V12Zero}) implies the existence of
$V_2^{(2)}$ and $Q_2$ such that (\ref{ConsistencyAtTheSecondOrder}). This means
that (\ref{W20IsExact}) is the necessary and sufficient condition
for the deformation to exist at the second order in $\varepsilon$.

\subsection{Comment about $Q_2$}
We will see in the next section that in our particular case 
(the beta-deformation) $W_2^{(0)}=0$ implies $Q_1^2=0$.
This implies that $Q_2=0$ (because Eq. (\ref{Q0Q1V12Zero}) holds
true off-shell; see the footnote before Eq. (\ref{Q0Q1OnV1})).

\subsection{Higher orders of perturbation theory}
\subsubsection{Going forward (necessary condition)}
Suppose that we have identified $V_p^{(2)}$ and $Q_p$ up to the order $n$, 
{\it i.e.} for $p=1,2,\ldots n$, so that:
\begin{eqnarray}\label{QVAndQQUPToN}
Q_0 V_p^{(2)} + Q_1 V_{p-1}^{(2)} + \ldots + Q_{p-1} V_1^{(2)} 
+Q_p {\cal L}^{(2)} & = & dX_p^{(1)}
\nonumber
\\[5pt]
Q_0 Q_p + Q_1 Q_{p-1} + \ldots + Q_{p-1} Q_1 + Q_p Q_0 & = & 0
\end{eqnarray}
Then:
\begin{eqnarray}
Q_1Q_0 V_n^{(2)} + Q_1Q_1 V_{n-1}^{(2)} + Q_1Q_2 V_{n-2}^{(2)} + 
\ldots + Q_1Q_{n-1} V_1^{(2)} 
+Q_1Q_n {\cal L}^{(2)} + 
\nonumber
\\[1pt]
+ 
Q_2Q_0 V_{n-1}^{(2)} + Q_2Q_1 V_{n-2}^{(2)} + \ldots + Q_2Q_{n-2} V_1^{(2)} 
+Q_2Q_{n-1} {\cal L}^{(2)} + 
\nonumber
\\[1pt]
+ 
Q_3Q_0 V_{n-2}^{(2)} + \ldots + Q_3Q_{n-3} V_1^{(2)} 
+Q_3Q_{n-2} {\cal L}^{(2)} + 
\nonumber
\\[1pt]
+ \ldots +
\nonumber
\\[1pt]
+ Q_n Q_0 V_1^{(2)} + Q_n Q_1 {\cal L}^{(2)} 
& \simeq &
\nonumber
\\[5pt]
\simeq \;
- Q_0( Q_1 V_n^{(2)} + Q_2 V_{n-1}^{(2)} + \ldots + Q_n V_1^{(2)} ) 
+ d I^{(1)}_{ Q_1Q_{n} + Q_2Q_{n-1} + \ldots + Q_{n}Q_1} \simeq
\nonumber
\\[5pt]
\simeq d(Q_1 X_n^{(1)} + Q_2 X_{n-1}^{(1)} + \ldots Q_n X_1^{(1)})
\label{QpXq}
\end{eqnarray}
This implies the existence of $W_{n+1}^{(0)}$ such that:
\begin{equation}\label{DefinitionOfWN}
Q_0( Q_1 X_n^{(1)} + \ldots + Q_n X_1^{(1)} - 
I^{(1)}_{ Q_1Q_{n} + \ldots + Q_{n}Q_1} ) = 
d W_{n+1}^{(0)}
\end{equation}
In other words, the validity of (\ref{QVAndQQUPToN}) for $p\in \{1,\ldots,n\}$
allows us to define $W_{n+1}^{(0)}$ by Eq. (\ref{DefinitionOfWN}).  Now, suppose
that we can construct $V_{n+1}^{(2)}$ and $Q_{n+1}$, so that
\begin{eqnarray}\label{VAndQEnPlusOne}
Q_0 V_{n+1}^{(2)} + Q_1 V_{n}^{(2)} + \ldots + Q_{n} V_1^{(2)} 
+Q_{n+1} {\cal L}^{(2)} & = & dX_{n+1}^{(1)}
\end{eqnarray}
Then this implies that $W^{(0)}_{n+1}$ satisfies some conditions.
Indeed, applying $Q_0$ to (\ref{VAndQEnPlusOne}) we get:
\begin{equation}
Q_0(Q_1V_n^{(2)} + \ldots + Q_n V_1^{(2)})
\simeq
dQ_0 ( X^{(1)}_{n+1} - I^{(1)}_{Q_{n+1}} ) 
\end{equation}
Now, returning to (\ref{QpXq}) we derive:
\begin{eqnarray}
d (Q_1 X_n^{(1)} + Q_2 X_{n-1}^{(1)} + \ldots Q_n X_1^{(1)} - 
I^{(1)}_{ Q_1Q_{n} + Q_2Q_{n-1} + \ldots + Q_{n}Q_1} ) & = &
\nonumber
\\[5pt]
= \quad 
- dQ_0(X^{(1)}_{n+1} - I^{(1)}_{Q_{n+1}})
\end{eqnarray}
and therefore there exists such a $T_{n+1}^{(0)}$ that:
\begin{eqnarray}
Q_1 X_n^{(1)} + Q_2 X_{n-1}^{(1)} + \ldots Q_n X_1^{(1)} - 
I^{(1)}_{ Q_1Q_{n} + Q_2Q_{n-1} + \ldots + Q_{n}Q_1}  & = &
\nonumber
\\[5pt]
= \quad 
- Q_0(X^{(1)}_{n+1} - I^{(1)}_{Q_{n+1}}) + dT_{n+1}^{(0)}
\end{eqnarray}
This implies:
\begin{equation}\label{EnthWExact}
W_{n+1}^{(0)} = Q_0 T_{n+1}^{(0)}
\end{equation}

\subsubsection{Going back (sufficient condition)}
Now suppose that (\ref{EnthWExact}) holds. Then 
\begin{equation}
Q_0( Q_1 X_n^{(1)} + \ldots + Q_n X_1^{(1)} - 
I^{(1)}_{ Q_1Q_{n} + \ldots + Q_{n}Q_1} - dT_{n+1}^{(0)}) = 0
\end{equation}
Assuming the triviality of the cohomology of $Q_0$ on the 1-forms
of the ghost number 2, we conclude the existence of
$X^{(1)}_{n+1}$ such that:
\begin{equation}
Q_1 X_n^{(1)} + \ldots + Q_n X_1^{(1)} - 
I^{(1)}_{ Q_1Q_{n} + \ldots + Q_{n}Q_1} = dT_{n+1}^{(0)} - Q_0 X_{n+1}^{(1)}
\end{equation}
This and (\ref{QpXq}) implies, under the assumption that
the cohomology of $Q_0$ on the 2-forms of the ghost number 1
is zero, the existence of $V^{(2)}_{n+1}$ such that:
\begin{equation}\label{EnthOnShell}
Q_1 V_n^{(2)} + Q_2 V_{n-1}^{(2)} + \ldots + Q_n V_1^{(2)} \simeq
dX^{(1)}_{n+1} - Q_0 V^{(2)}_{n+1}
\end{equation}
This means that $W^{(0)}_{n+1}$ being $Q_0$-exact is not only a necessary, but
also a sufficient condition to be able to extend the deformation to the order
$n+1$. The off-shell version of Eq. (\ref{EnthOnShell}):
\begin{equation}\label{EnthOffShell}
  Q_0 V^{(2)}_{n+1} + Q_1 V_n^{(2)} + Q_2 V_{n-1}^{(2)} 
+ \ldots + Q_n V_1^{(2)} + Q_{n+1}{\cal L} = dX^{(1)}_{n+1}
\end{equation}
This is the definition of $Q_{n+1}$. Notice that so defined $Q_{n+1}$
satisfies:
\begin{equation}
( Q_0 + \varepsilon Q_1 + \varepsilon^2 Q_2 
+ \ldots + \varepsilon^{n+1}Q_{n+1} )^2
= O(\varepsilon^{n+2})
\end{equation}
The proof goes by induction. We start with $Q_0^2 = 0$. The induction
hypothesis is 
$(Q_0 + \varepsilon Q_1 + \ldots + \varepsilon^n Q_n)^2 = O(\varepsilon^{n+1})$,
and this guarantees that $(Q_0+\varepsilon Q_1+\ldots + \varepsilon^{n+1}Q_{n+1})^2$
is also at least as small as $O(\varepsilon^{n+1})$: 
\begin{equation}\label{AtLeastAsSmallAs}
(Q_0 + \varepsilon Q_1 + \ldots + \varepsilon^{n+1}Q_{n+1})^2 
=
O(\varepsilon^{n+1})
\end{equation}
By construction:
\begin{equation}
\int 
(Q_0 + \varepsilon Q_1 + \ldots + \varepsilon^n Q_n + \varepsilon^{n+1}Q_{n+1}) 
({\cal L}^{(2)} + \varepsilon V^{(2)}_1 + \ldots
+ \varepsilon^{n+1} V^{(2)}_{n+1}) = O(\varepsilon^{n+2})
\end{equation}
Therefore:
\begin{equation}
\int 
(Q_0 + \varepsilon Q_1 + \ldots + \varepsilon^n Q_n + \varepsilon^{n+1}Q_{n+1})^2
({\cal L}^{(2)} + \varepsilon V^{(2)}_1 + \ldots
+ \varepsilon^{n+1} V^{(2)}_{n+1}) = O(\varepsilon^{n+2})
\end{equation}
This and (\ref{AtLeastAsSmallAs}) imply:
\begin{equation}\label{QSquareL}
\int 
(Q_0 + \varepsilon Q_1 + \ldots + \varepsilon^n Q_n + \varepsilon^{n+1}Q_{n+1})^2
{\cal L}^{(2)} = O(\varepsilon^{n+2})
\end{equation}
Let $P_{n+1}$ denotes the coefficient of $\varepsilon^{n+1}$:
\begin{equation}
(Q_0 + \varepsilon Q_1 + \ldots + \varepsilon^n Q_n + \varepsilon^{n+1}Q_{n+1})^2 =
\varepsilon^{n+1} P_{n+1} + \ldots
\end{equation}
Then (\ref{QSquareL}) implies that $P_{n+1}$ is a symmetry of the undeformed
theory: $\int P_{n+1}{\cal L} = 0$. 
Under the assumption that the pure spinor superstring in $AdS_5\times S^5$
does not have conservation laws of the ghost number two, it follows that
$P_{n+1}=0$. This completes the step of the induction.

\paragraph     {Conclusion} 
Provided that the deformation of the action is
defined up to the order $n$ in $\varepsilon$, the obstacle to defining
the deformation to the order $n+1$ is the $Q_0$ cohomology class of
$W^{(0)}_{n+1}$.

\section{Applying the general theory to beta-deformation}
\label{sec:ApplyingToBeta}
\subsection{Calculation of $Q_1$}
\label{sec:CalculationOfQ1}
The off-shell version of Eq. (\ref{DescentFromLambdaStep3}) for $Q V_1^{(2)}$ is:
\begin{equation}\label{QJJOffshell}
\epsilon Q \left( {1\over 2}B^{ab}\;j_{a}\wedge j_{b} \right) =
B^{ab}\; d\Lambda_{a}(\epsilon)\wedge j_b 
\end{equation}    
For the deformed action to be BRST invariant off-shell, we need to modify the BRST
transformation:
\begin{equation}
Q=Q_0 + Q_1 
\end{equation}
where $Q_0$ is the original (pure $AdS_5\times S^5$) BRST transformation, and
$Q_1$ is the modification.  

We will now argue that  $Q_1$ is in fact a $psu(2,2|4)$ transformation
with the {\em space-time-dependent parameter}. First of all, notice that under
the global rotations:
\begin{equation}
g\to gg_0 \;,\;\; g_0=\mbox{const}
\end{equation}
the Lagrangian is invariant. But what will happen if we allow $g_0$ to depend on
$\tau^{\pm}$? Consider an infinitesimal transformation: 
\begin{equation}\label{LocalizedSymmetry}
\delta g = g \alpha \quad \mbox{where }\alpha=\alpha(\tau^+,\tau^-)
\end{equation}
Then: 
\begin{equation}\label{DeltaAlphaJ}
\delta J = -g \; d\alpha \; g^{-1}
\end{equation}
and the variation of the Lagrangian is:
\begin{eqnarray}
\delta {\cal L}  & = &  {1\over 4}\; \mbox{Str} \; \left(
g^{-1} {dJ\over dl} g \; \wedge d\alpha \right) = 
\nonumber \\ 
& = & {1\over 4} \mbox{Str}\; 
\left[
(\hat{j}_+ + 4 g^{-1}w_{1+}^{\star}g)\partial_-\alpha -
(\hat{j}_- - 4 g^{-1}w_{3-}^{\star}g)\partial_+\alpha
\right] d\tau^+\wedge d\tau^-
\label{DeltaVariationOfL}
\end{eqnarray}
\commentstarts
{\small
In order to derive this formula, it is useful to rewrite the Lagrangian
(\ref{TheAction})
in the following interesting form:
\begin{eqnarray}
{\cal L} & = &  \mbox{Str}\left( 
{1\over 4} J_+ \left.{dJ_-\over dl}\right|_{l=0} 
+ w_{1+}\partial_-\lambda_3 + w_{3-}\partial_+\lambda_1 
- N_{0+}N_{0-} - w^{\star}_{1+} w^{\star}_{3-}  \right) =
\nonumber
\\
& = &  \mbox{Str}\left( 
- {1\over 4} \left.{dJ_+ \over dl}\right|_{l=0} J_- 
+ w_{1+}\partial_-\lambda_3 + w_{3-}\partial_+\lambda_1 
- N_{0+}N_{0-} - w^{\star}_{1+} w^{\star}_{3-}  \right) 
\end{eqnarray}
and then use (\ref{DeltaAlphaJ}).
}
\commentends

\vspace{8pt}

\noindent
We are now ready to calculate $Q_1$. To start, let us consider the following
infinitesimal transformation $\Xi_{\alpha}$: 
\begin{equation}
\Xi_{\alpha}: \;\;\;
\Xi_{\alpha} g  = g\alpha \;,\;\;
\Xi_{\alpha} w^{\star}_{3-} = 
{\cal P}_{31}(g \; \partial_-\alpha \; g^{-1})_3 \;,\;\;
\Xi_{\alpha} w^{\star}_{1+} = 
{\cal P}_{13}(g \; \partial_+\alpha \; g^{-1})_1
\end{equation}
This transformation combines the ``localized'' rotation (\ref{LocalizedSymmetry})
with the shift of the antifields $w^{\star}$.  Here the projectors ${\cal P}_{13}$
and ${\cal P}_{31}$ are defined by the formulas:
\begin{eqnarray}
{\cal P}_{13} A_1 & = & A_1 + [\lambda_3,\mbox{smth}_2] 
\nonumber
\\[1pt]
[\lambda_1, {\cal P}_{13} A_1] & = & 0
\nonumber
\\[8pt]
{\cal P}_{31} A_3 & = & A_3 + [\lambda_1,\mbox{smth}_2] 
\nonumber
\\[1pt] 
[\lambda_3, {\cal P}_{31} A_3] & = & 0
\end{eqnarray}
The purpose of these projectors in (\ref{QOneOfBareW}) is to 
 enforce the constraints (\ref{ConstraintsOnAntifields}). 

\paragraph     {On the definition of ${\cal P}$.} 
In this paragraph we will
prove the existence of the projector ${\cal P}_{13}$ satisfying
these properties. We will start with the following lemma: 

{\bf Lemma \ref{sec:CalculationOfQ1}.1:} 
If $\{[S_2,\lambda_3],\lambda_1\} = 0$ then $[S_2,\lambda_3]=0$. 

{\bf Proof:} Notice that
$\{[S_2,\lambda_3],\lambda_1\}\in {\bf C}\otimes {\bf g}_2$ (the complexification of ${\bf g}_2$). 
Using the spinor notations: 
\rem{(start-process "evince" "*scratch*" "evince" 
  "/home/andrei/a/Work/cohomology/vert.ps" "--page-label=8")}
$ \{[S_2,\lambda_3],\lambda_1\}_m = 
(\lambda_1,\Gamma_m \widehat{F} S_2^n\Gamma_n \; \lambda_3) $ where $\widehat{F}$ is the Ramond-Ramond
5-form field strength of $AdS_5\times S^5$ contracted with the Gamma-matrices. 
Then $\{[S_2,\lambda_3],\lambda_1\} = 0$
would imply that 
\begin{equation}\label{Lambda1XSLambda3}
(\lambda_1,X_2^m\Gamma_m \widehat{F} S_2^n\Gamma_n \; \lambda_3) = 0
\end{equation}
for any vector $X_2$. Let us introduce the notation $\overline{X}_2$ for the
vector with the components $X^m_2$ for $m\in \{0,\ldots,4\}$ and $-X^m_2$ for
$m\in \{5,\ldots,9\}$.  Let $\overline{X}_2$ run over the space of vectors
annihilating $\lambda_3$. Then for such $X_2$ (\ref{Lambda1XSLambda3})
becomes\footnote{Notice that $\mbox{Str}(\overline{X}_2 S_2)$ is the scalar
  product of $\overline{X}_2$ and $S_2$.}:
\begin{equation}
\mbox{Str}(\lambda_1\lambda_3)\; \mbox{Str}(\overline{X}_2S_2)
\end{equation}
Therefore
our assumption that $\{[S_2,\lambda_3],\lambda_1\}=0$ implies that
the scalar product of $S_2$ and $\overline{X}_2$ is zero for any $\overline{X}_2$ such that
$[\overline{X}_2,\lambda_3]=0$ ({\it i.e.} for $\overline{X}_2\in \mbox{Ann}(\lambda_3)$).
Because $\mbox{Ann}(\lambda_3)$ has the maximal
dimension possible of a null-plane,
this implies that $S_2$ itself belongs to the annihilator
of $\lambda_3$, {\it i.e.} $[S_2,\lambda_3]=0$. This proves the {\bf Lemma}.

\vspace{5pt}
\commentstarts
{\small If this was not the case, then the action would have 
a gauge symmetry $w_{1+}^{\star}\mapsto w_{1+}^{\star} + [S_{2+},\lambda_3]$
with $S_{2+}$ satisfying $\{[S_{2+},\lambda_3],\lambda_1\} = 0$.}
\commentends

\vspace{8pt}\noindent
We use the notation for the annihilator
of a pure spinor:
\begin{eqnarray}
\mbox{Ann}(\lambda_3) & = & \mbox{the subspace of }{\bf g}_2
\mbox{ consisting}
\nonumber
\\
&& \mbox{of vectors } X_2 \mbox{ such that }
[X_2,\lambda_3]=0
\end{eqnarray}
Notice that $\mbox{Ann}(\lambda_3)$ is a $5_{\bf C}$-dimensional null-subspace\footnote{A pure spinor
defines a null-plane of the maximal possible dimension.} of the 
complexification of ${\bf g}_2$. 
Our Lemma \ref{sec:CalculationOfQ1}.1 implies that:
\begin{equation}
\mbox{Ker}\left( 
X_2 \mapsto \{[X_2,\lambda_3],\lambda_1\}
\right) \; = \; \mbox{Ann}(\lambda_3)
\end{equation}
Therefore the image of the map
$X_2 \mapsto \{[X_2,\lambda_3],\lambda_1\}$ is also 
a $5_{\bf C}$-dimensional space (because ${\bf C}\otimes {\bf g}_2$
is $10_{\bf C}$-dimensional).  This image is  a subspace of $\mbox{Ann}(\lambda_1)$.
But notice that  $\mbox{Ann}(\lambda_1)$
 is itself $5_{\bf C}$-dimensional. This implies:
\begin{equation}
\mbox{Im}(X_2\mapsto \{[X_2,\lambda_3],\lambda_1\}) = \mbox{Ann}(\lambda_1)
\end{equation}
Given that $\{\lambda_1,A_1\}$ is in the annihilator of $\lambda_1$,
we conclude that there exists such $X_2$ that
$\{\lambda_1,A_1\} + \{\lambda_1,[\lambda_3,X_2]\} = 0$. This 
proves the existence of the projector ${\cal P}_{13}$. The existence
of ${\cal P}_{31}$ can be proven similarly.

\vspace{10pt}
\noindent
The Lagrangian depends on the antifields $w^{\star}$ through the last term in (\ref{TheAction});
therefore the part of the Lagrangian involving $w^{\star}$ is:
\begin{equation}
{\cal L}^{AF} = - \; \mbox{Str} (w_{1+}^{\star} w_{3-}^{\star})
\end{equation}
When we calculate the variation $\Xi_{\alpha}{\cal L}^{AF}$, the projectors
${\cal P}_{13}$ and ${\cal P}_{31}$ drop out 
because of the constraint (\ref{ConstraintsOnAntifields}). We get:
\begin{equation}\label{XiLAF}
\Xi_{\alpha} {\cal L}^{AF} = 
- \mbox{Str}(\partial_+\alpha \; g w_{3-}^{\star} g^{-1}) 
- \mbox{Str}(g w_{1+}^{\star} g^{-1} \;\partial_-\alpha)
\end{equation}
Combining (\ref{XiLAF}) and (\ref{DeltaVariationOfL}) we get:
\begin{equation}\label{XiL}
\Xi_{\alpha} {\cal L}  =   {1\over 4} \mbox{Str}\; 
\left[ j_+ \partial_-\alpha - j_- \partial_+\alpha \right] 
d\tau^+\wedge d\tau^- = - {1\over 4} \;\mbox{Str}\; d\alpha \wedge j
\end{equation}
Now we observe that $Q_1$ is an example of $\Xi_{\alpha}$ for a particular value of
$\alpha$, namely $\alpha=\Lambda_a B^{ab} t_b$:
\begin{align}
Q_1 = & \; 4 \left[ B^{ab} \Lambda_a t_b \; + \;
\partial_+\Lambda_a B^{ab}\;
\left({\cal P}_{13}(g t_b g^{-1})_1 \right)^{\dot{\alpha}}
\; {\delta\over\delta w_{1+}^{\star\dot{\alpha}}}  
+ \right.
\nonumber \\ 
&\phantom{\;\;4[ B^{ab} \Lambda_a t_b \;} + \;
\left.
\partial_-\Lambda_a B^{ab}\;
\left({\cal P}_{31}(g t_b g^{-1})_3 \right)^{\alpha} 
\; {\delta\over\delta w_{3-}^{\star\alpha}}  
\right]
\label{QOneOfBareW}
\end{align}
Indeed, Eqs. (\ref{XiL}) and (\ref{QJJOffshell}) imply that so defined 
$Q_1$ satisfies:
\begin{equation}
Q_1{\cal L} + Q_0V^{(2)}_1 = 0
\end{equation}
which is the defining equation of $Q_1$.

\commentstarts
{\small 
In the first term $B^{ab}\Lambda_at_b$ of 
Eq. (\ref{QOneOfBareW})  $t_b$ stands for the 
generators of the $psu(2,2|4)$ rotations,
which act only on the matter fields: $\delta_a g = gt_a$ (they do
not touch the ghosts and the antifields). 
}
\commentends

\subsection{Deformation of the BRST current}
\label{sec:DeformationOfBRSTCurrent}
\subsubsection{General procedure for calculating the current density}
To calculate the deformed charge density we will use a well-known
general procedure. Given the action $S[\phi]$ invariant under some global
symmetry transformation $\delta\phi^a = \xi^a$ we consider the 
position-dependent transformation $\delta_u\phi^a = u(\tau,\sigma) \xi^a$
and calculate the variation of the action. This should be proportional
to the derivatives of $u$ and (as any variation) should vanish on-shell:
\begin{equation}
\delta_u S = \int ( j_+ \partial_- u - j_- \partial_+ u ) \simeq 0
\end{equation}
Then it follows that the current $j_{\pm}$ is conserved:
$\partial_+j_- - \partial_-j_+ = 0$. 
\subsubsection{Particular case of BRST transformation}
\paragraph     {Original (undeformed) BRST current}
Let us start by deriving the BRST charge in the case of pure 
$AdS_5\times S^5$. Let us concentrate on $Q_L$. In this subsection
$\delta$ will stand for $Q_L$, and $\delta_u$ for $Q_L$ with
the replacements:
\begin{itemize}
\item $\delta g = \epsilon \lambda_3 g$ replaced with $\delta_u g = \epsilon u \lambda_3 g$ 
\item $\delta w_{1+} = -\epsilon J_{1+} - \epsilon w_{1+}^{\star}$ replaced with $\delta_u w_{1+} = -u\epsilon J_{1+} -u\epsilon w_{1+}^{\star}$ 
\item $\delta w_{3-}^{\star} = D_{0-}\epsilon\lambda_3 - [N_{0-},\epsilon \lambda_3]$ replaced with $\delta_u w_{3-}^{\star} = u D_{0-}\epsilon\lambda_3 - u [N_{0-},\epsilon\lambda_3]$ 
\end{itemize}
We can tautologically rewrite:
\begin{align}
\delta_u S[g,\lambda,w,w^{\star}] = & \phantom{+}
(\delta S)[g, u\lambda, u^{-1}w, w^{\star}] + 
\nonumber
\\
& + (\delta_u S[g, \lambda, w, w^{\star}] 
 -(\delta S)[g, u\lambda, u^{-1}w, w^{\star}])
\label{NoetherScaling}
\end{align}
We observe that $(\delta S)[g, u\lambda, u^{-1}w, w^{\star}] = 0$ because the action is 
BRST-invariant. On the other hand, in the second line of  
(\ref{NoetherScaling}), the  difference between
$\delta_u S[g, \lambda, w, w^{\star}] $ and $ (\delta S)[g, u\lambda, u^{-1}w, w^{\star}] $ is in two places:
\begin{itemize}
\item  the variation of the term $(w_{1+}\partial_-\lambda_3)$
\item  the variation of $-w_{1+}^{\star} w_{3-}^{\star}$
\end{itemize}
Therefore:
\begin{align}
  \epsilon\delta_u S[g, \lambda, w, w^{\star}] 
 -\epsilon(\delta S)[g, u\lambda, u^{-1}w, w^{\star}]
= \int d\tau d\sigma \; \partial_- u \; 
\mbox{Str}&  \Big( \epsilon(J_{1+} + w_{1+}^{\star}) \lambda_3 +
\nonumber
\\
& \; + w_{1+}^{\star} \epsilon \lambda_3 \Big) 
\end{align}
The terms containing $w_{1+}^{\star}$ cancel out, and we get:
\begin{align}
\epsilon j_{L+} = & -\mbox{Str}\big( J_{1+}  \epsilon\lambda_3 \big)
\\
\epsilon j_{L-} = & \;0
\end{align}
Notice that the deformation of the BRST transformation given by
Eq. (\ref{QOneOfBareW}) does not contribute to the deformation 
of the BRST current.

\paragraph     {Deformation of the BRST current}
The additional term in the action is:
\begin{equation}
S^{(2)} = \int d\tau d\sigma 
B^{ab}\left( g^{-1}\left( {dJ_+\over dl} - 4w_{1+}^{\star} \right) g \right)_a
\;   \left( g^{-1}\left( {dJ_-\over dl} +  4w_{3-}^{\star} \right) g \right)_b
\end{equation}
Remember that $Q w_{3-}^{\star} = D_{0-}\lambda_3 - [N_-,\lambda_3]$
and therefore we get:
\begin{align}
& \epsilon\delta_u S^{(2)}[g, \lambda, w, w^{\star}] 
 -\epsilon(\delta S^{(2)})[g, u\lambda, u^{-1}w, w^{\star}]
=
\\
& = - \int d\tau d\sigma \;
B^{ab} \; \left( g^{-1}\left( {dJ_+\over dl} - 4w_{1+}^{\star} \right) g \right)_a
\;   \left( g^{-1}  (4\partial_- u \epsilon\lambda_3)  g \right)_b
\end{align}
This means that the deformed BRST current is:
\begin{align}
\epsilon j_{L+} = & 
- \mbox{Str}\big( J_{1+} \epsilon\lambda_3 \big)
- 4 B^{ab} j_{a+} (g^{-1}\epsilon\lambda_3 g)_b
\label{JPlusDeformed}
\\
\epsilon j_{L-} = & \;0
\end{align}
\subsubsection{Conservation of the deformed current}
\paragraph     {Deformed equations of motion}
Under the variation $\delta_{\xi_3} g = \xi_3 g$ we get:
\begin{align}
\delta_{\xi} J_+ = & \; 
- D_+ (z^{-1} \xi_3) + (z^{-5} - z^{-1}) [N_+,\xi_3] 
+ (z^{-4} - 1) [J_{1+},\xi_3] 
\\
\delta_{\xi} J_- = & \;
- D_- (z^{-1} \xi_3) + (z^3 - z^{-1}) [N_-,\xi_3] 
+ (z^{-1} - z^3) D_{0-}\xi_3
\end{align}
In particular:
\begin{align}
\left.{d\over dl}\right|_{l=0} \delta_{\xi} J_+ = & \;
D_+\xi_3 - \left[ {dJ_+\over dl} , \xi_3 \right] 
- 4 [ N_+,\xi_3 ] - 4 [J_{1+},\xi_3]
\\
\left.{d\over dl}\right|_{l=0} \delta_{\xi} J_- = & \;
D_-\xi_3 - \left[{dJ_-\over dl}, \xi_3\right] 
+ 4 [ N_-,\xi_3 ] - 4 D_{0-} \xi_3
\end{align}
The variation of the undeformed action gives:
\begin{equation}
\delta_{\xi} S_0 = \int \mbox{Str}\left( 
\xi_3 (D_{0-} J_{1+} - [N_-,J_{1+}] + [J_{1-},N_+]) \;
\right)
\end{equation}
The variation of the ``small case currents'' is:
\begin{align}
\delta_{\xi} j_{+} = & \; 
g^{-1}\left( - 4 [N_+,\xi_3] - 4 [J_{1+},\xi_3] \right)g 
+ \partial_+(g^{-1}\xi_3 g)
\\
\delta_{\xi} j_-  = & \;
g^{-1}\left(   4 [N_-,\xi_3] + 4 [J_{1-}+J_{2-}+J_{3-}\; ,\; \xi_3]\right)g
- 3 \partial_-(g^{-1}\xi_3 g)
\end{align}
\remv{Calculation}\rem{(start-process "mirage" "*scratch*" "mirage" 
"photos/holomorphicity__var-curr.jpg" )}
Therefore:
\begin{align}
\delta_{\xi} \int d\tau d\sigma \;
B^{ab} j_{a+} j_{b-} = 
B^{ab}\left( g^{-1} \left(- 4 [N_+,\xi_3] - 4 [J_{1+},\xi_3]\right)g\right)_a j_{b-} +
\nonumber 
\\
+ B^{ab} j_{a+} 
\left(g^{-1}\left( 
4 [N_-,\xi_3] + 4 [J_{1-}+J_{2-}+J_{3-}\; ,\; \xi_3]
\right)g\right)_b +
\nonumber
\\
+ B^{ab}\left(
\partial_+ (g^{-1}\xi_3 g)_a j_{b-} - 3 j_{a+} \partial_- (g^{-1}\xi_3 g)_b
\right)
\end{align}
The last term is equivalent to 
$\int 4 (\partial_- j_{a+}) B^{ab} (g^{-1}\xi_3 g)_b $. 
We conclude that the deformed equation of motion for $J_{1+}$ is:
\begin{align}
& D_{0-} J_{1+} - [N_-,J_{1+}] + [J_{1-},N_+] +
\nonumber
\\
+ & \; 4 [N_+ + J_{1+}\;,\;gt_ag^{-1}]_1 B^{ab} j_{b-} 
     - 4 j_{a+}B^{ab} [N_- + J_{1-} + J_{2-} + J_{3-}\;,\; gt_bg^{-1}]_1 +
\label{DeformedEquationForJ1Plus}
\\
+ & \; 4 B^{ab} (\partial_- j_{a+}) (gt_bg^{-1})_1 = 0
\nonumber
\end{align}
We will also need the deformed equations of motion for $\lambda_3$, which is
obtained by varying the action with respect to $w_{1+}$:
\begin{equation}\label{DeformedEquationForLambda3}
D_{0-}\lambda_3 - [N_- , \lambda_3] -
4B^{ab} [ (gt_ag^{-1})_0, \lambda_3] j_{b-} = 0
\end{equation}
\paragraph     {Holomorphicity of the current}
Consider the derivative of the current given by Eq. (\ref{JPlusDeformed}):
\begin{align}
-\partial_- j_{L+} = & \;
\mbox{Str}\; \left( 
J_{1+} (D_{0-}\epsilon\lambda_3 - [N_-,\epsilon\lambda_3]) +
(D_{0-}J_{1+} - [N_-,J_{1+}]) \epsilon\lambda_3 \right) +
\nonumber
\\
& +
\partial_- (4 B^{ab} j_{a+} (g^{-1}\epsilon\lambda_3 g)_b)
\end{align}
Substitution of (\ref{DeformedEquationForJ1Plus}) and
(\ref{DeformedEquationForLambda3}) into this formula gives $\partial_-j_{L+}=0$.
\remv{Calculation}\rem{
(start-process "mirage" "*scratch*" "mirage" "photos/holomorphicity__dj-corrected.jpg" )}

\subsection{Relation between $W_2^{(0)}$ and the Schouten
bracket on $\Lambda^{\bullet}{\bf g}$}
\label{sec:Q1VAndGBracket}
Equation (\ref{W20IsExact}) tells us that a necessary condition
for the deformed theory to be BRST invariant to the order $\varepsilon^2$
is that $W_2^{(0)}$ is BRST exact.
Here we will explicitly calculate $W_2^{(0)}$ for the beta-deformation
and express it in terms of the
Schouten bracket on $\Lambda^2{\bf g}$. 

We start with the observation that actually:
\begin{equation}\label{NoTotalDerivative}
Q_0V_1^{(2)}+Q_1{\cal L}^{(2)} = 0
\end{equation}
In other words $X_1^{(1)}=0$.

\commentstarts
{\small
Notice that generally speaking we  only have 
Eq. (\ref{DefQ1WithRHS}), but in our particular case
$V= {1\over 2} B^{ab}j_a\wedge j_b$ we claim a stronger Eq. (\ref{NoTotalDerivative}).
}
\commentends

It remains to calculate $I_{Q_1^2}$.  Let us first calculate $Q_1^2$.  Let us
split $Q_1 = Q_1^{\bf F} + Q_1^{\bf AF}$ where $Q_1^{\bf F}=4 \Lambda_a B^{ab} t_b$ 
is the first term on
the right hand side of (\ref{QOneOfBareW}) and $Q_1^{\bf AF}$ is the sum of the remaining
two terms
({\it i.e.} $\bf F$ stands for fields and $\bf AF$ for antifields).  
We get:
\begin{eqnarray}
[Q_1^{\bf F}(\epsilon), Q_1^{\bf F}(\epsilon')] & = & \phantom{2\times} 16
\Lambda_a(\epsilon) B^{ap} \Lambda_b(\epsilon') B^{bq}
{f_{pq}}^rt_r + 
\nonumber
\\[1pt]
 & + & 
16\times 2\Lambda_a(\epsilon) B^{ap} {f_{pb}}^r \Lambda_r(\epsilon') B^{bq} t_q =
\nonumber
\\[5pt]
 & = &
16\times 3\; B^{p[a} {f_{pq}}^b B^{c]q} \Lambda_a(\epsilon) \Lambda_b(\epsilon') t_c
\end{eqnarray}
\rem{(link: "q1-square\.jpg" "beta" "mirage")}
This has similar structure to  $Q_1^{\bf F}\;\;$; namely it is a $psu(2,2|4)$-rotation,
but not a global symmetry because the parameter of the rotation
is space-time dependent. We have 
$[Q_1^{\bf AF}(\epsilon), Q_1^{\bf AF}(\epsilon')] = 0$,
and $[Q_1^{\bf F}(\epsilon), Q_1^{\bf AF}(\epsilon')]$ is given by:
\begin{eqnarray}
[Q_1^{\bf F}(\epsilon), Q_1^{\bf AF}(\epsilon')] w^{\star}_{1+} & = &
16 \; B^{cd} B^{eb} {f_{de}}^a 
\partial_+ (\Lambda_c(\epsilon) \Lambda_a(\epsilon'))
\;{\cal P}_{13} (g t_b g^{-1})_1 -
\nonumber
\\ 
& - & 
16 \; B^{cd} B^{eb} {f_{de}}^a 
\Lambda_c(\epsilon)\partial_+ \Lambda_b(\epsilon') 
\;{\cal P}_{13} (gt_a g^{-1})_1 =
\nonumber
\\[9pt] 
& = & 
16 \; B^{cd} {f_{de}}^a B^{eb} 
\partial_+ \Lambda_c(\epsilon) \Lambda_a(\epsilon')
\;{\cal P}_{13} (g t_b g^{-1})_1 - 
\nonumber
\\ 
& - & 
16 \; B^{cd} {f_{de}}^a B^{eb} 
\partial_+ \Lambda_a(\epsilon) \Lambda_c(\epsilon')
\;{\cal P}_{13} (g t_b g^{-1})_1 +
\nonumber
\\ 
& + &
16 \; B^{cd} {f_{de}}^a B^{eb} 
\partial_+ \Lambda_b(\epsilon) \Lambda_c(\epsilon')
\;{\cal P}_{13} (g t_a g^{-1})_1 =
\nonumber 
\\[9pt] 
& = & 3\times 16\; B^{d[c} {f_{de}}^a B^{b]e} 
\partial_+ \Lambda_c(\epsilon) \Lambda_a(\epsilon')
\;{\cal P}_{13} (g t_b g^{-1})_1 
\end{eqnarray}
This implies:
\begin{equation}
I^{(1)}_{Q_1^2} = 
3\times 16\; B^{p[a} {f_{pq}}^b B^{c]q} \Lambda_a(\epsilon) \Lambda_b(\epsilon') j_c
\end{equation}
and therefore:
\begin{equation}
W_2^{(0)} = 3\times 16\; B^{pa} {f_{pq}}^b B^{cq} \; \Lambda_a \Lambda_b \Lambda_c
\end{equation}
This means that $W_2^{(0)}$ is expressed in terms of the 
Schouten bracket on $\Lambda^{\bullet}{\bf g}$:
\begin{equation}\label{GBracketOnAlgebra}
{\gbl B_1 , B_2 \gbr}^{abc} = B_1^{[a|e|} {f_{ef}}^b B_2^{|f|c]}
\end{equation}

\subsection{Could a nonzero $\gbl B,B \gbr$ be harmless?}
\label{sec:Harmless}
\subsubsection{Operator $W_2^{(0)}$ may be $Q$-exact}
We have seen that the obstacle to extending the deformation to the second 
order in $\varepsilon$ is $\gbl B,B \gbr^{abc} \Lambda_a \Lambda_b \Lambda_c$. But a nonzero $\gbl B,B \gbr$ does
not yet mean that the deformation is obstructed, because 
$\gbl B,B \gbr^{abc} \Lambda_a \Lambda_b \Lambda_c$ can still be 
$Q_{BRST}$-exact:
\begin{equation}\label{BBQExact}
\gbl B,B \gbr^{abc} \Lambda_a \Lambda_b \Lambda_c 
\stackrel{?}{=}
Q_0 T
\end{equation}

\paragraph     {Example of a $Q$-exact expression of the ghost number 3}
Let us consider the following operator of the ghost number 2:
\begin{equation}\label{TExample}
T = A^{ma}\overline{\Lambda}_m \Lambda_a
\end{equation}
where $A^{ma}$ is some tensor which does not need to have any special symmetry
properties under the exchange $m\leftrightarrow a$. In this case we get:
\begin{equation}\label{QZeroTType}
Q_0 T = - A^{ma} {f_m}^{bc} \Lambda_b\Lambda_c\Lambda_a
\end{equation}
Therefore, if:
\begin{equation}\label{ConditionOnB}
\gbl B,B \gbr^{abc} = A^{m[a}{f_m}^{bc]}
\end{equation}
then such $\gbl B,B \gbr$ is harmless.
In particular, such harmless $\gbl B,B \gbr$ arise 
in the following situation. 
Consider $\gbl B_1, B_2 \gbr^{abc}\Lambda_a\Lambda_b\Lambda_c$ 
in the special case when $B_1^{ab}=G^l{f_l}^{ab}$.  We get:
\begin{equation}
G^l {f_l}^{e[a} {f_{ef}}^b B_2^{c]f} = - G^l[t^a ,[t^b,t_f]]_l B_2^{c]f}
= \frac{1}{ 2}{f_{fl}}^g G^l {f^{[ab}}_g   B_2^{c]f}
\end{equation}
\rem{(link: "BB-when-B-exact.jpg" "beta" "mirage")}
This expression is proportional to $f^{ab}_g$.  Therefore in this case 
$\gbl B_1, B_2 \gbr^{abc}\Lambda_a\Lambda_b\Lambda_d$ is BRST-exact: 
\begin{equation}\label{CorrectnessUnderEquivalenceOfB}
\gbl B_1, B_2 \gbr^{abc}\Lambda_a\Lambda_b\Lambda_c
=
Q({f^a}_{mn} G^m B_2^{nb}\;\overline{\Lambda}_a \Lambda_b)
\end{equation}
This implies that the condition that $\gbl B,B \gbr$ is exact is correctly
defined on the equivalence classes of $B^{ab}\sim B^{ab} + {f^{ab}}_c G^c$
in agreement with Sections \ref{sec:VertexAndItsDescent} and
\ref{sec:NotCovariant}.
Comparing (\ref{CorrectnessUnderEquivalenceOfB}) with (\ref{ConditionOnB})  
we see that in this
case $A^{ma}= {f^m}_{pq} G^p B^{qa}$. Notice that this $A^{ma}$ is not antisymmetric
in $a\leftrightarrow m$; but the {\em antisymmetrization} of $A$ is  a
Schouten bracket $\gbl G,B \gbr$.

\subsubsection{But $W^{(0)}$ is of ghost number 3; isn't it always $Q_0$-exact?}
\label{sec:GhostNumberThreeAlwaysQExact}
There is no nontrivial BRST cohomology in the ghost number 3, therefore
strictly speaking $W_2^{(0)}$ is always $Q_0$-exact. Since $Q_0W_2^{(0)} = 0$ 
we should always be able to find $T$ such that $W_2^{(0)} = Q_0 T$.
However, this is not true if we also impose some additional constraints
on $T$. There are two possible constraints on $T$:
\begin{enumerate}
\item Covariance, {\it i.e.} we demand $T$ to transform covariantly under
   $psu(2,2|4)$. Apriori it only transforms covariantly modulo $\mbox{Ker} \; Q_0$;
   see \cite{Mikhailov:2009rx}.
\item Absence of resonant terms; in other words $T$ is periodic in
   the global time of $AdS_5$. Notice that this  would  be automatically
   satisfied if we impose the covariance. 
\end{enumerate}
Considerations similar to \cite{Mikhailov:2009rx} show that there are
the following obstructions to the covariance of $T$:
\begin{eqnarray}
&& H^1\left( {\bf g}\; , \;
\mbox{Hom}_{\bf C}\left( {\bf g}\wedge {\bf g}\wedge {\bf g}\; ,\;
\left[ \begin{array}{c} 
\mbox{physical} \cr \mbox{states} \end{array}\right]
\right) \right)
\nonumber
\\
&& H^2\left( {\bf g}\; , \;
\mbox{Hom}_{\bf C}\left( {\bf g}\wedge {\bf g}\wedge {\bf g}\; ,\;
\left[ \begin{array}{c} 
\mbox{conserved} \cr \mbox{charges} \end{array}\right]
\right) \right)
\nonumber
\\
&& H^3\left( {\bf g}\; , \;
\mbox{Hom}_{\bf C}\left( {\bf g}\wedge {\bf g}\wedge {\bf g}\; ,\;
{\bf C}
\right) \right) 
\label{H1H2H3Obstructions}
\end{eqnarray}

\subsubsection{Open problem: how to tell if $W_2^{(0)}$ is BRST-exact in the covariant subcomplex?}
\label{sec:HarmlessConjecture}
We do not know the answer to this question. 

\vspace{10pt}
{\small \noindent
It appears to us that our example (\ref{ConditionOnB}) does not
exhaust all the possibilities for $\gbl B,B \gbr\Lambda\Lambda\Lambda$ 
to be $Q_0$-exact. Indeed we prove
in Appendix \ref{sec:AppendixProofOfLemma} that $\gbl \; B,B \; \gbr$ 
necessarily has some odd indices. But there are examples of solutions
(complex $\beta$-deformations) where $\gbl B,B \gbr$ only has even indices.}

\vspace{10pt}
The study of the obstructions
(\ref{H1H2H3Obstructions}) is one possible approach, but we will not do it
in this paper.
In this Section we will give a guess about what the image of $Q_0$ might be.

\vspace{8pt}
{\small
\noindent Consider the condition (\ref{BBQExact}):
\begin{equation}\label{RepeatBBQExact}
\gbl B,B \gbr^{abc} \Lambda_a \Lambda_b \Lambda_c 
\stackrel{?}{=}
Q_0 T_2^{(0)}
\end{equation}
with the restriction that $T_2^{(0)}$ 
transforms covariantly under the ${\bf psu}(2,2|4)$-rotation of $g$
and $B$.
 More precisely:
\begin{equation}\label{InducedFromSpinors}
T_2^{(0)} \in \mbox{ind}_{{\bf g}_0}^{\bf g} 
( \quad  
{\bf g}_3 \bullet {\bf g}_3 \quad \oplus \quad
{\bf g}_3 \otimes {\bf g}_1 \quad \oplus \quad
{\bf g}_1 \bullet {\bf g}_1 
\quad )
\end{equation}
Here $\bullet$ means the graded-symmetric product
(because ${\bf g}_1$ and ${\bf g}_3$ are odd subspaces, this is
actually the antisymmetric product in the usual sense). 
Moreover we are interested in
a {\em linear subspace} of (\ref{InducedFromSpinors}) consisting
of the $T_2^{(0)}$ such that $QT_2^{(0)}$ belongs to the linear
space generated by the expressions of the form:
\begin{equation}\label{SpaceOfThreeLambdas}
\Lambda_a \Lambda_b \Lambda_c
\end{equation}
Let ${\cal T}$ denotes the subspace of such $T_2^{(0)}$:
\begin{align}
{\cal T} \; \subset & \;\; \mbox{ind}_{{\bf g}_0}^{\bf g} 
( \quad  
{\bf g}_3 \bullet {\bf g}_3 \quad \oplus \quad
{\bf g}_3 \otimes {\bf g}_1 \quad \oplus \quad
{\bf g}_1 \bullet {\bf g}_1 
\quad )
\\
Q{\cal T} \; \subset & \;\; \mbox{expressions of the form } \;
W^{abc}\Lambda_a \Lambda_b \Lambda_c \mbox{ where } W \in \Lambda^3 {\bf g}
\end{align}
We do not have a complete description of ${\cal T}$. We want to point
out the following:
\begin{enumerate}
\item expressions of the form (\ref{TExample}) are in ${\cal T}$
\item but we think\footnote{because of the existence of
 complex $\beta$-deformations, see Section \ref{sec:AboutComplex}} that ${\cal T}$ is not exhausted by the expressions
   of the type (\ref{TExample})
\item the image of $Q$ is a linear subspace, therefore the condition
   on $B$ following from (\ref{RepeatBBQExact})  should be of the type: 
\begin{equation}
\gbl B,B \gbr \; \mbox{ belongs to a certain subspace } \;
S\subset \Lambda^3 {\bf g}
\end{equation}
\end{enumerate}
Let us call $S$ the ``harmless subspace''.

Let us look at some invariant subspaces subspaces in $\Lambda^3 {\bf g}$.
Consider $\widehat{\bf g}' = u(2,2|4)$.
The adjoint representation of ${\widehat{\bf g}}'$ 
is  a tensor product of the fundamental
and antifundamental representations; therefore 
$W\in \Lambda^3 \widehat{\bf g}'$ is represented as a matrix $W^{ikm}_{jln}$ , graded-antisymmetric
with respect to the exchange of pairs 
$\left({i\atop j}\right) \leftrightarrow 
 \left({k\atop l}\right) \leftrightarrow
 \left({m\atop n}\right) $. 
 Let us introduce the transposition operators
$(12)_{up}$ , $(13)_{up}$ , $(23)_{up}$ , $(12)_{dn}$ , $(13)_{dn}$ and $(23)_{dn}$ in the tensor product 
$\widehat{\bf g}'\otimes \widehat{\bf g}' \otimes \widehat{\bf g}'$,  for example:
\begin{equation}
(12)_{up}T^{ikm}_{jln} = T^{kim}_{jln} \quad, \quad
(13)_{up}T^{ikm}_{jln} = T^{mki}_{jln} \quad, \quad
(23)_{dn}T^{ikm}_{jln} = T^{ikm}_{jnl}
\end{equation}
For example, $\Lambda^3 \widehat{\bf g}'$ is the image of 
$\widehat{\bf g}'\otimes \widehat{\bf g}' \otimes \widehat{\bf g}'$ under the following pair-wise-antisymmetrization operator:
\[
1 - (12)_{up}(12)_{dn} - (13)_{up}(13)_{dn} - (23)_{up}(23)_{dn} +
(12)_{up}(23)_{up}(12)_{dn}(23)_{dn} + (23)_{up}(12)_{up}(23)_{dn}(12)_{dn}
\]
Let us denote this operator $\cal A$ (pairwise antisymmetrization). Then we denote:
\begin{align}
L_{(2,0)} = & \quad
{\cal A} (1+(23)_{up}) (1+(12)_{up}) \quad 
\widehat{\bf g}'\otimes \widehat{\bf g}' \otimes \widehat{\bf g}'
\\
L_{(0,2)} = & \quad
{\cal A} (1+(12)_{dn}) (1+(23)_{dn}) \quad 
\widehat{\bf g}'\otimes \widehat{\bf g}' \otimes \widehat{\bf g}'
\end{align}
Consider the complex structure ${\cal I}$ which we introduced 
in Section \ref{sec:ComplexStrucure}. Consider the expression
$\gbl \; e^{\phi{\cal I}}B \; , \; e^{\phi{\cal I}}B \; \gbr$. As a function of $\phi$ it  contains a part proportional to $e^{2i\phi}$,
a part proportional to $e^{-2i\phi}$ and a constant part. The part proportional
to  $e^{2i\phi}$ is in $L_{(2,0)}$, and the part proportional to $e^{-2i\phi}$ is in 
$L_{(0,2)}$. Therefore this condition:
\begin{equation}\label{BBInLL1}
\gbl\; B\; , \;B \;\gbr \in L_{(2,0)} + L_{(0,2)} + {\bf 1}\wedge {\bf g} \wedge {\bf g}
\end{equation}
is invariant under $B \to e^{\phi {\cal I}} B$. \rem{(start-process "mirage" "*scratch*" "mirage" 
"photos/symmetrizers.jpg" )}
Also, expressions of the type (\ref{QZeroTType})
are in $L_{(2,0)} + L_{(0,2)} + {\bf 1}\wedge {\bf g}\wedge {\bf g}$. 

{\small \noindent 
Indeed, let us take 
$X^{ikm}_{jln} = \delta^k_j A^{im}_{ln}$. Then ${\cal A}X$ is identified with (\ref{QZeroTType}). On the
other hand we observe:
\begin{align}
\left( [1-(23)_{up}(12)_{up}] X \right)^{ikm}_{jln} =
X^{ikm}_{jln} - \left( (12)_{up} X \right)^{imk}_{jln} =
X^{ikm}_{jln} - X^{mik}_{jln} = X^{ikm}_{jln} - \delta^i_j A^{mk}_{ln}
\end{align}
Taking into account that:
\begin{equation}
1= {1\over 2}(1-(23)_{up}(12)_{up}) + {1\over 4}(1+(23)_{up}) (1+(12)_{up})
+ {1\over 4}(1-(23)_{up}) (1-(12)_{up})
\end{equation}
this implies that ${\cal A}X$ is in 
$L_{(2,0)} + L_{(0,2)} + {\bf 1}\wedge {\bf g}\wedge {\bf g}$. 
}

It is natural to conjecture that (\ref{BBInLL1})
is sufficient for (\ref{RepeatBBQExact}). But this is only a guess.
}

\subsection{Calculation of $V^{(2)}_2$}
\label{sec:ConstructionOfV2Integrated}
\rem{(link: "toOscar1\.png" "beta" "mirage")}
We will start with calculating  $Q_1 j_{a+}$. We get:
\begin{eqnarray}
Q_1j_{a+} & = & 4\; \Lambda_b B^{bc} {f_{ca}}^e j_{e+} +
\nonumber
\\ 
& + &  
4\; \partial_+ \Lambda_b B^{bc}\; \mbox{Str}\left( t_a\;
g^{-1}\left( 
(gt_cg^{-1})_3 + 2(gt_cg^{-1})_2 + 3(gt_cg^{-1})_1  - \right.\right.
\nonumber
\\
& &\phantom{\partial_+ \Lambda_b B^{bc}\;g^{-1}\;\;\;\;\;\;\;} 
\left.\left. -4 {\cal P}_{13}(gt_cg^{-1})_1 \right) g\right)
\end{eqnarray}
\begin{eqnarray}
Q_1j_{a-} & = & 4\; \Lambda_b B^{bc} {f_{ca}}^e j_{e-} -
\nonumber
\\ 
& - &  
4\; \partial_- \Lambda_b B^{bc}\; \mbox{Str}\left( t_a\;
g^{-1}\left( 
3(gt_cg^{-1})_3 + 2(gt_cg^{-1})_2 + (gt_cg^{-1})_1  - \right.\right.
\nonumber
\\
& &\phantom{\partial_- \Lambda_b B^{bc}\;g^{-1}\;\;\;\;\;\;\;} 
\left.\left. -4 {\cal P}_{31}(gt_cg^{-1})_3 \right) g\right)
\end{eqnarray}
\rem{(link: "small-case-currents\.png" "beta" "mirage")}
This means:
\begin{eqnarray}
Q_1\;  B^{ab} j_{a+} j_{b-} & = &
4\; B^{cd} B^{ab} \Lambda_c ({f_{da}}^e j_{e+}  j_{b-} + {f_{db}}^e j_{a+} j_{e-}) +
\nonumber
\\[1pt]
& + & 
4\; B^{cd} B^{ab} \partial_+\Lambda_c \mbox{Str}\left(
gt_ag^{-1}\;\left( 
(gt_dg^{-1})_3 + 2(gt_dg^{-1})_2 + 3(gt_dg^{-1})_1  \right.\right. -
\nonumber
\\[1pt]
& &\phantom{\partial_+ \Lambda_b B^{bc}\;g^{-1}\;\;\;\;\;\;\;} 
-4 \left.\left. {\cal P}_{13}(gt_dg^{-1})_1 \right) \right)
\;\;j_{b-}\; +
\nonumber
\\[1pt]
& + & 
4\; B^{cd} B^{ab} j_{a+}\partial_-\Lambda_c \mbox{Str}\left(
gt_b g^{-1}\;\left( 
- 3(gt_dg^{-1})_3 - 2(gt_dg^{-1})_2 - (gt_dg^{-1})_1  \right.\right. +
\nonumber
\\[1pt]
& &\phantom{\partial_+ \Lambda_b B^{bc}\;g^{-1}\;\;\;\;\;\;\;} 
+ 4 \left.\left. {\cal P}_{31}(gt_dg^{-1})_3 \right) \right) =
\nonumber
\\[8pt]
& = &
4\; B^{cd} B^{ab} \Lambda_c ({f_{da}}^e j_{e+}  j_{b-} + {f_{db}}^e j_{a+} j_{e-}) +
\nonumber
\\[1pt]
& + & 
4\; B^{cd} B^{ab} Q_0(j_{+c}j_{b-}) \mbox{Str}\left(
gt_ag^{-1}\;\left( 
(gt_dg^{-1})_3 + 2(gt_dg^{-1})_2 + 3(gt_dg^{-1})_1  \right.\right. -
\nonumber
\\[1pt]
& &\phantom{\partial_+ \Lambda_b B^{bc}\;g^{-1}\;\;\;\;\;\;\;} 
-4 \left.\left. {\cal P}_{13}(gt_dg^{-1})_1 \right) \right)
\label{CalculatingQ1V1Step1}
\end{eqnarray}
Now we are going to use the condition (\ref{ConditionOnB}). Let us first assume
that $A=0$, and then consider the case when $A$ is nonzero.

\subsubsection{The case when $\gbl B,B \gbr = 0$}
\label{sec:CaseBBZero}
Taken into account that $\gbl B,B \gbr = 0$ we can transform in the first line of (\ref{CalculatingQ1V1Step1}):
\begin{equation}\label{ConsequenceOfGBZero}
B^{cd} B^{ab} \Lambda_c ({f_{da}}^e j_{e+}  j_{b-} + {f_{db}}^e j_{a+} j_{e-}) 
 = 
-B^{dc} B^{ab} \Lambda_e {f_{da}}^e j_{c+} j_{b-}
\end{equation}
\rem{(link: "Q1Bjj-transform-1st-line\.png" "beta" "mirage")}
Observe that:
\begin{eqnarray}
&& Q_{0L}\mbox{Str}\left(
gt_ag^{-1}\;\left( 
(gt_dg^{-1})_3 + 2(gt_dg^{-1})_2 + 3(gt_dg^{-1})_1  
-4  {\cal P}_{13}(gt_dg^{-1})_1 \right) \right)
=
\nonumber
\\[8pt]
&& = \mbox{Str}\left(
[\lambda_3, gt_ag^{-1}]\;\left(
(gt_dg^{-1})_3 + 2(gt_dg^{-1})_2 + 3(gt_dg^{-1})_1  
-4  {\cal P}_{13}(gt_dg^{-1})_1 \right) \right)
+
\nonumber
\\[1pt]
&& + \mbox{Str}\left(
gt_ag^{-1}\;\left(
[\lambda_3, (gt_dg^{-1})_0] + 2[\lambda_3, (gt_dg^{-1})_3] 
+ 3[\lambda_3, (gt_dg^{-1})_2]\right) \;\;\right)
\end{eqnarray}
Notice that $\mbox{Str}\left(
[\lambda_3, gt_ag^{-1}]
{\cal P}_{13}(gt_dg^{-1})_1 \right) = \mbox{Str}\left(
[\lambda_3, gt_ag^{-1}](gt_dg^{-1})_1 \right)$ and therefore:
\begin{eqnarray}
&& Q_{0L}\mbox{Str}\left(
gt_ag^{-1}\;\left( 
(gt_dg^{-1})_3 + 2(gt_dg^{-1})_2 + 3(gt_dg^{-1})_1  
-4  {\cal P}_{13}(gt_dg^{-1})_1 \right) \right)
=
\nonumber
\\[8pt]
&& = \mbox{Str}\left(
[\lambda_3, gt_ag^{-1}]\;\left(
(gt_dg^{-1})_3 + 2(gt_dg^{-1})_2 
-(gt_dg^{-1})_1 \right) \right)
+
\nonumber
\\[1pt]
&& + \mbox{Str}\left(
gt_ag^{-1}\;\left(
[\lambda_3, (gt_dg^{-1})_0] + 2[\lambda_3, (gt_dg^{-1})_3] 
+ 3[\lambda_3, (gt_dg^{-1})_2]\right) \;\;\right)
\end{eqnarray}
Now we split this up:
\begin{eqnarray}
&&   \mbox{Str}\left(
[\lambda_3, (gt_ag^{-1})_2](gt_dg^{-1})_3 + 
2[\lambda_3, (gt_ag^{-1})_3](gt_dg^{-1})_2 -
[\lambda_3, (gt_ag^{-1})_0](gt_dg^{-1})_1 \right. +
\nonumber
\\[1pt]
&& + \left.
(gt_ag^{-1})_1[\lambda_3, (gt_dg^{-1})_0] + 
2(gt_ag^{-1})_2[\lambda_3, (gt_dg^{-1})_3] + 
3(gt_ag^{-1})_3[\lambda_3, (gt_dg^{-1})_2]\right) =
\nonumber
\\[8pt]
& = &   \mbox{Str} \left(
- [\lambda_3, (gt_ag^{-1})_2] (gt_dg^{-1})_3 
- [\lambda_3, (gt_ag^{-1})_3] (gt_dg^{-1})_2 -
\right.
\nonumber
\\[1pt]
&&
\phantom{\mbox{Str}\;} \left.
- [\lambda_3, (gt_ag^{-1})_1] (gt_dg^{-1})_0
- [\lambda_3, (gt_ag^{-1})_0] (gt_dg^{-1})_1
\right)
\end{eqnarray}
Finally, this can be written as:
\begin{eqnarray}
&& Q_{0L}\; \mbox{Str}\left(
gt_ag^{-1}\;\left( 
(gt_dg^{-1})_3 + 2(gt_dg^{-1})_2 + 3(gt_dg^{-1})_1  
-4  {\cal P}_{13}(gt_dg^{-1})_1 \right) \right)
=
\nonumber
\\[2pt]
&&
= - \mbox{Str} (\lambda_3 [gt_ag^{-1}, gt_dg^{-1}]) = 
- {f_{ad}}^e (g^{-1}\lambda_3 g)_e 
\label{Q0LOnNonPolynomial}
\end{eqnarray}
Similarly, let us calculate the action of $Q_{0R}$ on the same expression:
\begin{eqnarray}
&& Q_{0R}\mbox{Str}\left(
gt_ag^{-1}\;\left( 
(gt_dg^{-1})_3 + 2(gt_dg^{-1})_2 + 3(gt_dg^{-1})_1  
-4  {\cal P}_{13}(gt_dg^{-1})_1 \right) \right)
=
\nonumber
\\[8pt]
&& = \mbox{Str}\left(
[\lambda_1, gt_ag^{-1}]\;\left(
(gt_dg^{-1})_3 + 2(gt_dg^{-1})_2 + 3(gt_dg^{-1})_1  
-4  {\cal P}_{13}(gt_dg^{-1})_1 \right) \right)
+
\nonumber
\\[1pt]
&& + \mbox{Str}\left(
gt_ag^{-1}\;\left(
[\lambda_1, (gt_dg^{-1})_2] + 2[\lambda_1, (gt_dg^{-1})_1] 
- [\lambda_1, (gt_dg^{-1})_0]\right) \;\;\right)
\end{eqnarray}
where we have used that ${\cal P}_{13} [\lambda_1,x_0] = [\lambda_1,x_0]$ for any $x$. We can write this in the 
following form:
\begin{eqnarray}
&& \mbox{Str}\left(
  [\lambda_1, (gt_ag^{-1})_0](gt_dg^{-1})_3 + 
2 [\lambda_1, (gt_ag^{-1})_1](gt_dg^{-1})_2 + 
3 [\lambda_1, (gt_ag^{-1})_2](gt_dg^{-1})_1  
\right)
+
\nonumber
\\[1pt]
&& + \mbox{Str}\left(
  (gt_ag^{-1})_1 [\lambda_1, (gt_dg^{-1})_2] + 
2 (gt_ag^{-1})_2 [\lambda_1, (gt_dg^{-1})_1] - 
  (gt_ag^{-1})_3 [\lambda_1, (gt_dg^{-1})_0]
\right)
=
\nonumber
\\[8pt]
&& = \mbox{Str}\left(
  [\lambda_1, (gt_ag^{-1})_0](gt_dg^{-1})_3 +
  [\lambda_1, (gt_ag^{-1})_3](gt_dg^{-1})_0 
\right) 
+
\nonumber
\\[1pt]
&& + \mbox{Str}\left(
  [\lambda_1, (gt_ag^{-1})_1](gt_dg^{-1})_2 +
  [\lambda_1, (gt_ag^{-1})_2](gt_dg^{-1})_1 
\right) 
\nonumber
\end{eqnarray}
Therefore:
\begin{eqnarray}
&& Q_{0R}\; \mbox{Str}\left(
gt_ag^{-1}\;\left( 
(gt_dg^{-1})_3 + 2(gt_dg^{-1})_2 + 3(gt_dg^{-1})_1  
-4  {\cal P}_{13}(gt_dg^{-1})_1 \right) \right)
=
\nonumber
\\[1pt]
&& = \mbox{Str}\left( g^{-1}\lambda_1 g [t_a,t_d] \right) 
   = {f_{ad}}^e (g^{-1}\lambda_1 g)_e
\label{Q0ROnNonPolynomial}
\end{eqnarray}
Combining (\ref{Q0LOnNonPolynomial}) and (\ref{Q0ROnNonPolynomial})
 with (\ref{CalculatingQ1V1Step1}) and (\ref{ConsequenceOfGBZero}) we get:
\begin{equation}\label{Q1V1EqualsQ0V2}
Q_1V_1^{(2)} = - Q_0 V_2^{(2)}
\end{equation}
where 
\begin{eqnarray}
V_2^{(2)} & = & - 4 \; B^{cd} B^{ab} \; j_{c+} j_{b-} \;
\mbox{Str} \left( 
  (gt_ag^{-1})_1 (gt_dg^{-1})_3 + 
2 (gt_ag^{-1})_2 (gt_dg^{-1})_2 + \right.
\nonumber
\\[1pt]
&& 
\phantom{B^{cd} B^{ab} j_{c+} j_{b-} \; \mbox{Str}}
\left.
+ 3 (gt_ag^{-1})_3 (gt_dg^{-1})_1 - 
4 (gt_ag^{-1})_3 {\cal P}_{13} (gt_dg^{-1})_1 \right)
\label{FormulaForV2}
\end{eqnarray}
Notice that $V_{2}^{(2)}$ is non-polynomial in pure spinors
because of ${\cal P}_{13}$. But we will see that the nonlocality
actually cancels out in the classical action if we substitute
the classical values for the antifields $w^{\star}$ (which are  non-zero after
the deformation). In other words, we  remove the terms with
${\cal P}$ by a shift of $w^{\star}$.

\subsubsection{The case when $\gbl B,B \gbr$ is of the form (\ref{ConditionOnB})}
\label{sec:CaseOfNonzeroA}
Now we have to explain what happens when $\gbl B,B \gbr$ is nonzero, but is
$Q$-exact in the sense of (\ref{ConditionOnB}). Then (\ref{ConsequenceOfGBZero}) fails and
consequently instead of (\ref{Q1V1EqualsQ0V2}) we are getting this:
\begin{equation}
Q_1V_1^{(2)} = - Q_0 V_2^{(2)} + 
3 A^{m[a}{f_m}^{bc]} \Lambda_a j_{b+} j_{c-} 
\end{equation}
The second term on the right hand side can be transformed as follows:
\begin{eqnarray}
&& A^{ma} {f_m}^{bc} (\Lambda_a \;j_b\wedge j_c  
+ \Lambda_b \;j_c\wedge j_a + \Lambda_c j_a\wedge j_b) \simeq
\nonumber
\\[5pt]
& \simeq &
- A^{ma}\Lambda_a d ( g^{-1}{d^2J\over dl^2} g )_m 
+ 2 A^{ma} [\Lambda, j]_m \wedge j_a =
\nonumber
\\[5pt]
& = &
- d \left( A^{ma} \Lambda_a (g^{-1} {d^2 J\over dl^2} g)_m\right) 
+ Q_0 \left(A^{ma} j_a (g^{-1} {d^2 J\over dl^2} g)_m\right) 
- A^{ma} j_a\wedge Q_0 (g^{-1} {d^2 J\over dl^2} g)_m +
\nonumber
\\[1pt]
&& 
+ 2 A^{ma} [\Lambda,j]_m \wedge j_a
\end{eqnarray}
Here we can use:
\begin{eqnarray}
\epsilon Q_0 (g^{-1} {d^2 J\over dl^2} g) & = &
  g^{-1} \left[ {d^2J\over dl^2} , \epsilon \lambda \right] g 
- g^{-1} {d^2\over dl^2} (D\epsilon\lambda) g = 
\nonumber
\\[5pt]
& = &
- 2 \left[ g^{-1}{dJ\over dl}g, g^{-1}{d\epsilon\lambda\over dl}g \right]
- g^{-1} D(\epsilon\lambda) g =
\nonumber
\\[5pt]
& = &
-2[\epsilon\Lambda, j] - d(\epsilon\overline{\Lambda})
\end{eqnarray}
and finally obtain:
\begin{eqnarray}
&& A^{ma} {f_m}^{bc} (\Lambda_a \;j_b\wedge j_c  
+ \Lambda_b \;j_c\wedge j_a + \Lambda_c\; j_a\wedge j_b) \simeq
\nonumber
\\[5pt]
& \simeq &
- d \left( 
   A^{ma} \Lambda_a (g^{-1} {d^2 J\over dl^2} g)_m
   + 
   A^{ma} \overline{\Lambda}_m j_a
\right) 
+ Q_0 \left(A^{ma} j_a (g^{-1} {d^2 J\over dl^2} g)_m\right) 
\end{eqnarray}
Therefore with $A\neq 0$ we get:
\begin{equation}
Q_1V_1^{(2)} \simeq - Q_0 \left( V_2^{(2)} + 
 A^{ma} j_a (g^{-1} {d^2 J\over dl^2} g)_m\right) 
+d(\mbox{smth})
\end{equation}

\subsection{Taking into account nonzero classical values of the antifields}
\label{sec:ClassicalValuesOfAntifields}
\subsubsection{Second order correction to the deformed action}
The terms in ${\cal L}+ V_1^{(2)}$ containing the antifields are the following:
\begin{equation}
- \mbox{Str}(w_{1+}^{\star} w_{3-}^{\star})
- 4 (g^{-1}w_{1+}^{\star}g)_a B^{ab} j_{b-}
+ 4 j_{a+} B^{ab} (g^{-1}w_{3-}^{\star}g)_b
\end{equation}
\rem{(link: "action-for-antifields\.png" "beta" "mirage")}
This means that the classical values of the antifields are:
\begin{align}
w_{3-}^{\star}|_{cl} = & - 4 {\cal P}_{31} (gt_ag^{-1})_3 B^{ab} j_{b-}
\label{W3Classical}
\\
w_{1+}^{\star}|_{cl} = & 4 {\cal P}_{13} j_{a+} B^{ab} (gt_bg^{-1})_1
\label{W1Classical}
\end{align}
When we substitute these classical values back into the action, we get:
\begin{equation}
w^{\star}_{1+}|_{cl} \; w^{\star}_{3-}|_{cl} = 
- 16 j_{c+} j_{b-} B^{cd} B^{ab} 
\mbox{Str}\left(
(gt_dg^{-1})_1 {\cal P}_{31} (gt_ag^{-1})_3
\right)
\end{equation}
Combining this with  (\ref{FormulaForV2}) we get:
\begin{eqnarray}
V_2^{(2)} + w_{1+}^{\star}|_{cl} w_{3-}^{\star}|_{cl} 
& = &
- 4 \; B^{cd} B^{ab} \; j_{c+} j_{b-} \;
\mbox{Str} \left( 
  (gt_ag^{-1})_1 (gt_dg^{-1})_3 +
\right.
\nonumber
\\[1pt]
&& 
\phantom{B^{cd} B^{ab} \; j_{c+} j_{b-}  \mbox{Str}\quad}
+ 2 (gt_ag^{-1})_2 (gt_dg^{-1})_2 +
\nonumber
\\[1pt]
&& 
\phantom{B^{cd} B^{ab} \; j_{c+} j_{b-}  \mbox{Str}\quad}
\left.
+ 3 (gt_ag^{-1})_3 (gt_dg^{-1})_1  \right)
\label{V2WithAntifields}
\end{eqnarray}
This formula describes the second order deformation of the classical
action.

\subsubsection{BRST transformation of the shifted antifields}
Taking into account (\ref{W3Classical}) and (\ref{W1Classical}) we
define the shifted antifields:
\begin{align}
\underline{w}_{3-}^{\star} \; = \; & 
w_{3-}^{\star} - w_{3-}^{\star}|_{cl} \; = \; 
w_{3-}^{\star} + 4 {\cal P}_{31} (gt_ag^{-1})_3 B^{ab} j_{b-}
\\
\underline{w}_{1+}^{\star} \; = \; &
w_{1+}^{\star} - w_{1+}^{\star}|_{cl} \; = \;  
w_{1+}^{\star} - 4 {\cal P}_{13} j_{a+} B^{ab} (gt_bg^{-1})_1
\end{align}
In terms of these shifted antifields the BRST transformation
$Q_0 + \varepsilon Q_1$ (where $Q_1$ is given by (\ref{QOneOfBareW}))
is:
\begin{align}
 (Q_0 + \varepsilon Q_1)\underline{w}_{1+}^{\star} = & \; 
D_{0+}\lambda_1 - [N_+,\lambda_1] + 4\varepsilon 
j_{a+} B^{ab} [ (gt_bg^{-1})_0 , \lambda_1 ]
\nonumber \\ 
 (Q_0 + \varepsilon Q_1)\underline{w}_{3-}^{\star} = & \;
D_{0-}\lambda_3 - [N_-,\lambda_3] + 4\varepsilon
j_{a-} B^{ab} [ (gt_bg^{-1})_0 , \lambda_3 ]
\end{align}
\remv{Calculation}\rem{(image-dired "photos/Q-on-shifted-antifields*.jpg")}

\subsection{Conclusion}
The action at the second order is given by:
\begin{align}
S =&\;   {R^2\over \pi} \int d^2 z\, \hbox{Str} \Big( {1\over 2} J_{2+}J_{2-} +
{3\over 4} J_{1+}J_{3-} + {1\over 4} J_{3+}J_{1-} + \nonumber \\[3pt]
&   \qquad      +
w_{1+}\partial_-\lambda_3 + w_{3-}\partial_+\lambda_1 +N_{0+}J_{0-}
+N_{0-}J_{0+}-N_{0+}N_{0-} + 
\nonumber
\\[3pt]
& \qquad + {1\over 2} \varepsilon B^{ab} j_{[a}\wedge j_{b]} - 
\\[7pt]
& \qquad - 4 \; \varepsilon^2\; B^{cd} B^{ab}  \; j_{c+} j_{b-} \;
\mbox{Str} \left( 
  (gt_ag^{-1})_1 (gt_dg^{-1})_3 +
\right.
\nonumber
\\[1pt]
&
\qquad \phantom{-4 B^{cd} B^{ab}\varepsilon^2 \; j_{c+} j_{b-}  \mbox{Str}}
+ 2 (gt_ag^{-1})_2 (gt_dg^{-1})_2 +
\nonumber
\\[1pt]
& 
\qquad \phantom{-4 B^{cd} B^{ab}\varepsilon^2 \; j_{c+} j_{b-}  \mbox{Str}}
\left.
+ 3 (gt_ag^{-1})_3 (gt_dg^{-1})_1  \right) - 
\nonumber
\\[3pt]
& 
\qquad - \underline{w}^{\star}_{1+} \underline{w}^{\star}_{3-} \Big)
\label{ActionToSecondOrder}
\end{align}

\subsection{Comments}
\subsubsection{About higher orders}
We have started with $V_1^{(2)} = B^{ab}j_{a+} j_{b-}$ and obtained
$V_2^{(2)} = -4 B^{ap} M_{pq} B^{qb} j_{c+} j_{b-}$ where
\begin{equation}
M_{pq}  =  \mbox{Str} \left( 
  (gt_ag^{-1})_1 (gt_dg^{-1})_3 + 2 (gt_ag^{-1})_2 (gt_dg^{-1})_2 
+  3 (gt_ag^{-1})_3 (gt_dg^{-1})_1  \right)
\end{equation}
Notice that while $V_1^{(2)}$ is parity-odd
$V_2^{(2)}$ is not. This corresponds to the fact that there is a nonzero
deformation of the metric at the second order. Also, notice the
schematic pattern in going from the first order vertex to the second
order vertex:
\begin{equation}
B^{ab} \longrightarrow B^{ap} M_{pq} B^{qb}
\end{equation}
We conjecture that higher orders follow the same pattern.
\subsubsection{About the gauge transformation $B^{ab}\mapsto B^{ab} + {f^{ab}}_c G^c$}
As we explained in Section \ref{sec:BProportionalToStructureConstants}
the gauge transformation 
\begin{equation}\label{GaugeTransformationInCommentsSection}
B^{ab}\mapsto B^{ab} + {f^{ab}}_c G^c 
\end{equation}
should be accompanied
by a field redefinition $G^a{\cal X}_a$. Therefore the condition of the gauge
invariance at the second order in $\epsilon$ is:
\begin{equation}\label{GaugeInvarianceAtSecondOrder}
{f^{ab}}_c G^c {\delta\over\delta B^{ab}} V^{(2)}_2 +
G^a{\cal X}_a V^{(2)}_1 = d(\mbox{smth})
\end{equation}
This means that $V^{(2)}_2$ is not invariant under the gauge transformation
(\ref{GaugeTransformationInCommentsSection}) in the naive sense, but rather in the
sense of Eq. (\ref{GaugeInvarianceAtSecondOrder}).

\section{Properties of the Schouten bracket
on $\Lambda^{\bullet}{\bf g}$}
\label{sec:PropertiesOfSchouten}

\subsection{Projection to ${\bf g}\otimes {\bf g}$}
Given $a\wedge b\wedge c \subset \Lambda^3{\bf g}$
we consider:
\begin{equation}
\label{ProjectionToTensorProduct}
[a\wedge b\wedge c] = [a,b]\otimes c - [a,c]\otimes b + [b,c]\otimes a
\quad \in {\bf g}\otimes {\bf g}
\end{equation}
If the internal commutator of $B$ (defined in Section 
\ref{sec:AdditionalConstraintOnB}) vanishes, then:
\begin{equation}
[ \; \gbl \; B,B \; \gbr \; ] \in {\bf g} \bullet {\bf g}
\end{equation}
where $\bullet$ means the symmetric product. More precisely, this is
 ${f^a}_{kl} {f^b}_{mn} B^{km} B^{ln}$.

\subsection{From the $r$-matrix point of view}

Suppose that $B$ satisfies $\gbl B,B \gbr =0$. 
Then we can think of $B$ as a classical $r$-matrix. It defines the Poisson bracket
on ${\bf g}$, and therefore the structure of the Lie algebra on ${\bf g}^*$, 
in the following way \cite{Faddeev:1987ph}:
\begin{equation}\label{RMatrixLieBracket} 
[X,Y]^{(1)}_B = \iota(B)\; d(X\wedge Y) = \mbox{ad}^*_{X_b B^{bc} t_c} Y - (X\leftrightarrow Y)
\end{equation}
In this formula $d(X\wedge Y)$ is the differential on $\Lambda^{\bullet}{\bf g}^*$,
which is the same $d$ as defines the Lie algebra cohomology.
This differential is ``dual'' to the Lie bracket on ${\bf g}$, in the following sense.
Remember that in our notations the coordinates of an element $\xi\in {\bf g}$ are
enumerated with the upper indices:
\begin{equation}
\xi = \xi^a t_a \in {\bf g}
\end{equation}
The commutator is $ [\xi,\eta]=\xi^a\eta^b {f_{ab}}^c t_c $. Therefore the elements of ${\bf g}^*$
have lower indices, so the pairing of $X\in {\bf g}^*$ and $\xi \in {\bf g}$
is $ \langle X , \xi \rangle  =  X_a \xi^a$. The structure of a Lie algebra on ${\bf g}$ determines
the differential on $\Lambda^{\bullet}{\bf g}^*$, which is continued by the polylinearity from:
\begin{equation}
(d X)_{ab} = {f_{ab}}^c X_c
\end{equation}
When $B$ is decomposable, {\it i.e.} $B=b_1\wedge b_2$, we can rewrite (\ref{RMatrixLieBracket})
as follows:
\begin{equation}
[X,Y]^{(1)}_B = \iota_{b_{[1}} \mbox{ad}^*_{b_{2]}} \; X\wedge Y
\end{equation}
 The Jacobi identity for $[,]_B^{(1)}$ gives:
\begin{equation}
[[X,Y]_B^{(1)},Z]_B^{(1)} \pm \mbox{(cycl)} =
(\iota_{b_{[1}} \mbox{ad}^*_{b_2]})^2 \;\; (X\wedge Y\wedge Z) =
\gbl B,B \gbr^{abc} \iota_{t_a} \iota_{t_b} \mbox{ad}^*_{t_c} \;\; (X\wedge Y\wedge Z)
\end{equation}
where the last equality is true also if $B$ is not decomposable.
This means that $[,]_B$ satisfies the Jacobi identity iff $\gbl B,B \gbr = 0$.

But what happens if $\gbl B,B \gbr$ is not zero, but is $Q$-exact in the sense
of (\ref{ConditionOnB})? Then $ \gbl B, B \gbr^{abc} = A^{[a|m|}\; {f_m}^{bc]} $ and 
we get:
\begin{eqnarray}
&& 
\langle \;\; [[X,Y]_B^{(1)},Z]_B^{(1)} \pm \mbox{(cycl. $X,Y,Z$)} \; , \; t_q \; \rangle = 
6 A^{[a|m|} {f_m}^{bc]} X_{[a} Y_b Z_{c']} {f^{c'}}_{cq} =
\nonumber
\\[5pt]
&& \qquad
= 4 A^{am}{f_m}^{bc} X_{[a} Y_b Z_{c']} {f^{c'}}_{cq} 
+ 2 A^{cm}{f_m}^{ab} X_{[a} Y_b Z_{c']} {f^{c'}}_{cq} =
\nonumber
\\[5pt]
&& \qquad
= 4 A^{am}{f_m}^{[b|c|} X_{[a} Y_b Z_{c']} {f^{c']}}_{cq} 
+ 2 A^{cm}{f_m}^{ab} X_{[a} Y_b Z_{c']} {f^{c'}}_{cq} =
\nonumber
\\[5pt]
&& \qquad
= 2 A^{am}{f_{mq}}^{c} X_{[a}Y_b Z_{c']} {f^{c'}}_c{}^b
+ 2 A^{cm}{f_m}^{ab} X_{[a} Y_b Z_{c']} {f^{c'}}_{cq} =
\nonumber
\\[5pt]
&& \qquad
= {1\over 3} A^{am} {f_{mq}}^c X_a [Y,Z]_c 
+ {1\over 3} A^{cm} [X,Y]_m Z_{c'} {f^{c'}}_{cq} \; \pm \mbox{(cycl. $X,Y,Z$)}
\end{eqnarray}
where $[X,Y]_a = {f_a}^{bc} X_b Y_c$ (so defined $[X,Y]$ is a bracket on ${\bf g}^*$
which turns ${\bf g}^*$ into a Lie algebra isomorphic to ${\bf g}$).  Suppose that
$A$ is antisymmetric: $A^{am}=-A^{ma}$. In this case, let us
define a new operation $[,]_A^{(2)}: {\bf g}^*\wedge {\bf g}^* \to {\bf g}^*$; in coordinates:
\begin{eqnarray}
([X,,Y]_A^{(2)})_q & = & A^{am}X_a {f_{mq}}^b Y_b + A^{cm}{f_{cq}}^a X_a Y_m
- (X\leftrightarrow Y) =
\nonumber
\\[5pt]
& = & \langle (X\wedge Y)\;,\; \gbl t_q, A\gbr \rangle
\end{eqnarray}
\paragraph     {Conclusion} If $\gbl B,B \gbr$ is $Q$-exact in the sense of 
(\ref{ConditionOnB}) with antisymmetric $A$ then the following bracket on ${\bf g}^*$:
\begin{equation}
[,] + \varepsilon [,]_B^{(1)} + \varepsilon^2 [,]_A^{(2)}
\end{equation}
is satisfies the Jacobi identity up to the order $\epsilon^2$.

\subsection{The space of solutions to $\gbl B,B \gbr = 0$.}
Unfortunately we do not have an explicit description of the 
space of solutions to $\gbl B,B \gbr = 0$. Here we will discuss
an a subspace which corresponds to the Maldacena-Lunin
solution. Then we will argue that this subspace does not exhaust
all the solutions. In other words, there are beta-deformations other
than the Maldacena-Lunin solution. 
\subsubsection{The solutions of Maldacena-Lunin type}
Let us introduce the basis  in  $gl(m|n)$ consisting of the
$m|n$ matrices $E^i_j$, which have $0$ in all positions except
for $1$ in the $i$-th row and $j$-th column. For example,
for $E^2_3\in gl(3)$ is this:
\[
E^2_3 = \left( 
\begin{array}{ccc}
0 & 0 & 0 \cr
0 & 0 & 1 \cr
0 & 0 & 0
\end{array}
\right)
\]
Notice that $E^j_j$ is a diagonal matrix. 
 
It is straightforward to see that these matrices satisfy $\gbl B,B \gbr = 0$:
\begin{equation}\label{CanonicalBGL}
B = \sum_{1\leq i<j \leq 8} h^{ij} E_i^i\wedge E_j^j
\end{equation}
In fact, for such $B$ even stronger identity is true:
\begin{equation}\label{BfBZero}
B^{ae}{f_{eg}}^bB^{gc} = 0
\end{equation}
--- this is true even if we do not antisymmetrize  $a,b,c$.

\subsubsection{Solutions of more general type}
Let us consider the deformations of (\ref{CanonicalBGL}) of the
following type:
\begin{eqnarray}
B & \rightarrow & B + \delta B
\nonumber
\\[1pt]
\delta B & = & \sum a_{ij} E^i_j\wedge E^j_i
\end{eqnarray}
Notice that $\gbl B, \delta B \gbr =0$, but
 $B^{ae}{f_{eg}}^b\delta B^{gc}$ is nonzero. If this deformation remains unobstructed at
higher orders, then
we must conclude that $\gbl B,B \gbr =0 $  {\em does not}
imply (\ref{BfBZero}). As we discussed in Section \ref{sec:IntroRealBeta} the
existing mathematical results on the classification of solutions of the
classical Yang-Baxter equation seem to imply that indeed there are 
solutions of $\gbl \; B,B \; \gbr =0$ which do not imply (\ref{BfBZero}).

\subsection{Calculation of the bracket for elements of $6^{u(3)}_{\bf C}\subset 45_{\bf C}^{su(4)}$}
\label{sec:CalculationForSubspace}
The $B$-tensors corresponding to $6^{u(3)}_{\bf C}\subset 45_{\bf C}^{su(4)}$ are of the following form:
\begin{equation}
B^{ik}_{jl} = u^{ikp}\epsilon_{pjl} + u^*_{pjl}\epsilon^{pik}
\end{equation}
We have  explicitly calculated $\gbl B , B \gbr$ for such $B$:
\begin{eqnarray}
\gbl B, B \gbr^{\bf ikm}_{\bf jln} & = & 
- 2\; \delta^{\bf i}_{\bf l} \; u^{{\bf km}p} u^*_{{\bf jn}p} 
+ 2\; \epsilon^{{\bf ik}p} \epsilon_{{\bf ln}q} u^{r{\bf m}q} u^*_{{\bf j}rp} +
\nonumber
\\
&& + 2\; u^{{\bf ik}p} \epsilon_{{\bf j} rp} u^{r{\bf m}q} \epsilon_{{\bf ln}q} 
   + 2\; \epsilon^{{\bf ik}p} u^*_{{\bf j} rp} \epsilon^{r{\bf m}q} u^*_{{\bf ln}q} +
\nonumber
\\ 
&&
+\left[
\;
{\mbox{terms dictated}\atop \mbox{by antisymmetry}}
\;\;\; \left({{\bf i}\atop {\bf j}}\right)\leftrightarrow 
\left({{\bf k}\atop {\bf l}}\right)
\leftrightarrow \left({{\bf m}\atop {\bf n}}\right)
\; 
\right]
\label{GBracketOnSixC}
\end{eqnarray}
\rem{(image-dired "photos/g-bracket-su3.jpg")}
(we boldfaced the uncontracted indices for convenience).
\rem{(image-dired "photos/reduced-bracket*.jpg"
;; The comparison of those two pictures clearly shows that the
;; contraction of $i$ and $j$ gives just zero
)}

\section{Complex $\beta$-deformation}
\label{sec:AboutComplex}

\subsection{Complex structure in $45_{\bf C}$}
\label{sec:ComplexStructureIn45}
There is an injective map from $45_{\bf C}^{su(4)}$ to the space
of $B$-fields on $S^5$. For each $B\in 45_{\bf C}^{su(4)}$ we
get the corresponding $B$-field which we call ${\cal B}$ and which is
given by Eq. (\ref{BField}):
\begin{equation}
{\cal B} = B^{ab}\;
\mbox{tr} ((dg g^{-1})_2\; g t_a g^{-1}) \wedge
\mbox{tr} ((dg g^{-1})_2\; g t_b g^{-1}) 
\end{equation}
or, more explicitly, in terms of the coordinates $X^a$ describing
the embedding of $S^5$ into ${\bf R}^6$ by Eq. (\ref{BInSDirections}):
\begin{equation}\label{BFieldInTermsOfX}
{\cal B} =  B^{[kl][mn]} 
 X_{[k} dX_{l]} \wedge X_{[m} dX_{n]}
\end{equation}
This shows that the space of 2-forms on $S^5$ has a subspace
which transforms as $45_{\bf C}$ under $so(6)$. We would like to
stress that these $B$-fields are {\em real}; but the representation
in which they transform happens to have a complex structure; therefore
the subindex ${\bf C}$ in $45_{\bf C}$.
This complex structure  is explained in  Section \ref{sec:AppendixRepresentationTheory}, but here
we want to discuss a physical explanation of it.

It turns out that if the field ${\cal B}$ is defined by (\ref{BFieldInTermsOfX})
then $*_5{\cal B}$ is exact:
\begin{equation}\label{DefBTilde}
*_5{\cal B} = d\widetilde{\cal B}
\end{equation}
Indeed, notice that $*_5{\cal B}$ is closed. Indeed, $*_5d*_5{\cal B}$ is
a 1-form on $S^5$, in other words:
\begin{eqnarray}
*_5d*_5{\cal B} \in \mbox{ind}_{so(5)}^{so(6)} {\bf R}^5
\end{eqnarray}
But there is no $45_{\bf C}^{so(6)}$ in $\mbox{ind}_{so(5)}^{so(6)} {\bf R}^5$:
\begin{eqnarray}
\mbox{Hom}_{so(6)} \left( 
45_{\bf C} \;,\;\; \mbox{ind}_{so(5)}^{so(6)} {\bf R}^5\right) 
=
\mbox{Hom}_{so(5)} \left( 
{({\bf R}^5\wedge {\bf R}^5\;\oplus {\bf R}^5)^{\wedge 2}
\over
{\bf R}^5\wedge {\bf R}^5\;\oplus {\bf R}^5}\;,\;\;
{\bf R}^5 \right) 
= 0
\end{eqnarray}
and this shows that $d*_5{\cal B} = 0$.
Strictly speaking Eq. (\ref{DefBTilde}) only defines $\widetilde{\cal B}$ up to a total derivative.
But this ambiguity is fixed by the requirement that the correspondence
${\cal B} \mapsto \widetilde{\cal B}$ commutes with the symmetries. We can demonstrate this in the
following way. Let us start by fixing $B$ so that:
\begin{equation}
\Box_{6} {\cal B} = 0
\end{equation}
Then we get:
\begin{equation}
*_5 d *_5 d {\cal B} = \iota_E *_6 d \iota_E *_6 d {\cal B} =
\iota_E *_6 {\cal L}_E *_6 d {\cal B} = - 16 {\cal B}
\end{equation}
Let us therefore denote:
\begin{equation}
{\cal I}{\cal B} = {1\over 4} *_5 d{\cal B}
\end{equation}
The operator ${\cal I}$ is the complex structure, ${\cal I}^2=-1$.  We have:
\begin{equation}
*_5{\cal B} = d\left( -{1\over 4} {\cal I}{\cal B}\right)
\end{equation}
Roughly speaking, the complex structure exchanges the NSNS and the RR
$B$-fields.

\subsection{Comparison with the quadratic obstruction found
by Aharony,  Kol,  and Yankielowicz \cite{Aharony:2002hx}}
\label{sec:ComparisonWithAKY}
\subsubsection{Formula for obstruction suggested in 
\cite{Aharony:2002hx}}
The deformations considered in \cite{Aharony:2002hx} correspond on the
Yang-Mills side to the following deformation of the superpotential:
\begin{equation}
W=\frac{1}{ 3}h_{ijk} \mbox{tr}(\Phi^i\Phi^j\Phi^k)
\end{equation}
The coefficients $h_{ijk}$ transform in the symmetric tensor
product of three fundamental representations of $u(3)\subset so(6)$. 
In \cite{Aharony:2002hx} the following condition on $h_{ijk}$ was
obtained:
\begin{equation}\label{FromAKY}
h_{ipq}\overline{h}^{jpq} 
- \frac{1}{ 3} \delta_i^j h_{pqr}\overline{h}^{pqr} = 0
\end{equation}
This is Eq. (3.5) in \cite{Aharony:2002hx}.

\subsubsection{Comparison with $\gbl B,B \gbr$}
To compare with our results we consider the case when $B$ is
in $6^{u(3)}_{\bf C}\subset 45_{\bf C}^{su(4)}$. We have considered this case in 
Section \ref{sec:CalculationForSubspace}. Notice that (\ref{FromAKY}) is  much weaker
than the condition of vanishing of $\gbl B,B \gbr$ which on $6^{u(3)}_{\bf C}$ is
given by (\ref{GBracketOnSixC}). Indeed, the complex $\beta$-deformations do not
satisfy $\gbl B,B \gbr = 0$, and not even (\ref{SchoutenBracketInLDelta}). 
As we have explained in Section \ref{sec:Harmless}, we have to take into
account the possibility that $\gbl\; B,B \;\gbr$ is nonzero
but $\gbl \; B,B \; \gbr^{abc} \Lambda_a \Lambda_b \Lambda_c$ is $Q_0$-exact. 
The careful analysis of the SUGRA equations presented in
\cite{Aharony:2002hx} shows that the log terms (the resonant terms,
corresponding to the
anomalous dimension) actually
appear only at the third order in $\epsilon$. This suggests that 
$Q_0^{-1}\gbl \; B,B \; \gbr^{abc} \Lambda_a \Lambda_b \Lambda_c$ exists
and does not contain log terms, although it might be not strictly
covariant (maybe only covariant up to $\mbox{Ker}\; Q_0$). 

\subsubsection{Supersymmetric extension of (\ref{FromAKY})}
\label{sec:SUSYExtensionOfAKY}
It is natural to ask the following questions:
\begin{enumerate}
\item what is  the supersymmetric extension of (\ref{FromAKY})?
\item is it possible to express it in terms of the Schouten
   bracket $\gbl B,B \gbr$?
\end{enumerate}
The Schouten bracket $\gbl B,B \gbr$ is an element of 
${\bf g}\wedge {\bf g}\wedge {\bf g}$. But let us {\em pick a representative} $\widehat{B}$ for $B$ in
$\widehat{g} = su(2,2|4)$ and consider  
$\gbl \widehat{B},\widehat{B} \gbr$
as an element of $\widehat{\bf g}\wedge \widehat{\bf g}\wedge \widehat{\bf g}$. Notice that $\widehat{\bf g}$ is a subspace
in the tensor product of the fundamental and the antifundamental representation,
defined by the tracelessness condition. Therefore we parametrize
$x\in \widehat{g}$ by a $(4|4)\times (4|4)$-matrix $x^a_b$; the upper index parametrizes 
the fundamental representation, and the lower index the antifundamental. 
With these notations a tensor
$X\in \widehat{\bf g}\wedge \widehat{\bf g}\wedge \widehat{\bf g}$ has six indices $X^{ace}_{bdf}$. Let us consider the following 
contraction:
\begin{equation}
|\; \gbl \widehat{B}, \widehat{B} \gbr \; |^a_b
\; \stackrel{\mbox{\tiny def}}{=} \;
\gbl \widehat{B}, \widehat{B} \gbr^{apq}_{pqb}
\end{equation}
\remv{Definition}\rem{(image-dired "photos/dbl-contr-def.jpg")}
It turns out that $ |\; \gbl \widehat{B},\widehat{B} \gbr \;| $
does not change if we change 
$\widehat{B} \mapsto \widehat{B} + {\bf 1}\wedge x$ 
({\it i.e.} pick a different representative for $B$ in $\widehat{\bf g}$),
and also does not change if we change 
$B^{ab}\mapsto {f^{ab}}_c G^c$. 
There is a simplified formula for $|\; \gbl \widehat{B},\widehat{B} \gbr \;|$:
\begin{equation}
|\; \gbl \widehat{B},\widehat{B} \gbr \;|^a_b = 
\widehat{B}^{pa}_{qr} \widehat{B}^{rq}_{pb}
\end{equation}
A direct calculation shows that $|\; \gbl \widehat{B},\widehat{B} \gbr \;|^a_b$ 
is invariant under 
$\widehat{B}^{ab}_{cd} \mapsto \widehat{B}^{ab}_{cd} + \delta^a_c X^b_d - \delta^b_d X^a_c$ and under 
$\widehat{B}^{ab}_{cd}\mapsto \widehat{B}^{ab}_{cd} + \delta^a_d G^b_c - \delta^b_c G^a_d$. Therefore 
 $|\; \gbl \widehat{B},\widehat{B} \gbr \;|^a_b$ can be considered a function
of $B$, and moreover is correctly defined on the equivalence classes of $B$
with respect to (\ref{EquivalenceRelation}). Also, it follows from symmetries that
$|\; \gbl \widehat{B},\widehat{B} \gbr \;|$ is traceless:
\begin{equation}
|\; \gbl \widehat{B},\widehat{B} \gbr \;|^a_a = 0
\end{equation}
Also:
\begin{equation}
|\; \gbl \widehat{B}, I\widehat{B} \gbr \;| = 0
\end{equation}
\remv{Correctness w.r.to equivalence relations}\rem{
(image-dired "photos/dbl-contr-check-def.jpg")}
Moreover, one can check that when $B\in {\bf 6}_{\bf C}$:
\begin{equation}
{1\over 6} |\; \gbl \widehat{B},\widehat{B} \gbr \;|^j_i = 
u_{ipq}\overline{u}^{jpq} 
- \frac{1}{ 3} \delta_i^j h_{pqr}\overline{h}^{pqr} 
\end{equation}
\remv{Reduction to $su(3)$}\rem{
(image-dired "photos/dbl-contr-of-su3-sector.jpg")}
This means that the following condition:
\begin{equation}\label{SupersymmetricAKY}
|\; \gbl \widehat{B},\widehat{B} \gbr \;|^a_b = 0
\end{equation}
is the supersymmetric version of Eq. (\ref{FromAKY}).

Let us now return to the question: ``is it possible to
formulate the supersymmetric analogue of the condition (\ref{FromAKY}) suggested 
in \cite{Aharony:2002hx} in terms of the Schouten bracket 
$\gbl B,B \gbr$?'' We find ourselves in the following interesting situation.
Our condition (\ref{SupersymmetricAKY}) is expressed in terms of
the Schouten bracket in $\widehat{\bf g}$, which requires the choice of $\widehat{B}$,
which is a lift of $B$ from ${\bf g}$ up to $\widehat{\bf g}$. The projection of the bracket
$|\; \gbl \widehat{B},\widehat{B} \gbr \;|$ does not depend on the choice of $\widehat{B}$
for a given $B$, and therefore is a well defined quadratic
function of $B$. But the bracket $\gbl \widehat{B},\widehat{B} \gbr$ itself
does depend on the choice of a lift. Notice that our  $\gbl \widehat{B},\widehat{B} \gbr$
takes values\footnote{
When we calculate $\gbl , \gbr$ on $B$ we consider it taking values
in  ${\bf g}\wedge {\bf g}\wedge {\bf g}$. 
But when we calculate $\gbl , \gbr$ on $\widehat{B}$ we consider it
taking values in $\widehat{\bf g}\wedge \widehat{\bf g}\wedge \widehat{\bf g}$.
In the first case, this is the Schouten bracket on $\Lambda^{\bullet}{\bf g}$,
and in the second case this is the Schouten bracket on 
$\Lambda^{\bullet} \widehat{\bf g}$. Both are denoted $\gbl,\gbr$.
} in
$\widehat{\bf g}\wedge \widehat{\bf g}\wedge \widehat{\bf g}$
rather than ${\bf g}\wedge {\bf g}\wedge {\bf g}$. Then we apply the ``projection'' $|\ldots |$.
This projection is defined on $\widehat{\bf g}\wedge \widehat{\bf g}\wedge \widehat{\bf g}$:
\begin{equation}
|\ldots | \; : \; \widehat{\bf g}\wedge \widehat{\bf g}\wedge \widehat{\bf g} 
\to \widehat{\bf g}'
\end{equation}
But there is no such operation as $|\ldots |$ on 
${\bf g}\wedge {\bf g}\wedge {\bf g}$  because the operation
$|x\wedge y \wedge z |$ is not correctly defined with respect
to the identification $x^a_b\simeq x^a_b + c\delta^a_b$.

\section{Reading the supergravity fields from the vertex}
\label{sec:ReadingSupergravityFields}
In this section we will restrict ourselves to the case when
the nonzero components of $B^{ab}$ are only in the direction of $S^5$.
We will consider the deformed action (\ref{LinearizedBeta}) and
read the supergravity fields from this action. 

The only nonzero perturbations at the first order in the deformation
parameter are the NSNS and RR $B$-fields. Both can be read from the
deformed action (\ref{LinearizedBeta}). We will start in Section \ref{sec:ActionOfBH} with
briefly reviewing the the general action of Berkovits and Howe 
\cite{Berkovits:2001ue}. We then compare our deformed action (\ref{LinearizedBeta})
with their general action.
In Section \ref{sec:NSNS} we  consider 
the ``bosonic part of the action'', {\it i.e.}
the action at $\theta=0$, and read the NSNS $B$-field from the term
$B_{mn}\partial x^m \overline{\partial} x^n$. The measurement of the RR $B$-field is a bit
more subtle, it requires the analysis of the fermionic terms in the
action. We discuss the fermionic terms and the RR $B$-field in
Section \ref{sec:RRBfield}.

\subsection{The action of Berkovits and Howe}
\label{sec:ActionOfBH}
\paragraph     {Brief partial review of  \cite{Berkovits:2001ue} }
The most general action with BRST symmetry is \cite{Berkovits:2001ue}:
\begin{multline}
S = \frac{1}{2\pi\alpha'} \int d^2 z \bigg( \half \big(G_{MN}(Z) 
+ B_{MN}(Z)\big) \partial Z^M \overline\partial Z^N 
+ E^\alpha_M (Z) d_\alpha \overline\partial Z^M + \bigg.\\
\qquad + E^{\hat\alpha}_M (Z) \tilde{d}_{\hat\alpha} \partial Z^M 
+ \Omega_{M\alpha}{}^\beta(Z) \lambda^\alpha  w_\beta \overline\partial Z^M 
+ \hat\Omega_{M\hat\alpha}{}^{\hat\beta}(Z) \tilde\lambda^{\hat\alpha} 
\tilde w_{\hat\beta} \partial Z^M + \\
\qquad + P^{\alpha\hat\beta}(Z) d_\alpha \tilde d_{\hat\beta} 
+ C^{\beta\hat\gamma}_{\alpha}(Z) \lambda^{\alpha}  w_{\beta} \tilde d_{\hat\gamma} 
+ \hat C^{\hat\beta\gamma}_{\hat\alpha}(Z) \tilde\lambda^{\hat\alpha} 
\tilde w_{\hat\beta} d_{\gamma} + \\
\bigg. + S^{\beta\hat\delta}_{\alpha\hat\gamma}(Z) 
\lambda^{\alpha}  w_{\beta} \tilde \lambda^{\hat\gamma} \tilde w_{\hat\delta} 
+ \half \alpha' \Phi(Z)r\bigg) + S_\lambda + S_{\hat\lambda}
\label{ActionOfBerkovitsAndHowe}
\end{multline}
In this action $Z^M$ are the coordinates of
the $(10|32)$-dimensional supermanifold (the space-time). They are
the {\em worldsheet} fields, {\it i.e.} $Z^M = Z^M(z,\bar{z})$. There are also the pure
spinor worldsheet fields $\lambda(z,\bar{z}), \tilde{\lambda}(z,\bar{z}), w(z,\bar{z}), \tilde{w}(z,\bar{z})$ and the worldsheet
auxiliary fields $d(z,\bar{z}),\tilde{d}(z,\bar{z})$. 
It is important that fields $\lambda,w,d$ are of two kinds:
``left'' and ``right'', for example $\lambda$ and $\tilde{\lambda}$. 
They become left- and right-movers in the flat space limit.
The ``right'' fields are marked by tilde. There is a left ghost number
$U(1)_L$ under which $\lambda^{\alpha}$ has charge $1$ and
$\tilde{\lambda}^{\hat{\alpha}}$ charge $0$, and a similar right ghost number $U(1)_R$.

\vspace{5pt}

\commentstarts
{\small In the general curved spacetime there is no separation of
the worldsheet dynamics into ``left'' and ``right'' sectors, like it
was in flat space. But some aspects of it survive in curved spacetime. 
There is a separate left and right ghost number.
The conserved current
corresponding to the left BRST transformation is holomorphic, and
the right one is antiholomorphic.
}
\commentends

\vspace{5pt}

\noindent
The pure spinor variables $\lambda$ and their momenta $w$ take values
in the spin bundle, therefore greek letters are used for their indices.
The {\em target space} fields are:
\begin{equation}\label{BasicSuperFields}
G_{MN}, \; B_{MN}, \; E^{\alpha}_M,\; E^{\hat{\alpha}}_M,\; 
\Omega_{M\alpha}^{\beta},\; \hat{\Omega}_{M\hat{\alpha}}^{\hat{\beta}},\;
P^{\alpha\hat{\beta}},\; C_{\alpha}^{\beta\hat{\gamma}},\; 
\hat{C}_{\hat{\alpha}}^{\hat{\beta}\gamma},\; S_{\alpha\hat{\gamma}}^{\beta\hat{\delta}},\;
\Phi
\end{equation}
The leading components of the superfields ${\Omega_{M\alpha}}^{\beta}$ and
${\hat{\Omega}_{M\hat{\alpha}}}^{\hat{\beta}}$ are the spin connections. 
The space-time metric $G_{MN}$ is degenerate, of the rank $(10|0)$.
(A non-degenerate metric would have rank $(10|32)$.) 
The BRST transformations act on the coordinate fields $Z^M$ in the
following way:
\begin{equation}\label{BRSTvsDegenerateMetric}
Q Z^M = 
\lambda^{\alpha} E_{\alpha}^M 
+
\tilde{\lambda}^{\hat{\alpha}}E_{\hat{\alpha}}^M 
\end{equation}
where $E_{\hat{\alpha}}^M$ and  $E_{\alpha}^M$ are vector fields
which span the kernel of the metric:
\begin{equation}
E_{\alpha}^MG_{MN} = E_{\hat{\alpha}}^MG_{MN} =0 
\end{equation}
and also satisfy the following equations:
\begin{equation}
E^{\alpha}_M E^M_{\beta} = \delta^{\alpha}_{\beta} \;,\;\;
E^{\alpha}_M E^M_{\hat{\beta}} = 0 \;,\;\;
E^{\hat{\alpha}}_M E^M_{\beta} = 0 \;,\;\;
E^{\hat{\alpha}}_M E^M_{\hat{\beta}} = \delta^{\hat{\alpha}}_{\hat{\beta}}
\end{equation}
(Therefore, $E^M_{\alpha}$ and $E^M_{\hat{\alpha}}$ are determined
in terms of the basic fields (\ref{BasicSuperFields}).)

\vspace{5pt}

\commentstarts {\small
The metric $G_{MN}$ is degenerate, but the degeneracy
is non-physical. It is non-physical because the vector field corresponding to
$Q_{BRST}$ traverses the kernel of $G_{MN}$ as  $\lambda$ and $\tilde{\lambda}$ 
traverse the pure spinor cones. In this sense, the degeneration is
BRST trivial. }
\commentends

\vspace{5pt}

\noindent
After integrating out the 
auxiliary fields $d$ and $\tilde{d}$ the effective metric becomes
non-degenerate:
\begin{equation}
G_{MN} - 2 E^{\alpha}_{(M} P_{\alpha\hat{\beta}} E^{\hat{\beta}}_{N)}
\end{equation}
The pure spinor kinetic terms $S_{\lambda}$ and $S_{\tilde{\lambda}}$ are the same as in flat space:
\begin{equation}\label{BHKineticTerm}
S_{\lambda} = \int d^2z (w_{\alpha +},\partial_-\lambda^{\alpha}) 
\;\;,\;\;\;\;
S_{\tilde{\lambda}} = \int d^2z (\tilde{w}_{\hat{\alpha} -},
\partial_+\tilde{\lambda}^{\hat{\alpha}})
\end{equation}
Since $\lambda$ and $\tilde{\lambda}$ are constrained to satisfy
$\lambda^{\alpha} \Gamma_{\alpha\beta}^m \lambda^{\beta} = 
\tilde{\lambda}^{\hat{\alpha}} 
\Gamma_{\hat{\alpha}\hat{\beta}}^m \tilde{\lambda}^{\hat{\beta}} = 0$,
the conjugate momenta $w_{\alpha +}$ and $\tilde{w}_{\hat{\alpha}-}$ are subject to the
following gauge transformations with arbitrary vector parameters 
$k_{m+}$ and $\hat{k}_{m-}$:
\begin{equation}\label{BHGaugeTransformationsOfw}
\delta_k w_{\alpha + } = k_{m +} \Gamma^m_{\alpha\beta} \lambda^{\beta}
\;,\;\;
\delta_{\hat{k}} \tilde{w}_{\hat{\alpha} - } 
= \hat{k}_{m -} \Gamma^m_{\hat{\alpha}\hat{\beta}} \tilde{\lambda}^{\hat{\beta}}
\end{equation}
The kinetic terms (\ref{BHKineticTerm}) 
are preserved by the separate $so(1,9)$ rotations
of ``left'' and ``right'' $(\lambda,w)$: 
\begin{equation}\label{TwoLocalLorentz}
\begin{array}{lcl}
\delta\lambda^{\alpha} = A^{\alpha}_{\beta}\lambda^{\beta} 
& , &
\delta w_{\alpha +} = - A^{\beta}_{\alpha}w_{\beta +}
\cr
\delta\tilde{\lambda}_{1}^{\hat{\alpha}} = \hat{A}^{\hat{\alpha}}_{\hat{\beta}} \tilde{\lambda}_1^{\hat{\beta}}
& , &
\delta \tilde{w}_{\hat{\alpha}-} = - \hat{A}^{\hat{\beta}}_{\hat{\alpha}}\tilde{w}_{\hat{\beta}-}
\end{array}
\end{equation}
which act as gauge transformations
on $E^{\alpha}_M,\; E^{\hat{\alpha}}_M,\; 
\Omega_{M\alpha}^{\beta},\; \hat{\Omega}_{M\hat{\alpha}}^{\hat{\beta}},\;
P^{\alpha\hat{\beta}},\; C_{\alpha}^{\beta\hat{\gamma}},\; 
\hat{C}_{\hat{\alpha}}^{\hat{\beta}\gamma},\; S_{\alpha\hat{\gamma}}^{\beta\hat{\delta}}$. In this sense, there are 
two independent local $so(1,9)$ symmetries $A$ and $\hat{A}$, {\it i.e.} 
 $so(1,9)\oplus so(1,9)$. 

\vspace{5pt}

\commentstarts {\small
At first sight it might seem that we can rotate with
$A$ and $\hat{A}$ in $gl(16,{\bf C})$; the kinetic terms
(\ref{BHKineticTerm}) would be invariant. However only the 
transformations with $A$ and $\hat{A}$ in $so(1,9)\subset gl(16,\bf{C})$
respect the gauge  transformations (\ref{BHGaugeTransformationsOfw}). }
\commentends

\vspace{5pt}

\noindent
The two Lorentz symmetries (\ref{TwoLocalLorentz}) are field redefinitions
on the worldsheet, which look like gauge transformations from the
target space point of view. In the target space $E^M_{\alpha}, E^M_{\hat{\alpha}}$
correspond to the basis in the tangent space to $\mbox{Ker}\; G$. Therefore the
field redefinition (\ref{TwoLocalLorentz}) corresponds geometrically
to the change of a basis in $\mbox{Ker}\;G$ consistent with the
action of the $U(1)_{L-R}$ ghost symmetry ({\it i.e.} $E_{\alpha}^M$ cannot be
mixed with $E_{\hat{\alpha}}^M$). 

The authors of \cite{Berkovits:2001ue} suggested to partially fix this
gauge symmetry, down to the conventional Lorentz gauge symmetry of
the gravity theory:
\begin{equation}\label{BHGaugeFix}
so(1,9)\oplus so(1,9) \longrightarrow so(1,9)
\end{equation}
They did it in the following way.  Note that
$\mbox{Ker}\;G$ is not an integrable distribution, there is
a  Frobenius map:
\begin{equation}
[,]\;:\;\;\mbox{Ker}\;G\; \wedge\; \mbox{Ker}\; G\; \to \mbox{Im}\;G
\end{equation}
This map is given by the commutator of the vector fields; it is also
called ``torsion''. Note that $\mbox{Im}\;G$ has an orthogonal structure 
(given by $G$ itself). Let us therefore fix some orthogonal basis 
in $\mbox{Im}\; G$; the ambiguity of such a fix is parametrized by
$so(1,9)$  in (\ref{BHGaugeFix}). Let us denote such a basis $E_m^M$. Here the
vector index $m$ runs from $0$ to $9$. Then, the partial gauge
fixing of \cite{Berkovits:2001ue} consists of requiring the
Frobenius map to satisfy\footnote{We multiplied the odd vector fields $E_{\alpha}$ and $E_{\beta}$ by
formal anticommuting parameters $\epsilon,\epsilon'$, to ease the notations}:
\begin{equation}
[\epsilon E_{\alpha}, \epsilon' E_{\beta}]^M 
= \epsilon\epsilon' \Gamma^m_{\alpha\beta} E_m^M
\end{equation}
It is a nontrivial fact, proven in \cite{Berkovits:2001ue}, that
this choice of gauge is possible.
We will discuss this gauge fixing specifically for our example of $\beta$-deformation
in Section \ref{sec:TorsionComponents}.

\paragraph     {RR bispinor}
The leading component of $P^{\alpha\hat{\beta}}$
is the bispinor of the RR field strengths:
\begin{equation}\label{RRBispinor}
P^{\alpha\hat{\beta}} = F_m (\Gamma^m)^{\alpha\hat{\beta}} 
+ F_{klm} (\Gamma^{klm})^{\alpha\hat{\beta}} 
+ F_{klmnp} (\Gamma^{klmnp})^{\alpha\hat{\beta}} 
\end{equation}
Notice that these gauge transformations act on $P^{\alpha\hat{\beta}}$
and that it has a nonzero constant ``background'' value in the pure
 $AdS_5\times S^5$:
\begin{equation}
P_0^{\alpha\hat{\beta}} = F_{klmnp}^{(0)} (\Gamma^{klmnp})^{\alpha\hat{\beta}}
\end{equation}
where $F^{(0)}$ is the RR flux of the pure $AdS_5\times S^5$.
Since the background value $P_0^{\alpha\hat{\beta}}$
 transforms nontrivially under the two local
Lorentz transformations, we have to be careful in defining
the fluctuation. 
When we write (\ref{RRBispinor}) we have to explain how we break the gauge
from $so(1,9)\oplus so(1,9)$ down to the diagonal $so(1,9)$. 
This is discussed in Section \ref{sec:TorsionComponents}.

\remv{Link to their paper}\rem{(start-process "gv" "*scratch*" "gv" 
"--spartan" "literature/BerkovitsHowe.pdf")}

\paragraph     {Notations specific to $AdS_5\times S^5$}
Because we expand around AdS, rather than flat space, we will choose a
normalization of $w$ and $\lambda$ which is slightly different
from what Berkovits and Howe used. We use the kinetic term:
\begin{equation}\label{OurKineticTerm}
S_{\lambda_3} = \int d^2z \;\mbox{Str}\;(w_{1+}\partial_-\lambda_{3})
\;\;,\;\;\;
S_{\lambda_1} = \int d^2z \;
\mbox{Str}\;(w_{3-}\partial_+\lambda_{1})
\end{equation}
rather than (\ref{BHKineticTerm}). In the flat space notations our kinetic terms
(\ref{OurKineticTerm}) gives instead of (\ref{BHKineticTerm}) the following
expression:
\begin{align}
S_{\lambda}  = & \int d^2z 
(w^{\hat{\alpha}}_{+},(\Gamma_{56789}\partial_-\lambda)_{\hat{\alpha}}) 
\label{BHKineticTermLeft}
\\
S_{\tilde{\lambda}}  = & \int d^2z 
(\tilde{w}^{\alpha}_-,
(\Gamma_{56789}\partial_+\tilde{\lambda})_{\alpha})
\label{BHKineticTermRight}
\end{align}
where $\Gamma_{56789}$ is the product of gamma-matrices in the directions 
tangent\footnote{We use $\{0,\ldots,4\}$ to enumerate the tangent
space directions of $AdS_5$, and $\{5,\ldots,9\}$ for the tangent
space directions of $S^5$; this should not be confused with
the enumeration of $X_1,\ldots,X_6$ in (\ref{EmbeddedBeta}), where
$X_i$ are the flat coordinates of the auxiliary ${\bf R}^6$ which
we use to embed $S^5\subset {\bf R}^6$.}
to $S^5$. The relation between (\ref{OurKineticTerm}) and
(\ref{BHKineticTermLeft}), (\ref{BHKineticTermRight}) can be summarized in the
formula:
\begin{equation}\label{RelationBetweenStrAndFlatScalarProduct}
\mbox{Str}(w_1\lambda_3) = 
(w_1^{\hat{\alpha}}C_{\hat{\alpha}\beta} \; \lambda_3^{\beta})
\end{equation}
where $C_{\hat{\alpha}\beta}$ is some bispinor. In fact $C_{\alpha\hat{\beta}}$ the inverse of the
Ramond-Ramond bispinor field $P_0^{\alpha\hat{\beta}}$ in $AdS_5\times S^5$. In agreement with
(\ref{ActionOfBerkovitsAndHowe}) we denote $P^{\alpha\hat{\beta}}$ the 
Ramond-Ramond bispinor field, and $P_0^{\alpha\hat{\beta}}$ the background
value of  $P^{\alpha\hat{\beta}}$ in the pure $AdS_5\times S^5$. 

\subsection{NS-NS $B$-field}
\label{sec:NSNS}
\paragraph     {Measuring the NSNS $B$-field}
The NS-NS $B$ field can be read from (\ref{LinearizedBeta}) in a
rather straightforward way, from the bosonic ($\theta=0$) terms in the
deformed action. Notice that the bosonic part of the undeformed action
(\ref{TheAction}) is the $\theta=0$ part of  ${1\over 2} \int \mbox{Str}\left(J_{2+} J_{2-}\right)$; it
is parity-even ({\it i.e.} invariant under $+\leftrightarrow -$). This means, of
course, that pure $AdS_5\times S^5$ has zero NSNS $B$ field. Let us now consider
the deformation (\ref{LinearizedBeta}):
\[
S \longrightarrow S + {1\over 2}B^{ab} \int j_{[a}\wedge j_{b]}  
\]
The $\theta=0$ part of $j_a$ is $2\;\mbox{Str}((dgg^{-1})_2 g t_a g^{-1})$. Therefore the $j\wedge j$
term gives a parity-odd piece:
\begin{equation}
 2 \int\; B^{ab}\;
\mbox{tr} ((dg g^{-1})_2\; g t_a g^{-1}) \wedge
\mbox{tr} ((dg g^{-1})_2\; g t_b g^{-1}) +\ldots
\label{BFieldInDeformedAction}
\end{equation}
which corresponds to the NSNS $B$-field. Unfortunately we have
a conflict of notations: the letter $B$ already stands for the
deformation parameter $B^{ab}$. To avoid confusion we will
denote the NSNS $B$ field with the calligraphic letter ${\cal B}$.
Eq. (\ref{BFieldInDeformedAction}) implies:
\begin{equation}\label{calligraphicB}
{\cal B}_{\mu\nu} = B^{ab}\;
\mbox{Str} (t^2_{\mu}\; g t_a g^{-1}) \;
\mbox{Str} (t^2_{\nu}\; g t_b g^{-1})
\end{equation}
In particular when $B^{ab}$ is tangent 
to the sphere, there is another way to write this:
\begin{equation}\label{BField}
{\cal B}_{[\mu\nu]} = B^{ab} [(gt_ag^{-1})_2 , (gt_bg^{-1})_2]_{[\mu\nu]}
\end{equation}
\paragraph     {Comparing with the formulas of Maldacena and Lunin}
We should compare with the formula from \cite{Lunin:2005jy}:
\begin{equation}\label{BFromML}
B= \mu_1^2\mu_2^2 d\phi_1\wedge d\phi_2 
 + \mu_2^2\mu_3^2 d\phi_2\wedge d\phi_3 
 + \mu_3^2\mu_1^2 d\phi_3\wedge d\phi_1
\end{equation}
where $\mu$ and $\phi$ are described in  Eq. ((3.11) of \cite{Lunin:2005jy}) which in our
language translates:
\begin{align}
X_1  =  & \mu_1\cos\phi_1\;,\;\; X_2  =  \mu_1\sin\phi_1 
\nonumber
\\
X_3  =  & \mu_2\cos\phi_2\;,\;\; X_4  =  \mu_2\sin\phi_2 
\label{EmbeddedBeta}
\\
X_5  =  & \mu_3\cos\phi_3\;,\;\; X_6  =  \mu_3\sin\phi_3
\end{align}
Here $X_1,\ldots,X_6$ give the embedding of $S^5$ into ${\bf R}^6$:
$
X_1^2 + X_2^2 + \ldots + X_6^2 = 1
$
In terms of this embedding the bosonic part of the global conserved
current is:
\begin{equation}
j_{[AB]} = 2 (dg g^{-1})_{[AB]} = 2 X_{[A} dX_{B]}
\end{equation}
It follows that indeed (\ref{BFromML}) is of the form:
\begin{equation}
B^{[kl][mn]} \;
 X_{[k} dX_{l]} \wedge X_{[m} dX_{n]}
\end{equation}
where $B^{[kl][mn]}$ has the following nonvanishing components:
\begin{equation}\label{NonzeroComponents}
B^{[12][34]} = B^{[34][56]} = B^{[56][12]} = 1
\end{equation}
(And those which follow from them by the antisymmetry of $B$,
{\it e.g.} $B^{[34][12]}=-1$.) This is in agreement with our Eq. (\ref{BField}).

\subsection{RR $B$-field}\label{sec:RRBfield}
The RR field is more subtle to reconstruct. It requires the analysis
of the fermionic terms in the action. An additional complication
is the  nonzero value of the background RR 5-form in 
$AdS_5\times S^5$. This implies that the derivation of the RR 3-form
from the fermionic terms in the action requires a careful treatment
of the local symmetry breaking.

\subsubsection{How we measure the RR field strength}
The general formula for the BRST charges is:
\begin{equation}
Q_L = \int d\tau^+ \epsilon\lambda^{\alpha} d_{\alpha +} \;\; , \;\;\;
Q_R = -\int d\tau^- \epsilon\tilde{\lambda}^{\hat{\alpha}} \tilde{d}_{\hat{\alpha} -}
\end{equation}
where $d_{\alpha +}$ and $\hat{d}_{\hat{\alpha} -}$ are some composite fields, which are interpreted
as auxiliary fields in \cite{Berkovits:2001ue}. We have denoted them $j_{L+}$ and  $j_{R-}$
in  Section \ref{sec:DeformationOfBRSTCurrent}:
\begin{equation}\label{IdentifyBRSTAsd}
j_{L+} = d_+ \;\;,\;\;\; j_{R-}=d_-
\end{equation}
 Integrating out $d_{\alpha}$ in (\ref{ActionOfBerkovitsAndHowe}) gives:
\begin{equation}\label{DAlphaPlus}
d_{\alpha +} = P_{\alpha\hat{\beta}} E_M^{\hat{\beta}} \partial_+ Z^M + \ldots
\end{equation}
where $\ldots$ stands for terms containing $w$ and $\lambda$. On the other hand
it follows from  ((62) of \cite{Berkovits:2001ue}) that the
BRST operator $Q_R$ acts on the matter fields $Z^M$ in the following
way:
\begin{equation}\label{QROnZM}
Q_R Z^M = \tilde{\lambda}^{\hat{\alpha}}E_{\hat{\alpha}}^M
\end{equation}
It follows from (\ref{DAlphaPlus}) and (\ref{QROnZM}) that:
\begin{equation}
\left. d_{\alpha +} \right|^{\mbox{\tiny substitute:}}_{\partial_+ Z^M \to Q_R Z^M} = 
P_{\alpha\hat{\beta}} \tilde{\lambda}^{\hat{\beta}}
\end{equation}
In other words:
\begin{equation}\label{Substitution}
\epsilon\lambda^{\alpha} P_{\alpha\hat{\beta}} \epsilon'\tilde{\lambda}^{\hat{\beta}} =
\left. \epsilon\lambda^{\alpha}d_{\alpha+}  
\right|^{\mbox{\tiny substitute:}}_{\partial_+ Z^M \to \epsilon' Q_R Z^M} 
\end{equation}

\paragraph     {Procedure for measuring the RR field strength:}
\begin{enumerate}
\item Calculate $d_{\alpha+}$ using (\ref{IdentifyBRSTAsd}) and (\ref{JPlusDeformed}).
\item Calculate $P_{\alpha\hat{\beta}}$ from (\ref{Substitution}). In fact it will turn out that 
   the $\theta=0$ component of $P_{\alpha\hat{\beta}}$ is 
   the same as in pure $AdS_5\times S^5$.
\item Examine the torsion and fix the proper gauge, 
   Section \ref{sec:TorsionComponents}. At this
   stage we do some field redefinition, an $so(1,9)$ rotation of
   $w$ and $\lambda$, and the rotation in the opposite direction of $\tilde{w}$
   and $\tilde{\lambda}$. This affects $P_{\alpha\hat{\beta}}$, generating
   a 3-form piece corresponding to the RR 3-form.
\end{enumerate}

\subsubsection{Steps 1 and 2: calculate the deformation of  $P_{\alpha\dot{\beta}}$}
\label{sec:DeformationOfP}
From (\ref{Substitution}) and (\ref{JPlusDeformed}) we read:
\begin{equation}\label{LambdaPLambda}
\epsilon\lambda^{\alpha} P_{\alpha\hat{\beta}} \epsilon'\tilde{\lambda}^{\hat{\beta}} =
\mbox{Str}\big( \epsilon'\lambda_1 \epsilon\lambda_3 \big)
- 16 (g^{-1} \epsilon'\lambda_1 g)_a B^{ab} (g^{-1}\epsilon\lambda_3 g)_b
\end{equation}
\paragraph     {The $\theta=0$ component of $P_{\alpha\hat{\beta}}$ is undeformed}
The first term $\mbox{Str}\big( \epsilon'\lambda_1 \epsilon\lambda_3 \big)$
on the right hand side of (\ref{LambdaPLambda}) corresponds 
to the background RR 5-form in the undeformed $AdS_5\times S^5$,
and the second term 
$- 16 (g^{-1} \epsilon'\lambda_1 g)_a B^{ab} (g^{-1}\epsilon\lambda_3 g)_b$ 
is the deformation. But we observe that the $\theta=0$ component
of the deformation $- 16 (g^{-1} \epsilon'\lambda_1 g)_a B^{ab} (g^{-1}\epsilon\lambda_3 g)_b$ 
is zero\footnote{Remember
that in this section we are only consider the case when the only nonzero
components of $B^{ab}$ are those with $a$ and $b$ both bosonic indices,
{\it i.e.} both $a$ and $b$ are in ${\bf g}_{even} = {\bf g}_0 + {\bf g}_2$.
Therefore the expansion of $(g^{-1}\epsilon\lambda g)$ starts with
$[\epsilon\lambda,\theta]$.}.
Naively this would imply that the RR field strength is undeformed.
However, for the purpose of measuring the RR field this $P_{\alpha\hat{\beta}}$
is {\it a priori} in the {\bf wrong gauge}. 
To understand what is the proper gauge choice we will 
look at the torsion, in Section \ref{sec:TorsionComponents}.

\subsubsection{Digression: integrating in $d$ and $\hat{d}$}

{\bf\small (this section is not needed for the main line of argument)}

\vspace{7pt}

\noindent
In this paper we are using the formulation of the worldsheet theory
without the auxiliary fields $d,\hat{d}$; the difference with the Berkovits--Howe
action (\ref{ActionOfBerkovitsAndHowe}) is that $d$ and $\hat{d}$
have been integrated out. In this section we explain how  to restore, or
``integrate in'' these auxiliary fields, and present the action
in the form (\ref{ActionOfBerkovitsAndHowe}).

Let us calculate $d_{\alpha+}P^{\alpha\hat{\beta}} d_{\hat{\beta}-}$. We have:
\begin{align}
d_{1+} = & J_{1+} + 4 B^{ab} j_{a+} (g t_b g^{-1})_1
\\
d_{3-} = & J_{3-} + 4 B^{ab} j_{a-} (g t_b g^{-1})_3
\end{align}
Combining this with (\ref{LambdaPLambda}) we get:
\begin{align}d_{\alpha+}P^{\alpha\hat{\beta}} d_{\hat{\beta}-} = & \;
\mbox{Str}(J_{1+}J_{3-}) + 
4  B^{ab} j_{a+} (g^{-1} J_{3-}g)_b -
4  B^{ab} (g^{-1}J_{1+} g)_a j_{b-} +
\nonumber
\\
& 
+ 16 B^{ab} (g^{-1}J_{1+} g)_a (g^{-1}J_{3-} g)_b
\end{align}
\remv{Calculation}\rem{(start-process "mirage" "*scratch*" "mirage" 
"photos/integrating-in.jpg" )}
\remv{Str conventions}\rem{
(find-file "photos/integrating-in_str-conv.jpg")}

Consider the Lagrangian with $d$ and $\hat{d}$ integrated in:
\begin{align}
\mbox{Str}& \Big( 
{1\over 2} J_{2+}J_{2-} +
{3\over 4} J_{1+}J_{3-} + {1\over 4} J_{3+}J_{1-} \Big) 
+ B^{ab} j_{a+} j_{b-}
- 
\\[3pt]
  - \mbox{Str}& \Big(  J_{1+} J_{3-} \Big)
- 4 B^{ab} j_{a+} (g^{-1}J_{3-}g)_b + 4 B^{ab} (g^{-1} J_{1+} g)_a j_{b-}
- 16 B^{ab} (g^{-1} J_{1+} g)_a (g^{-1} J_{3-} g)_b 
+
\nonumber
\\[3pt]
&+\mbox{(terms linear and quadratic in $d,\hat{d}$)} 
+ \mbox{ (terms with ghosts) }
\nonumber
\end{align}
Let us denote $k_{1a} = (g^{-1} J_1 g)_a$, $k_{2a} = (g^{-1}J_2g)_a$,
and $k_{3a} = (g^{-1} J_3 g)_a$. We get:
\begin{align}
\mbox{Str}& \Big( 
{1\over 2} J_{2+}J_{2-} -
{1\over 4} J_{1+}J_{3-} + {1\over 4} J_{3+}J_{1-} \Big) +
\nonumber
\\[3pt]
& +   B^{ab} k_{1a+} k_{1b-} + 2 B^{ab} k_{1a+} k_{2b-} - B^{ab} k_{1a+} k_{3b-} -
\\
& - 2 B^{ab} k_{2a+} k_{1b-} - 4 B^{ab} k_{2a+} k_{2b-} + 2 B^{ab} k_{2a+} k_{3b-} -
\nonumber
\\
& -   B^{ab} k_{3a+} k_{1b-} - 2 B^{ab} k_{3a+} k_{2b-} + B^{ab} k_{3a+} k_{3b-}
\nonumber
\\[3pt]
&+\mbox{(terms linear and quadratic in $d,\hat{d}$)} 
+ \mbox{ (terms with ghosts) }
\nonumber
\end{align}
\remv{Calculation}\rem{
(start-process "mirage" "*scratch*" "mirage" "photos/integrating-in_k.jpg" )}
Consider the parity-even part of the
deformed action, {\it i.e.} the part symmetric under $(+\leftrightarrow -)$. 
It is equal to:
\begin{align}\label{DeformedGTerm}
\mbox{Str}\left( 
{1\over 2}J_{2+}J_{2-} 
\right) 
+ 2B^{ab}( k_{1a+} k_{2b-} + k_{1a-} k_{2b+} )
+ 2B^{ab}( k_{2a+} k_{3b-} + k_{2a-} k_{3b+} )
\end{align}
This should be identified with the term $1/2 G_{MN} \partial_+ Z^M \partial_- Z^N$ in the
action (\ref{ActionOfBerkovitsAndHowe}) of Berkovits and Howe.
The corresponding deformed metric $G_{MN}$ has an important property. 
Namely, it remains degenerate to the first order in $B^{ab}$. It is important
in \cite{Berkovits:2001ue} that $G_{MN}$ is a degenerate metric.
More precisely, it should have rank $(10|0)$ while the maximal
possible rank would be $(10|32)$. If there was  a term proportional
to $B^{ab}(k_{1a+}k_{3b-} + k_{1a-}k_{3b+})$ in (\ref{DeformedGTerm}), then
this would mean that the deformed metric is not sufficiently degenerate.
But there is no such term, therefore the deformed metric is as degenerate
as it should be according to \cite{Berkovits:2001ue}. 

\subsubsection{Step 3: examine the torsion components $T_{\alpha\beta}^m$ and $\hat{T}_{\hat{\alpha}\hat{\beta}}^m$ and do the
necessary field redefinitions of $w$ and $\lambda$}
\label{sec:TorsionComponents}
The torsion is defined using the commutator of the covariant
derivatives:
\begin{equation}\label{DefTorsion}
\{D_{\alpha}, D_{\beta}\} = T^m_{\alpha\beta} D_m + 
T^{\gamma}_{\alpha\beta} D_{\gamma} + T^{\hat{\gamma}}_{\alpha\beta} D_{\hat{\gamma}}
+ \mbox{ (terms without derivative)}
\end{equation}
In this formula the covariant derivatives $D_{\alpha}$  and $D_{\hat{\alpha}}$ 
can be read from the BRST transformation of the matter fields:
\begin{equation}\label{MeasureD}
Q_{BRST} Z^M = 
(\lambda^{\alpha} D_{\alpha} + \lambda^{\hat{\alpha}} D_{\hat{\alpha}}) Z^M
+\ldots
\end{equation}
where $\ldots$ stands for terms containing $w$ and $\lambda$. 
As explained in \cite{Berkovits:2001ue}, the worldsheet theory of (\ref{ActionOfBerkovitsAndHowe})
has three independent Lorentz gauge groups: one acting on the spinor indices
$\alpha,\beta,\gamma,\ldots$, the other acting on the hatted spinor
indices $\hat{\alpha},\hat{\beta},\hat{\gamma},\ldots$, and the third
one acting on the vector indices $m,n,\ldots$.  These three Lorentz gauge
groups are fixed down to one ``usual'' Lorentz gauge group by requesting
that the $T_{\alpha\beta}^m$ and $T_{\hat{\alpha}\hat{\beta}}^m$ are equal to the
gamma-matrices. In our notations, they can be identified with the
structure constants:
\begin{equation}\label{TorsionConstraints}
T_{\alpha\beta}^m = f_{\alpha\beta}^m 
\;\;\; , \;\;\;
\hat{T}_{\hat{\alpha}\hat{\beta}}^m = f_{\hat{\alpha}\hat{\beta}}^m
\end{equation}
It is {\em in this gauge} that the field $P^{\alpha\dot{\beta}}$ is identified
with the Ramond-Ramond bispinor.
Our action (\ref{PerturbedActionSeries}) is {\em not automatically} in this form. In fact,
we have to make the infinitesimal field redefinitions:
\begin{equation}\label{InfinitesimalRotation}
\begin{array}{lcl}
\delta\lambda_3 = [\Phi,\lambda_3] 
& , &
\delta w_{1+} = [\Phi,w_{1+}]
\cr
\delta\tilde{\lambda}_{1} = - [\Phi,\tilde{\lambda}_1]
& , &
\delta \tilde{w}_{3-} = - [\Phi,\tilde{w}_{3-}]
\end{array}
\end{equation}
with some matrix $\Phi$ in order to satisfy (\ref{TorsionConstraints}).
Indeed, let us calculate $T^m_{\alpha\beta}$ in our case using (\ref{DefTorsion}) and
(\ref{MeasureD}). 
Consider the action of $Q = Q_0 + \varepsilon Q_1$ on $g$. Let
$R_{t_a}$ denotes the the right shift $R_{t_a} g=gt_a$, then Eq. (\ref{QOneOfBareW})
implies:
\begin{equation}
Q = R_{g^{-1}\lambda_3g} + 4\varepsilon\Lambda_a B^{ab} R_{t_b}
\end{equation}
(We are here only interested in the action on $g$, so we only
keep the first term in (\ref{QOneOfBareW}).)
Therefore:
\begin{align}
D_{\alpha} = & 
R_{g^{-1}t^3_{\alpha} g} +  
4 \varepsilon\; (g^{-1} t_{\alpha}^3\; g)_a 
B^{ab} R_{t_b}
\end{align}
Notice that, when acting on scalar functions, 
$R_{t_b}\simeq R_{g^{-1}(gt_bg^{-1})_{\perp}g}$. If we look at the
$\theta=0$ component, $(gt_bg^{-1})_{\perp} = (gt_bg^{-1})_2$.
(Remember that in this Section, $B^{ab}$ has only bosonic indices.)
This implies that:
\begin{align}
\{ D_{\alpha}, D_{\beta} \}  =  & {f_{\alpha\beta}}^m
\left(R_{g^{-1}t^2_m g} + 
8\varepsilon\; (g^{-1}t_m^2 g)_a B^{ab} R_{g^{-1}(gt_bg^{-1})_2g}\right)
\\
=  & {f_{\alpha\beta}}^m
\left(R_{g^{-1}t^2_m g} + 
8\varepsilon\; (gt_a g^{-1})_m B^{ab} R_{g^{-1}(gt_bg^{-1})_2g}\right)
\end{align}
We gauge away the term proportional to $\varepsilon$ by
rotating $\lambda$ and $w$ as in (\ref{InfinitesimalRotation}) with $\Phi$ given by:
\begin{equation}\label{PhiViaB}
\Phi = 4B^{ab} [(gt_ag^{-1})_2 \;,\; (gt_bg^{-1})_2]
\end{equation}

\subsubsection{Conclusion: RR field strength is $*$-dual in $TS^5$ to NSNS
$B$-field}
We have two observations:
\begin{itemize}
\item
   Explicit evaluation of $P_{\alpha\hat{\beta}}$ in Section \ref{sec:DeformationOfP}
   shows that the $\theta=0$ component is undeformed, {\bf but:}
\item 
   The phase rotation (\ref{InfinitesimalRotation}) of $w$ and $\lambda$
   with the parameter $\Phi$ given by (\ref{PhiViaB}) is needed to
   bring the torsion components to the canonical form. 
   In a similar way,
   one can see that $\tilde{w}$ and $\tilde{\lambda}$ should be rotated
   in the opposite direction, {\it i.e.} $\tilde{\Phi} = - \Phi$.
\end{itemize}
Comparing Eq. (\ref{PhiViaB}) to Eq. (\ref{BField}) of Section
\ref{sec:NSNS}, we see that $\Phi$ is equal to the NSNS $B$ field contracted
with two gamma-matrices:
\begin{equation}\label{PhiAndB}
\Phi = 8 \;\iota(\Gamma) \iota(\Gamma) {\cal B}
\end{equation}
As we explained in Section \ref{sec:ActionOfBH}, this phase
rotation corresponds to the local Lorentz transformation of the
target space fields. 
After this $\Phi$-rotation the $\theta=0$ component of the bispinor
field becomes:
\begin{equation}
P_{(0)} + 2\Phi P_{(0)}
\end{equation}
Given the relation (\ref{RRBispinor}) between $P$ and the
RR field strengths this implies:
\begin{equation}
F_{m_1m_2m_3} = 16\; {\cal B}^{pq} \; F_{pqm_1m_2m_3}
\end{equation}
Therefore we conclude that the RR field strength is
$*$-dual in the tangent space to $S^5$ to the NSNS $B$-field.

\subsubsection{Digression: the coupling of $ w\lambda $ to the NSNS 3-form}
\label{sec:BRSTOffShell}
{\bf \small (this section is not needed for the main line of argument)}
\paragraph     {What do we expect the kinetic term to be?}
Consider the terms in (\ref{LinearizedBeta}) which are quadratic
in the pure spinor variables $w$ and $\lambda$. 
What can we say about the part of the action quadratic in pure spinors?
Looking at (\ref{ActionOfBerkovitsAndHowe}) we should expect it 
to be of the form:
\begin{equation}\label{curvedPSaction}
\int d^2z \; \left[ (w_+ (\partial_- + \Omega_-) \lambda) +
             (\tilde{w}_- (\partial_+ + \widetilde{\Omega}_+) \tilde{\lambda})
\right]
\end{equation}
 where $\Omega$ is the spin connection, which is the sum of the
``geometrical'' spin connection and the 
NSNS field strength\footnote{Recall that we are using a calligraphic letter ${\cal B}$
for the NSNS B-field to distinguish it from the 
deformation parameter $B$; see Eqs. (\ref{calligraphicB}) and (\ref{BFieldVsBParameter}).}
$H=d{\cal B}$ contracted with two gamma-matrices.

\paragraph     {Notice that there is $H_{NSNS}$ in the covariant derivative}
{\small 
The reason for including the NSNS field strength inside the
spin connection is because given the supergravity constraints derived
by Berkovits and Howe as conditions for classical BRST invariance and
the solution of those constraints, we can use the Bianchi identity 
$dH =0$; more specifically the components $(\nabla H)_{ab\alpha\beta}$
of this Bianchi identity, to find the relations $T_{ab}{}^c = H_{abd}\eta^{dc}$ and
$\tilde T_{ab}{}^{c} = - H_{abd}\eta^{dc}$, where $T_{ab}{}^c$ and $\tilde T_{ab}{}^c$
are the torsion components constructed with the spin connections
$\Omega$ and $\tilde \Omega$ respectively. Now, given those relations
between torsions and the NSNS field strengths, one can make the
redefinitions
\begin{eqnarray}\label{connectionredefinitions}
\Omega_{ab}{}^c &\to & \Omega_{ab}{}^c - {1\over 2}H_{abd}\eta^{dc}, \\
\nonumber
\tilde\Omega_{ab}{}^c &\to & \tilde \Omega_{ab}{}^c + {1\over
2}H_{abd}\eta^{dc},
\end{eqnarray}
which means that the bosonic components of the redefined torsions are
set to zero.
Also, we remind that because of the pure spinor condition, the spin
connections which appear in the Berkovits and Howe action can be
decomposed as
\begin{align}\label{connectionsplitting}
\Omega_{M\alpha}{}^{\beta} =& \Omega_M^{(s)}\delta_\alpha {}^\beta +
{1\over 4}\Omega_{Mab}(\Gamma^{ab})_\alpha {}^\beta \\ \nonumber
\tilde\Omega_{M\hat\alpha}{}^{\hat\beta} =&
\tilde\Omega_M^{(s)}\delta_{\hat\alpha} {}^{\hat\beta} +
{1\over 4}\tilde\Omega_{Mab}(\Gamma^{ab})_{\hat\alpha} {}^{\hat\beta}.
\end{align}
Furthermore, as argued in \cite{Bedoya:2006ic}, the constraints
$T_{a\alpha}{}^\alpha = T_{a\hat\alpha}{}^{\hat\alpha} =0$ imply that
$\Omega^{(s)}_a = \tilde\Omega^{(s)}_a =0$. Then, this last fact, the
replacements (\ref{connectionredefinitions}) and equations 
(\ref{connectionsplitting}) replaced in (\ref{curvedPSaction}) show
that indeed there is a three-form in the connection contracted with
two gamma matrices. 
}

\paragraph     {What we get from our construction}
There are terms coming
from the undeformed action, pure $AdS_5\times S^5$:
\begin{equation}\label{FreeWLambda}
\int d^2 z \; \mbox{Str}\left(
w_{1+}D_{0-}\lambda_3 + w_{3-}D_{0+}\lambda_1
\right)
\end{equation}
and terms coming from the deformation term $\int V_1^{(2)}$:
\begin{equation}\label{NotASpinConnection}
\int d^2z \; B^{ab}\left( 
- 4 j_{a-} \mbox{Str} ( \{ w_{1+} , \lambda_3 \} (g t_b g^{-1})_0) 
- 4 j_{a+} \mbox{Str} ( \{ w_{3-} , \lambda_1 \} (g t_b g^{-1})_0)
\right)
\end{equation}
\rem{(link: "verify_part-of-Qj\.jpg" "beta" "mirage")}
Therefore, the part of the action quadratic in $w$ and $\lambda$ is 
the total (\ref{FreeWLambda}) + (\ref{NotASpinConnection}).
Let us consider (\ref{NotASpinConnection}), with the fermionic
fields $\theta$ turned off. In this case $j$ becomes
$2g^{-1} (*dg g^{-1})_2 g$ and (\ref{NotASpinConnection}) reads:
\begin{eqnarray}
\int d^2z && 8 B^{ab}\left( \;
\mbox{Str} ((\partial_-g g^{-1})_2\; g t_a g^{-1})\;\;
\mbox{Str} ( \{ w_{1+} , \lambda_3 \} g t_b g^{-1}) \;\;-
\right.
\nonumber \\
&&- \left. \;\;\;\;\;
\mbox{Str} ((\partial_+g g^{-1})_2 \;g t_a g^{-1})\;\;
\mbox{Str} ( \{ w_{3-} , \lambda_1 \} g t_b g^{-1})
\;\right)
\label{BosonicPartOfPureSpinorsCoupling}
\end{eqnarray}
\underline{This has to be compared to (\ref{curvedPSaction}).}
What do we expect $\Omega$ to be in (\ref{curvedPSaction})?
We know that {\em the metric} is undeformed at the first order in the
deformation parameter, therefore there should be no correction to the
``geometrical'' part of the spin connection. 
But {\em the $B$ field} is nonzero and the corresponding
2-form ${\cal B}_{mn}\; dx^m\wedge dx^n$ is given by:
\begin{equation}
{\cal B} = 2 B^{ab} \;\;
\mbox{Str} \left( (dg g^{-1})_2 (g t_a g^{-1}) \right)
\wedge
\mbox{Str} \left( (dg g^{-1})_2 (g t_b g^{-1}) \right)
\end{equation}
\rem{Coeff $ 2 $: $ 4 $ from $ j = 2dgg $ and $ 1/2 $ from $ 1/2 Bj\wedge j $
Then Coeff $ 1/4 $ to normalize the $ B $ field relative to the kinetic
term, because we define the action to be:
\[
  1/2 G_{mn} \partial_+ x^m \partial_- x^n  
+ 1/2 B_{mn} \partial_+ x^m \partial_- x^n
\]
and the second term of the last formula is actually 
$ 1/4 B_{mn} dx^m \wedge dx^n  =  1/4 B $
But don't forget also that the kinetic term in fact also has $Str$ bringing
another factor of $4$. To conclude:
\[
4 \times 1/4 B = 1/2 Bj\wedge j
\]
}
The field strength is:
\begin{eqnarray}\label{BFieldVsBParameter}
H = d{\cal B} & = &
-8\;\mbox{Str}\left((dg g^{-1})_2 (dgg^{-1})_2 (gt_ag^{-1})_0\right)
B^{ab} \; \mbox{Str}\left((dg g^{-1})_2 (gt_bg^{-1})_2\right)
\end{eqnarray}
\rem{(link: "d-on-tr-dgg-gtg.jpg" "beta" "mirage")}
Obviously, $H$ has three tangent space indices, corresponding to the
three $dgg^{-1}$. The contraction of the tangent space indices with the gamma-matrices
works as follows:
\begin{equation}
\Gamma^m \iota_{R_{t_m^2}} 
\quad \mbox{where }  R_{t_m^2} 
 \mbox{ is the right invariant vector field:} \quad 
R_{t_m^2}g = t_m^2g 
\end{equation}
In other words, the contraction with he gamma-matrices can be schematically
presented as the following rule:
\begin{equation}
dg g^{-1} \mapsto \Gamma^m\otimes t^2_m
\end{equation}
For example, let us contract $\mbox{tr}\left((dg g^{-1})_2 (dgg^{-1})_2 (gt_ag^{-1})_0\right)$
with two gamma-matrices. We get:
\begin{equation}\label{ExampleContraction}
- \Gamma^{m}\Gamma^{n} \;\mbox{Str} \left( [t^2_m , t^2_n] (gt_ag^{-1})_0 \right)
\end{equation}
This expression can be simplified in the following way. Notice that we are
considering $B$ with only nonzero components in $so(6)\wedge so(6) \subset {\bf
  g}\wedge {\bf g}$. This means that the index $a$ in (\ref{ExampleContraction})
only runs in $so(6)_2\subset {\bf g}_2$.  Besides that, we turned off all the
thetas, so $g\in SO(2,4)\times SO(6)$.  
This means that $(gt_ag^{-1})_0\in
so(5)\subset so(6)\subset so(2,4)\oplus so(6)$.  
In the spinor representation, the action of $t^2_m$ on ${\bf g}_{odd}$ is:
\begin{eqnarray}
[t^2_m \;,\; \psi^{\alpha}t_{\alpha}^3] 
& = &
{1\over 2}( \widehat{F}_+ \Gamma_m \psi )^{\hat{\beta}} t_{\hat{\beta}}^1 
\nonumber
\\[5pt]
[t^2_m \;,\; \psi^{\hat{\alpha}}t_{\hat{\alpha}}^1] 
& = &
{1\over 2}( \widehat{F}_+ \Gamma_m \psi )^{\beta} t_{\beta}^3 
\end{eqnarray}
where $ \widehat{F}_+ = \Gamma_0\Gamma_1\Gamma_2\Gamma_3\Gamma_4 +
\Gamma_5\Gamma_6\Gamma_7\Gamma_8\Gamma_9 $ is the bispinor associated to the RR
field strength.  \rem{(start-process "evince" "*scratch*" "evince"
  "/home/andrei/a/Work/cohomology/vert.ps" "--page-label=8")} This means that:
\begin{equation}
[t^2_m,t^2_n] = \overline{\Gamma}_{[m}\Gamma_{n]}
\end{equation}
where
\begin{equation}
\overline{\Gamma}_m = \left\{
\begin{array}{rl}
 \Gamma_m & \mbox{ for $m\in \{0,\ldots,4\}$ } \cr
-\Gamma_m & \mbox{ for $m\in \{5,\ldots,9\}$ }
\end{array}
\right.
\end{equation}
Therefore, when $(gt_ag^{-1})\in so(6)$, we get
\begin{equation}\label{GammaContractedSingleTrace}
\iota(\Gamma)\iota(\Gamma) 
\mbox{Str} \left( (dgg^{-1})_2 (dgg^{-1})_2 (gt_ag^{-1})_0 \right) =
-\Gamma^{m}\Gamma^{n} \mbox{Str} \left( [t^2_m , t^2_n] (gt_ag^{-1})_0 \right)
= -2 (gt_ag^{-1})_0 
\end{equation}
\vspace{10pt}

{\small 
\commentstarts
Note that
our supertrace, which we denote Str, is in the fundamental representation
of $su(2,2|4)$; therefore the trace in (\ref{GammaContractedSingleTrace})
should be understood as the trace in the spinor representation of
$so(6)$, which is the fundamental of $su(4)$. Therefore $\mbox{tr} \; 1=4$.
Both $t_m$ and $[t_m,t_n]$ are conjugate to 
$\mbox{diag}(1/2,1/2,-1/2,-1/2)$. Therefore $-\mbox{tr}(t_mt_n)=\delta_{mn}$ and
$-\mbox{tr}([t_m,t_n][t_k,t_l])=\delta_{mk}\delta_{nl} - \delta_{ml}\delta_{nk}$.
In particular, $\Gamma^m\Gamma^n\mbox{Str}([t^2_m,t^2_n] [t^2_k,t^2_l])
= [\Gamma_k,\Gamma_l] = 2 [ t^2_k , t^2_l ]$.
\commentends
}

\vspace{10pt}

\noindent
Similarly,  the contraction of $H$ with two gamma-matrices gives:
\begin{eqnarray}
\iota(\Gamma)\iota(\Gamma)H
& = & \phantom{-} \; 8\times 2\;(gt_ag^{-1})_0 B^{ab} \mbox{Str}
\left( (dg g^{-1})_2 (gt_bg^{-1})_2 \right) +
\nonumber
\\[1pt]
& & + \; 8\times 2\; 
B^{ab} [ [ (gt_ag^{-1})_0 , (dg g^{-1})_2],  (gt_bg^{-1})_2] 
=
\nonumber
\\[7pt]
& = & \phantom{-} \; 8 \times 2\;(gt_ag^{-1})_0 B^{ab} \mbox{Str}
\left( (dg g^{-1})_2 (gt_bg^{-1})_2 \right) -
\nonumber
\\[1pt]
& & - \; 8 \;D_0(B^{ab} [ (gt_ag^{-1})_2 ,  (gt_bg^{-1})_2])
=
\nonumber
\\[7pt]
& = & \phantom{-} \;8\;(gt_ag^{-1})_0 B^{ab} 
*j_b -
\nonumber
\\[1pt]
& & - \; 8\; D_0(B^{ab} [ (gt_ag^{-1})_2 ,  (gt_bg^{-1})_2])
\label{CouplingToH}
\end{eqnarray}
\rem{(link: "D0_B\.jpg" "beta" "mirage")}
Now we see that:
\begin{itemize}
\item the first term in (\ref{CouplingToH}) reproduces the bosonic part of the
   coupling (\ref{BosonicPartOfPureSpinorsCoupling})
\item the second term is missing from (\ref{BosonicPartOfPureSpinorsCoupling}). 
\end{itemize}
But in fact, that second term is a total derivative: 
$ D_0(-8 \; B^{ab} [(gt_ag^{-1})_2
, (gt_bg^{-1})_2]) $ 
and can be absorbed into the redefinition of $\lambda$ and $w$.
More precisely, the pure spinor kinetic term:
\begin{eqnarray}
{\cal L}_{w\lambda} & = & 
\phantom{+}\mbox{Str}\left(
w_{1+}(D_{0-}\lambda_3  + 4 j_{a-} B^{ab} [(g t_b g^{-1})_0 , \lambda_3])
\right) + 
\nonumber 
\\
&& + \mbox{Str}\left(
w_{3-}(D_{0+}\lambda_1  + 4 j_{a+} B^{ab} [(g t_b g^{-1})_0 , \lambda_1])
\right)
\end{eqnarray}
can be rewritten as follows:
\begin{eqnarray}
{\cal L}_{w\lambda} & = & \mbox{Str}\left( w_{1+} \left( D_{0-}+
{1\over 2}(\iota(\Gamma)\iota(\Gamma)H_-) 
+ 4 D_0(B^{ab}[(gt_ag^{-1})_2, (gt_bg^{-1})_2])
\right) \lambda_3 \right) +
\nonumber
\\
& + & \mbox{Str}\left( w_{3-} \left( D_{0-} -
{1\over 2}(\iota(\Gamma)\iota(\Gamma)H_+) 
- 4 D_0(B^{ab}[(gt_ag^{-1})_2, (gt_bg^{-1})_2])
\right) \lambda_1 \right)
\end{eqnarray}
Therefore the term with the total derivative $D_0(B^{ab}[(gt_ag^{-1})_2, (gt_bg^{-1})_2])$
can be removed by the following infinitesimal field redefinition:
\begin{equation}
\begin{array}{lcl}
\delta\lambda_3 = [\Phi,\lambda_3] 
& , &
\delta w_{1+} = [\Phi,w_{1+}]
\cr
\delta\tilde{\lambda}_{1} = - [\Phi,\tilde{\lambda}_1]
& , &
\delta \tilde{w}_{3-} = - [\Phi,\tilde{w}_{3-}]
\end{array}
\end{equation}
where the parameter $\Phi$ is:
\begin{equation}\label{PhiIsBTimesCommutator}
\Phi = 4 B^{ab} [ (gt_ag^{-1})_2 , \; (gt_bg^{-1})_2 ] 
\end{equation}

\section{Relation to the description by Alday--Arutyunov--Frolov}
\label{sec:RelationToAAF}
Generally speaking, let us consider a theory with some action $S_0$
and symmetry algebra ${\bf g}$. Suppose that we act by the symmetry 
transformations with some constant parameters $\epsilon^a$. Then the action
is invariant:
\begin{equation}
S[\phi + \epsilon^a t_a \phi] = S[\phi]
\end{equation}
Now suppose that $\epsilon^a$ is not a constant but depends on 
$\tau^+$ and $\tau^-$. Then the variation of the action is:
\begin{equation}\label{NoetherTheorem}
S[\phi + \epsilon^at_a\phi] = S[\phi] 
+ \int\int d\tau^+ d\tau^- \left(
   j_{a+}\partial_-\epsilon^a 
   + j_{a-}\partial_+\epsilon^a \right)
\end{equation}
(This formula holds off-shell.) In particular, let us consider $\epsilon^a$
satisfying these equations:
\begin{eqnarray}
\partial_+\epsilon^a & = &   \frac{1}{ 2} B^{ab} j_{b+} 
\label{DPlusEpsilon}\\
\partial_-\epsilon^a & = & - \frac{1}{ 2} B^{ab} j_{b-}
\label{DMinusEpsilon}
\end{eqnarray}
Then Eq. (\ref{NoetherTheorem}) tells us that the change of 
variables $\phi\mapsto \tilde{\phi}$:
\begin{equation}
\phi = \tilde{\phi} + \epsilon^a t_a \tilde{\phi} + \ldots
\end{equation}
transforms $S$ into the deformed action:
\begin{equation}
S[\phi] = S[\tilde{\phi}] + \int\int d\tau^+ d\tau^- \;
B^{ab} j_{a+} j_{b-}
\end{equation}
When we solve for $\epsilon$ satisfying (\ref{DPlusEpsilon}) and
(\ref{DMinusEpsilon}) the resulting $\epsilon$ will not be
periodic in $\sigma$. Indeed, the deviation from the periodicity
will accumulate, and is given by the integral:
\begin{equation}
\int d\sigma B^{ab} j_{b\tau} = 
\epsilon^a(\tau,\sigma + 2\pi)  - \epsilon^a(\tau,\sigma)
\end{equation}
This results to the twisted boundary conditions
of \cite{Alday:2005ww}.

Note that we define $\epsilon$ by Eqs. (\ref{DPlusEpsilon}) and
(\ref{DMinusEpsilon}), but these equations are compatible only
on-shell. Indeed, the compatibility condition of 
(\ref{DPlusEpsilon}) and (\ref{DMinusEpsilon}) is:
\begin{equation}
\partial_-j_{a+} + \partial_+ j_{a-} = 0
\end{equation}
which is precisely the currents conservation, and only holds on-shell. 
Therefore, this is only limited to calculating the value of the
action on the classical solution.

Our approach, on the other hand, is not limited to the classical
configurations. 

\section{General relation between NSNS and RR fields
in the $\beta$-deformed solution}
\label{sec:RelationBetweenNSNSAndRR}
In this section we will discuss the relation between the NSNS and the
RR fields of the Maldacena-Lunin solution.  It will turn out that 
the RR 3-form $dC_2$ is in fact Hodge dual to the
$B^{NS}$:
\begin{equation}\label{HodgeDual}
B^{NS}_{ij}= c \; {\epsilon_{ij}}^{klm} \partial_k C_{lm}
\end{equation}
where $c$ is some coefficient, which is constant at the linearized
level, but becomes a function in the full nonlinear solution.

\subsection{At the linearized level}
We will start by discussing the relation (\ref{HodgeDual}) at the
first order in the deformation parameter.  Notice that Maldacena and
Lunin identify $\varphi_1$ and $\varphi_2$ as cyclic coordinates
corresponding to their two $U(1)$ symmetries; those two $U(1)$ act as
$\frac{\partial}{\partial \varphi_1}$ and
$\frac{\partial}{ \partial\varphi_2}$. On the other hand Eq. (3.2) of
their paper implies that in terms of $\phi_1$ and $\phi_2$ they act
as:
\begin{equation}
\frac{\partial }{ \partial \varphi_1} =
\frac{\partial }{ \partial \phi_2} - \frac{\partial}{\partial \phi_3}
\;,\;\;
\frac{\partial }{ \partial \varphi_2} =
- \frac{\partial }{ \partial \phi_1} + \frac{\partial }{ \partial \phi_2}
\end{equation}
For any vector field $v^{\mu}$, we can contract it with the metric 
and get a one form $g(v)$, defined as: $g(v)_{\mu} = g_{\mu\nu} v^{\nu}$. 
Then Eq. (3.11) from Maldacena-Lunin implies:
\begin{eqnarray}
g\left(\frac{\partial }{ \partial \varphi_1}\right) & = &
\mu_2^2 d\phi_2 - \mu_3^2 d\phi_3 
\\
g\left(\frac{\partial }{ \partial \varphi_2}\right) & = &
-\mu_1^2d\phi_1 + \mu_2^2 d\phi_2
\end{eqnarray}
Now we see:
\begin{equation}\label{OurBNS}
B^{NS}=\hat{\gamma}R^2\; g\left(\frac{\partial}{\partial\varphi_1}\right)
\wedge g\left(\frac{\partial}{\partial\varphi_2}\right)
\end{equation}
Then we can use that for any two vector fields $v$ and $u$:
\begin{equation}
* (g(v)\wedge g(u)) = \iota_v \iota_u \;(\mbox{volume-form})
\end{equation}
The volume form of $S^5$ in terms of  
$(\alpha,\theta,\psi,\varphi_1,\varphi_2)$ is:
\begin{equation}
s_{\alpha}^3 c_{\alpha} c_{\theta} s_{\theta} \;\;
d\alpha \wedge d\theta \wedge d\psi \wedge d\phi_1 \wedge d\phi_2
\end{equation}
This equation with our Eq. (\ref{OurBNS}) and equation for $C_2$ on p.9
of Maldacena and Lunin imply our Eq. (\ref{HodgeDual}).

\subsection{Exact relation for the full solution}
The deformed metric is:
\begin{multline}
ds^2_{\gamma} = d\alpha^2 + s^2_\alpha d\theta^2 + G(1 + 9\widehat{\gamma}^2 s_\alpha^4 c_\alpha^2 s_\theta^2 c_\theta^2)d\psi^2 + \\ 
+ G s^2_\alpha d\p^2 + G(s^2_\alpha c^2_\theta + c^2_\alpha) d\pp^2 + 2 G s^2_\alpha (c^2_\theta - s^2_\theta) d\psi d\p + \\
+ 2 G (s^2_\alpha c^2_\theta - c^2_\alpha)d\psi d\pp + 2 G s^2_\alpha c^2_\theta d\p d\pp
\end{multline}
where\be\label{G}
G = \frac{1}{1+\widehat{\gamma}^2(\mu_1^2\mu_2^2 + \mu_1^2\mu_3^2 + \mu_2^2\mu_3^2)} = \frac{1}{1+\widehat{\gamma}^2 s_\alpha^2(c_\alpha^2 + s_\alpha^2 s_\theta^2 c_\theta^2)}.
\ee

We may write it as:
\begin{multline}
ds^2_{\gamma} = d\alpha^2 + (\mu_2^2 + \mu_3^2) d\theta^2 + G(1 + 9\widehat{\gamma}^2 \mu_1^2\mu_2^2\mu_3^2) d\psi^2 + G (\mu_2^2 + \mu_3^2) d\p^2 + \\
+ G (\mu_1^2 + \mu_2^2) d\pp^2 + 2 G (\mu_2^2 - \mu_3^2) d\psi d\p + \\
+ 2 G (\mu_2^2 - \mu_1^2)d\psi d\pp + 2 G \mu_2^2 d\p d\pp.
\end{multline}

The determinant of the  metric above is 
\be
g_{\gamma} = 9\mu_1^2\mu_2^2\mu_3^2 G^2 (\mu_2^2 + \mu_3^2),
\ee
and the non-zero components of the inverse matrix are:
\bea
& g^{\alpha\alpha}_{\gamma} = 1 \nn
& g^{\theta\theta}_{\gamma} = \frac{1}{\mu_2^2 + \mu_3^2} \nn
& g^{\psi\psi}_{\gamma} = \frac{1}{9\mu_1^2\mu_2^2\mu_3^2} (\mu_1^2\mu_2^2 + \mu_1^2\mu_3^2 + \mu_2^2 \mu_3^2) \nn
& g^{\psi\p}_{\gamma} = \frac{1}{9\mu_1^2\mu_2^2\mu_3^2} (\mu_1^2\mu_3^2 + \mu_2^2 \mu_3^2 - 2\mu_1^2\mu_2^2) \nn
& g^{\psi\pp}_{\gamma} = \frac{1}{9\mu_1^2\mu_2^2\mu_3^2} (\mu_1^2\mu_2^2 + \mu_1^2\mu_3^2 - 2\mu_2^2 \mu_3^2) \nn
& g^{\p\p}_{\gamma} = (\mu_1^2 + \mu_2^2)\widehat{\gamma}^2 + \frac{\mu_1^2 + \mu_2^2 - (\mu_1^2 - \mu_2^2)^2}{9\mu_1^2\mu_2^2\mu_3^2}\nn
& g^{\pp\pp}_{\gamma} = (\mu_2^2 + \mu_3^2)\widehat{\gamma}^2 + \frac{\mu_2^2 + \mu_3^2 - (\mu_2^2 - \mu_3^2)^2}{9\mu_1^2\mu_2^2\mu_3^2}\nn
& g^{\p\pp}_{\gamma} = -\mu_2^2 \widehat{\gamma}^2 + \frac{\mu_2^4 + \mu_1^2\mu_2^2 - \mu_2^2 \mu_3^2 - \mu_2^2}{9\mu_1^2\mu_2^2\mu_3^2} \nonumber
\eea

The RR field strength $F_3=dC_2$ is given by:
\be
F_3 = \frac{1}{3!} F_{\mu\nu\rho} dx^\mu \wedge dx^\nu \wedge dx^\rho
\ee
where the only non-zero component is $F_{\alpha\theta\psi} = -12\gamma \pi N s^2_\alpha s_{2\alpha} s_{2\theta}$. Its Hodge dual on the deformed sphere $S^5_\gamma$ is:
\begin{multline}
{}^* F_3 = \sqrt{g_{\gamma}} g^{\alpha\alpha}_{\gamma} g^{\theta\theta}_{\gamma} F_{\alpha\theta\psi} \bigg( g^{\psi\psi}_{\gamma} d\p \wedge d\pp - g^{\p\psi}_{\gamma} d\psi \wedge d\pp - g^{\pp\psi}_{\gamma} d\p \wedge d\psi \bigg) \\
 = - 16\gamma\pi N G \bigg( (\mu_1^2\mu_2^2 + \mu_1^2\mu_3^2 + \mu_2^2\mu_3^2) d\p \wedge d\pp + \bigg.\\
\qquad\qquad\qquad + \bigg. (\mu_1^2 \mu_2^2 - 2\mu_2^2\mu_3^2 + \mu_3^2\mu_1^2) d\psi \wedge d\p + \bigg.\\
+ \bigg. (2\mu_1^2 \mu_2^2 - \mu_2^2\mu_3^2 - \mu_3^2\mu_1^2) d\psi \wedge d\pp \bigg).
\end{multline}

From $B_2 = c \ \ {}^*F_3$ and
\bea\label{b2}
B_2 = \widehat{\gamma} R^2 G & \bigg( & (\mu_1^2\mu_2^2 + \mu_1^2\mu_3^2 + \mu_2^2\mu_3^2) d\p \wedge d\pp + \bigg.\nn
&& + \bigg. (\mu_1^2 \mu_2^2 - 2\mu_2^2\mu_3^2 + \mu_3^2\mu_1^2) d\psi \wedge d\p + \bigg.\nn
&& + (2\mu_1^2 \mu_2^2 - \mu_2^2\mu_3^2 - \mu_3^2\mu_1^2) d\psi \wedge d\pp \bigg),
\eea
we get
\be\label{c}
c  = -\frac{R^4}{16\pi N}
\ee

\appendix

\section{Antisymmetric tensor product of two adjoint representations
of $su(4)$}
\label{sec:AppendixRepresentationTheory}
\subsection{As a representation of $su(4)$}
Consider the antisymmetric tensor product of two adjoint representations
of $su(4)$. The adjoint of $su(4)$ is the traceless part of the
tensor product of the fundamental and the complex conjugate of the
fundamental. The elements of the adjoint of $su(4)$ can be written
as $u^i_j$ where the upper index $i$ is the fundamental of $su(4)$, and
the lower index is the complex conjugate of the fundamental. The 
anti-hermiticity condition is:
\begin{equation}
\left(u^i_j\right)^* = - u_i^j
\end{equation}
and the traceless condition is $u^i_i=0$. 

The antisymmetric tensor product of two adjoint representation is a
subspace in the space of tensors $b^{ik}_{jl}$ such that:
\begin{equation}\label{AntisymmetryOfTwoAdjoint}
b^{ik}_{jl} = - b^{ki}_{lj}
\end{equation}
satisfying the trace condition:
\begin{equation}
b^{ik}_{il} = 0
\end{equation}
and the reality condition:
\begin{equation}\label{RealityOfTensor}
\left( b^{ik}_{jl} \right)^* = b^{jl}_{ik}
\end{equation}
The adjoint representation is a subspace:
\begin{equation}\label{AdjointSubspace}
\mbox{ad}\subset \mbox{ad}\wedge \mbox{ad} 
\end{equation}
consisting of the tensors of the following form:
\begin{equation}
b^{ik}_{jl} = \delta^i_l b^k_j - \delta^k_j b^i_l
\end{equation}
where $b^i_j$ can be chosen to satisfy: $b^i_i = 0$. To project on the orthogonal
subspace we impose the additional trace condition on $b$:
\begin{equation}\label{AdditionalTraceCondition}
b^{ip}_{pj} = 0
\end{equation}
Every tensor satisfying (\ref{AntisymmetryOfTwoAdjoint}) can be represented
in a unique way as a sum of a tensor antisymmetric in lower indices, and
symmetric in upper indices, and a tensor antisymmetric in upper indices, and
symmetric in lower indices:
\begin{eqnarray}
&& b^{ik}_{jl} = x^{ik}_{jl} + y^{ik}_{jl} \\
&& x^{ik}_{jl} = x^{ki}_{jl} = - x^{ik}_{lj} := b^{(ik)}_{jl}\\
&& y^{ik}_{jl} = - y^{ki}_{jl} = y^{ik}_{lj} := b^{[ik]}_{jl}
\end{eqnarray}
To project on the subspace orthogonal to (\ref{AdjointSubspace}) we
impose the constraint:
\begin{equation}\label{XYTraceless}
x^{pi}_{pj} = y^{pi}_{pj} = 0
\end{equation}
This is equivalent to the additional trace condition (\ref{AdditionalTraceCondition}) 
on $b$.
Moreover, the reality condition (\ref{RealityOfTensor}) relates $y$ to $x$:
\begin{equation}
y^{ik}_{jl} = \left( x^{jl}_{ik} \right)^*
\end{equation}
\paragraph     {Lemma:} There is a one-to-one correspondence between complex
tensors $x^{ik}_{jl}$ symmetric in the upper indices $ik$ and antisymmetric
in the lower indices $jl$, and tensors $b^{ik}_{jl}$ satisfying
the antisymmetry condition (\ref{AntisymmetryOfTwoAdjoint}) and the
{\em reality condition} (\ref{RealityOfTensor}).

Notice that there is no reality condition on $x^{ik}_{jl}$.
This means that the irreducible component in the 
 antisymmetric product of two adjoints can be
described in terms of the {\em complex} tensor $x^{ik}_{jl}$
satisfying the following symmetry and tracelessness conditions:
\begin{eqnarray}
&& x^{ik}_{jl} = x^{ki}_{jl} = - x^{ik}_{lj} 
\label{XSymmetry}
\\
&& x^{pi}_{pj} = 0
\label{XTracelessness}
\end{eqnarray}
There are 45 linearly independent (over ${\bf C}$) tensors $x^{ik}_{jl}$
satisfying (\ref{XSymmetry}) and (\ref{XTracelessness}), therefore
this representation is called $45_{\bf C}$. 

\paragraph     {Conclusion:} The antisymmetric tensor product of two adjoint
representations contains a 90-dimensional irreducible component, which is actually
defined over ${\bf C}$, and is denoted $45_{\bf C}$.

\subsection{As a representation of $u(3)\subset su(4)$}
Let us agree that the superindex of the representation symbol indicates
a representation of which algebra we are considering. So, the $45_{\bf C}$ of
$su(4)$ will be denoted:
\[
45_{\bf C}^{su(4)}
\]
This representation splits into several irreducible representations
of $u(3)\subset su(4)$. For example, there is a representation
$6_{\bf C}^{su(3)}$ which is realized on the symmetric complex tensors 
$u^{ijk}$ where the indices $i,j,k$ enumerate the fundamental representation
of $u(3)$.
We observe:
\begin{equation}
6_{\bf C}^{u(3)}\subset 45_{\bf C}^{su(4)}
\end{equation}
In terms of $x^{ik}_{jl}$ this is:
\begin{equation}
x^{ik}_{jl} = u^{ikp}\epsilon_{pjl}
\end{equation}
\section{BRST operator in the near flat space expansion}
\label{sec:NearFlatBRST}
We will use the ``most symmetric'' gauge:
\begin{equation}\label{MostSymmetricGauge}
g=e^X
\end{equation}
where $X=x_2+\theta_3+\theta_1$. The BRST operator acts on $g$ in the following way:
\begin{equation}
\epsilon Q g = (\epsilon\lambda_3 + \epsilon\lambda_1) g =
\left(e^{X + \epsilon Q X} - e^X\right) + \omega_0(\epsilon) e^X
\end{equation}
--- this equation is the definition of $\epsilon Q X$ and $\omega_0(\epsilon)$.
\begin{equation}
\epsilon\lambda - \omega_0(\epsilon) = 
{e^{\mbox{ad}(X)}-1\over \mbox{ad}(X)} \epsilon QX
\end{equation}
This gives us the following recursive recursive formulas for $QX$ and $\omega_0$:
\begin{align}
\epsilon QX = & \; 
\epsilon \lambda - {1\over 2} [X , \epsilon QX]_{\perp} 
- {1\over 6} [X, [X, \epsilon QX]]_{\perp} 
- {1\over 24} [X, [X, [X, \epsilon QX]]]_{\perp}
-\ldots
\\
\omega_0(\epsilon) = & \; \phantom{\epsilon \lambda}
- {1\over 2} [X , \epsilon QX]_0 
- {1\over 6} [X, [X, \epsilon QX]]_0 
- {1\over 24} [X, [X, [X, \epsilon QX]]]_0
- \ldots
\end{align}
This gives us:
\begin{align}
\epsilon QX = & \; \epsilon \lambda - {1\over 2} [X,\epsilon\lambda]_{\perp} +
\nonumber
\\
& + {1\over 12} [ X, [ X, \epsilon\lambda ]_{\perp} ]_{\perp}
  - {1\over  6} [ X, [ X, \epsilon\lambda ]_0 ]_{\perp} +
\\
& + {1\over 24} \Big(
   [ X, [ X, [ X, \epsilon\lambda ]_0 ]_{\perp} ]_{\perp} 
   + [ X, [ X, [ X, \epsilon\lambda ]_{\perp} ]_0 ]_{\perp} 
   \Big) + \ldots
\nonumber
\end{align}
\remv{Calculation}\rem{(start-process "mirage" "*scratch*" "mirage" 
"photos/Qx.jpg" )}
We observe that to preserve the gauge (\ref{MostSymmetricGauge}) we need
to combine the action of $Q_{BRST}$ with the gauge transformation
with the parameter $-\omega_0$:
\begin{equation}
\delta_{\epsilon}g = -\omega_0(\epsilon) g
\end{equation}
We will include this ``compensating'' gauge transformation into the definition
of $Q_{BRST}$. Then, in particular, our ``combined'' BRST transformation 
{\em acts on the pure spinor}:
\begin{equation}
\epsilon Q \epsilon'\lambda = [\epsilon'\lambda,\omega_0(\epsilon)] 
\end{equation}

\section{Proof of a technical Lemma used in
Section \ref{sec:HarmlessConjecture}}
\label{sec:AppendixProofOfLemma}
Here we will prove that any tensor of the form $A^{p[a} {f_p}^{bc]}$ should
necessarily have some fermionic indices. It is useful to write this in the matrix
notations. Let $i,j,\ldots$ denote the indices of the upper left square (the
fundamental representation of $su(2,2)$), and $\alpha,\beta,\ldots$ denote the
indices of the lower right square (the fundamental representation of
$su(4)$). Then the generators $E^i_j$ and $E^{\alpha}_{\beta}$ are bosonic, and
the generators $E^i_{\alpha}$ and $E^{\alpha}_j$ are fermionic. We want to see if
there exists $A^{pa}$ such that $Q(A^{pa} \bar{c}_p c_a) = A^{pa} {f_p}^{bc} c_a c_b
c_c$ contains 
only $c^i_j$ and $c^{\alpha}_{\beta}$ but not $c^i_{\alpha}$ and $c^{\alpha}_i$. In matrix notations:
\begin{equation}
A^{pa} \bar{c}_p c_a = A{}^i_j{}^k_l \bar{c}_i^j c_k^l
+ A{}^i_{\alpha}{}^{\beta}_k \bar{c}_i^{\alpha} c_{\beta}^k + \ldots
\end{equation}
Then:
\begin{align}
Q(A^{pa}\bar{c}_p c_a)|_{\rm mixed} = \quad &
A{}_i^j{}^l_k c^i_{\afbox{\alpha}} c^{\afbox{\alpha}}_j c^k_l + 
A{}_{\alpha}^{\beta}{}_k^l c^{\alpha}_{\afbox{m}} c^{\afbox{m}}_{\beta} c^k_l +
\label{QccLine1}
\\[1pt]
+ & 
A{}_i^j{}^{\beta}_{\alpha} c^i_{\afbox{\gamma}} c^{\afbox{\gamma}}_j c^{\alpha}_{\beta} +
A{}^{\delta}_{\gamma}{}^{\beta}_{\alpha} 
  c^{\gamma}_{\afbox{m}} c^{\afbox{m}}_{\delta} c^{\alpha}_{\beta} +
\label{QccLine2}
\\[1pt]
+ &
A{}^{\alpha}_i{}^{\beta}_j 
  (c^i_{\afbox{\gamma}}c^{\afbox{\gamma}}_{\alpha}c^j_{\beta} +
   c^i_{\afbox{m}}c^{\afbox{m}}_{\alpha} c^j_{\beta}) +
A{}^i_{\alpha}{}^j_{\beta} 
  (c^{\alpha}_{\afbox{m}}c^{\afbox{m}}_i c^{\beta}_j + 
   c^{\alpha}_{\afbox{\gamma}}c^{\afbox{\gamma}}_i c^{\beta}_j) +
\label{QccLine3}
\\[1pt]
+ &
A{}^i_{\alpha}{}^{\beta}_j 
  (c^{\alpha}_{\afbox{m}} c^{\afbox{m}}_i c^j_{\beta} +
   c^{\alpha}_{\afbox{\gamma}} c^{\afbox{\gamma}}_i c^j_{\beta}) +
A{}^{\alpha}_i{}^j_{\beta}
  (c^i_{\afbox{m}}c^{\afbox{m}}_{\alpha} c^{\beta}_j +
   c^i_{\afbox{\gamma}}c^{\afbox{\gamma}}_{\alpha} c^{\beta}_j)
\end{align}
where we put boxes around the summation indices, just to make them clearly
visible. We require that these mixed terms all cancel. 
Because of the different tensor structure of different terms, this actually
implies  separate cancellations:
\begin{align}
0 = &
A{}_i^j{}^l_k c^i_{\afbox{\alpha}} c^{\afbox{\alpha}}_j c^k_l + 
A{}_{\alpha}^{\beta}{}_k^l c^{\alpha}_{\afbox{m}} c^{\afbox{m}}_{\beta} c^k_l +
A{}^i_{\alpha}{}^{\beta}_j c^{\alpha}_{\afbox{m}} c^{\afbox{m}}_i c^j_{\beta} +
A{}^{\alpha}_i{}^j_{\beta} c^i_{\afbox{m}}c^{\afbox{m}}_{\alpha} c^{\beta}_j 
\label{AZeroLine1}
\\[5pt]
0 = &
A{}_i^j{}^{\beta}_{\alpha} c^i_{\afbox{\gamma}} c^{\afbox{\gamma}}_j c^{\alpha}_{\beta} +
A{}^{\delta}_{\gamma}{}^{\beta}_{\alpha} 
  c^{\gamma}_{\afbox{m}} c^{\afbox{m}}_{\delta} c^{\alpha}_{\beta} +
A{}^i_{\alpha}{}^{\beta}_j 
   c^{\alpha}_{\afbox{\gamma}} c^{\afbox{\gamma}}_i c^j_{\beta} +
A{}^{\alpha}_i{}^j_{\beta}
   c^i_{\afbox{\gamma}}c^{\afbox{\gamma}}_{\alpha} c^{\beta}_j
\label{AZeroLine2}
\\[5pt]
0 = &
A{}^{\alpha}_i{}^{\beta}_j 
c^i_{\afbox{\gamma}}c^{\afbox{\gamma}}_{\alpha}c^j_{\beta} 
\label{AZeroLine3}
\\[5pt]
0 = &
A{}^{\alpha}_i{}^{\beta}_j 
c^i_{\afbox{m}}c^{\afbox{m}}_{\alpha} c^j_{\beta}
\label{AZeroLine4}
\\[5pt]
0 = &
A{}^i_{\alpha}{}^j_{\beta} 
c^{\alpha}_{\afbox{m}}c^{\afbox{m}}_i c^{\beta}_j 
\label{AZeroLine5}
\\[5pt]
0 = &
A{}^i_{\alpha}{}^j_{\beta} 
   c^{\alpha}_{\afbox{\gamma}}c^{\afbox{\gamma}}_i c^{\beta}_j
\label{AZeroLine6}
\end{align}
Eqs. (\ref{AZeroLine3}) --- (\ref{AZeroLine6})  imply that:
\begin{equation}\label{AUpDownZero}
A{}^{\alpha}_i{}^{\beta}_j 
=
A{}^{\alpha}_i{}^{\beta}_j 
= 
A{}^i_{\alpha}{}^j_{\beta} 
=
A{}^i_{\alpha}{}^j_{\beta} = 0
\end{equation}
For example, $A^{\alpha\beta}_{ij}=0$ can be proven as follows. We can think
$A^{\alpha\beta}_{ij}=0$ as a tensor in $L_{dn}\otimes L_{dn} \otimes
L_{up}'\otimes L_{up}'$. Suppose that we can
find a nonzero $A$ such that (\ref{AZeroLine3}) is satisfied. 
The space of such $A$ is a subspace in 
$L_{dn}\otimes L_{dn} \otimes L_{up}'\otimes L_{up}'$ closed under the
action of $su(2,2)\oplus su(4)$. But in fact there are only four
such invariant subspaces:
\begin{equation}
\Lambda^2 L_{dn}\otimes \Lambda^2 L_{up}' \;,\;\;
S^2 L_{dn}\otimes \Lambda^2 L_{up}' \;,\;\;
\Lambda^2 L_{dn}\otimes S^2 L_{up}' \;,\;\;
S^2 L_{dn}\otimes S^2 L_{up}' 
\end{equation}
and we can show by a direct examination that neither of these
solves (\ref{AZeroLine3}). For example the first one corresponds
to considering the antisymmetrization of $ij$ and $\alpha\beta$: 
\[
c^{[i|}_{\afbox{\gamma}}c^{\afbox{\gamma}}_{[\alpha}c^{|j]}_{\beta]}
\]
which does not give zero. The other three possibilities ({\it e.g.}
symmetrizing $ij$ and antisymmetrizing $\alpha\beta$ also do not
give zero); therefore the only solution to (\ref{AZeroLine3}) is 
$A^{\alpha\beta}_{ij}=0$. The other three identities in (\ref{AUpDownZero}) can be
proven in a similar way. 

The analysis of (\ref{AZeroLine1}) and (\ref{AZeroLine2}) is slightly more
complicated, because besides symmetrization/antisymmetrization one can
also use the Kronecker  $\delta^i_j$ and $\delta^{\alpha}_{\beta}$. Just symmetrization/antisymmetrization
does not work, but there are nonzero solutions for $A$ involving the
Kronecker delta. Indeed, let us write the most general ansatze with the
Kronecker delta:
\begin{eqnarray}\label{AnsatzeForAcc}
A^{pa}\bar{c}_p c_a & = & 
  \mbox{tr} (\bar{c}^u_ux^u_uc^u_u) +
  \mbox{tr} (c^u_u \tilde{x}^u_u \bar{c}^u_u) +
\nonumber \\
& + & 
  \mbox{tr} (\bar{c}^u_dy^d_dc^d_u) +
  \mbox{tr} (c^u_d \tilde{y}^d_d \bar{c}^d_u) +
\nonumber \\
& + & 
  \mbox{tr} (\bar{c}^d_u z^u_u c^u_d) +
  \mbox{tr} (c^d_u \tilde{z}^u_u \bar{c}^u_d) +
\nonumber \\
& + &
  \mbox{tr} (\bar{c}^d_d w^d_d c^d_d) +
  \mbox{tr} (c^d_d \tilde{w}^d_d \bar{c}^d_d)
\end{eqnarray}
where we introduced the $4\times 4$-matrices of ghosts:
\begin{equation}
c^u_u = \sum\limits_{i,j=1}^4 c^i_j E^j_i  \;\; ,\;\;\;
c^u_d = \sum\limits_{i,\alpha=1}^4 c^i_{\alpha} E^{\alpha}_i \;\;,\;\;
\mbox{etc.}
\end{equation}
and $x,y,z,w,\tilde{x},\tilde{y},\tilde{z},\tilde{w}$ are some coefficients
To exemplify our notations, we write explicitly some terms:
\begin{eqnarray}
  \mbox{tr} (\bar{c}^u_ux^u_uc^u_u) & = &
\bar{c}^i_j x^j_k c^k_i \nonumber \\[3pt]
  \mbox{tr} (\bar{c}^d_u z^u_u c^u_d) & = &
\bar{c}^{\alpha}_i z^i_j c^j_{\alpha}
\end{eqnarray}
Let us impose the condition that $QA^{pa}\bar{c}_p c_a$ does not contain any
fermionic ghosts.  This leads to the following equations:
\begin{eqnarray}
\mbox{from } c^u_u c^u_d c^d_u: && x^u_u + \tilde{z}^u_u  = 0
\nonumber
\\ 
\mbox{from } c^u_d c^d_u c^u_u: && \tilde{x}^u_u - z^u_u  = 0
\nonumber
\\ 
\mbox{from } c^d_u c^u_u c^u_d: && y^d_d - \tilde{y}^d_d  = 0
\nonumber
\\ 
\mbox{from } c^d_u c^u_d c^d_d: && -y^d_d - \tilde{w}^d_d = 0
\nonumber
\\ 
\mbox{from } c^d_d c^d_u d^u_d: && \tilde{y}^d_d + w^d_d  = 0
\nonumber
\\ 
\mbox{from } c^u_d c^d_d c^d_u: && z^u_u - \tilde{z}^u_u  = 0
\end{eqnarray}
This means that (\ref{AnsatzeForAcc}) collapses to this:
\begin{align}
&
\mbox{tr}\left( 
   x^u_u c^u_u \bar{c}^u_u + x^u_u \bar{c}^u_u c^u_u + 
   x^u_u \bar{c}^u_d c^d_u + x^u_u c^u_d \bar{c}^d_u \right) +
\label{xcc}
\\ 
+ &
\mbox{tr}\left( 
   y^d_d c^d_d \bar{c}^d_d + y^d_d \bar{c}^d_d c^d_d + 
   y^d_d \bar{c}^d_u c^u_d + y^d_d c^d_u \bar{c}^u_d \right)
\label{ycc}
\end{align}
and one can see that $Q$ of this is actually zero. (Not only $Q$ of
(\ref{xcc}) and (\ref{ycc}) does not contain $c^u_d$ and $c^d_u$, but
actually it is just 0.)  This proves the {\bf Lemma}.

\section*{Acknowledgments}
We want to thank Y.~Aisaka, N.J.~Berkovits, A.~Losev, L.~Mazzucato and
A.~Tseytlin for many useful discussions.
O.A.B. would like to thank FAPESP grant 09/08893-9 for financial support,
as well as the Aspen Center of Physics for hospitality during the workshop
``Unity of String Theory''.
The research of L.I.B. is supported by CNPq Grant No. 141546/2006-9.
The research of A.M. was supported by the Sherman Fairchild 
Fellowship and by the
World Premier International Research Center Initiative 
(WPI Initiative), MEXT, Japan, and in part
by the RFBR Grant No. 10-02-01315.
The work of V.O.R. is supported 
by CNPq grant No. 304495/2007-7, FAPESP grant
No. 2008/05343-5 and PROSUL grant No. 490134/2006-8.


\providecommand{\href}[2]{#2}\begingroup\raggedright\endgroup

\end{document}